%% file: main.tex
\documentclass[
reprint
nofootinbib,
amsmath,amssymb,
aps,
prx,
floatfix,
longbibliography,
author-year      
]{revtex4-2} 

\usepackage{pgfplots}
\pgfplotsset{compat=1.18} 

\input{packages}

\input{commands}

\newcommand{\laguerre}{l}
\newcommand{\hermite}{m}
\newcommand{\nhermite}{N_{\hermite}}
\newcommand{\nlaguerre}{N_{\laguerre}}
\newcommand{\ntheta}{N_\parallel}

\newcommand{\Dhyper}{D_{\mathrm{hyper}}} 
\newcommand{\lref}{L_\text{ref}}
\newcommand{\outero}{\mathrm{o}}
\newcommand{\ompar}{\omega_\parallel}
\newcommand{\omlin}{\gamma_{k_y}}
\newcommand{\omlinouter}{\gamma_{k_y^{\outero}}}
\newcommand{\tnl}{t_{\mathrm{nl}}}
\newcommand{\aspect}{\mathcal{A}}
\newcommand{\aspecto}{\mathcal{A}^{\outero}}
\newcommand{\Lpar}{L_\parallel}
\newcommand{\dphipotamp}{\delta \phipot_{k_y}}
\newcommand{\dphipotampouter}{\delta \phipot_{k_y^{\outero}}}
\newcommand{\kparmin}{\kpar^{\mathrm{min}}}
\newcommand{\curvature}{\mathcal{C}}
\newcommand{\npol}{N_{\mathrm{pol}}}
\newcommand{\corr}{G}
\newcommand{\Lparcorr}{{\Lpar^{\text{corr}}}}

\newcommand{\Lpareff}{{\Lpar^{\text{eff}}}}
\newcommand{\Rzonal}{{R_\text{zonal}}}
\newcommand{\nfp}{N_{\text{fp}}}

\renewcommand{\Fos}{F_\s}
\renewcommand{\qs}{Z_\s e}

\newcommand{\figscale}{0.92} 
\graphicspath{{figures/}}

\begin{document}
	
\title[]{Asymptotic scaling theory of electrostatic turbulent transport in magnetised fusion plasmas}

\author{T. Adkins}
\email{tadkins@pppl.gov} 
\affiliation{
	Princeton Plasma Physics Laboratory, Princeton, New Jersey, 08540, USA
}

\author{I.~G.~Abel}
\affiliation{
	Department of Physics, University of Maryland, College Park, Maryland, 20742, USA
}

\author{M.~Barnes}
\affiliation{
	Rudolf Peierls Centre for Theoretical Physics, University of Oxford, Oxford, OX1 3PU, UK
}

\author{S.~Buller}
\affiliation{
	Department of Astrophysical Sciences, Princeton University, Princeton, New Jersey, 08543, USA
}


\author{W.~Dorland}
\email{Deceased Sept. 22, 2024} 
\affiliation{
	Department of Physics, University of Maryland, College Park, Maryland, 20742, USA
}	
\affiliation{
	Institute for Research in Electronics and Applied Physics, Energy Research Facility, 8279 Paint Branch Drive, College Park, MD 20742
}


\author{P.~G.~Ivanov}
\affiliation{
	United Kingdom Atomic Energy Authority, Culham Campus, Abingdon, OX14 3DB, UK
}

\author{R.~Meyrand}
\affiliation{
	Department of Physics \& Astronomy, University of New Hampshire, Durham, New Hampshire, 03824, USA
}

\author{F.~I.~Parra}
\affiliation{
	Princeton Plasma Physics Laboratory, Princeton, New Jersey, 08540, USA
}
	
\author{R.~Nies}
\affiliation{
	Rudolf Peierls Centre for Theoretical Physics, University of Oxford, Oxford, OX1 3PU, UK
}

\author{A.~A.~Schekochihin}
\affiliation{
	Rudolf Peierls Centre for Theoretical Physics, University of Oxford, Oxford, OX1 3PU, UK
}

\author{J.~Squire}
\affiliation{
	Department of Physics, University of Otago, Dunedin, 9016, New Zealand
}

\date{\today}

\begin{abstract}
Turbulent transport remains one of the principal obstacles to achieving efficient magnetic confinement in fusion devices. Two of the dominant drivers of the turbulence are microscale instabilities fuelled by electron- and ion-temperature gradients (ETG and ITG), whose nonlinear saturation determines the cross-field transport of particles and energy. We present a simple asymptotic scaling theory that unifies ETG- and ITG-driven turbulence within a common framework. By balancing the fundamental time scales of linear free energy injection, nonlinear decorrelation, and parallel propagation, the theory isolates the dependence of the heat flux on equilibrium parameters to two key quantities: the parallel system scale and the outer-scale aspect ratio. We show that these quantities encapsulate the essential physics of saturation, leading to distinct predictions for ETG and ITG transport: a cubic scaling with the temperature gradient in the electron channel, and a linear scaling in the ion channel, the latter recovering the ITG scaling recently proposed and numerically verified in axisymmetric geometry by Nies et al. [Phys. Rev. Res. 8, 013295 (2026)]. Extensive nonlinear gyrokinetic simulations confirm that these theoretical predictions in fact hold irrespective of the magnetic geometry (slab, tokamak, or stellarator), including the first numerical confirmation of the cubic ETG scaling anticipated by earlier studies. Our theory depends only on the parallel system scale and the outer-scale aspect ratio, and hence provides a physics-based foundation for fast, geometry-aware transport models, offering a pathway toward reactor optimisation in both tokamaks and stellarators.
\end{abstract}

\maketitle

\newpage






	
\section{Introduction}
\label{sec:introduction}
Magnetic-confinement-fusion devices such as tokamaks and stellarators aim to sustain plasmas at temperatures and densities high enough for the energy produced by fusion reactions to outpace the losses. The dominant source of these losses comes from turbulence --- a chaotic, nonlinear ensemble of Larmor-scale fluctuations driven by the gradients of the plasma density and temperature --- that transports particles and heat across the confining magnetic field. Since the overall power balance, and thus viability, of the device is governed by this turbulent transport, the design of future fusion reactors needs reliable, predictive, physics-based models that connect the parameters of the plasma and magnetic equilibria to the resulting turbulent fluxes of particle and energy. Today's most accurate tools, first-principles gyrokinetic simulations \citep{frieman82,abel13,catto19,kotschenreuther95GS2,jenko00GENE,candy16,mandell24}, are computationally intensive, while widely used reduced models (e.g., \texttt{TGLF} \cite{staebler07,staebler10} or \texttt{QuaLiKiz} \cite{bourdelle07,bourdelle16}) rely on closures and fits that limit extrapolation to the novel parameter regimes that future fusion reactors will inhabit. This motivates a complementary approach: the development of simple yet fundamental scaling theories that distil the essential mechanisms that determine the steady-state turbulent transport, thereby providing reliable guidance for reactor design.

A large part of the energy losses is due to the turbulence driven by electrostatic instabilities associated with either the electron-temperature gradient (ETG) \citep{liu71,lee87,dorland00,jenko00} or the ion-temperature gradient (ITG) \citep{waltz88,cowley91,kotschenreuther95}, which typically live on scales comparable to the electron and ion Larmor radii, respectively. Though it has been long understood that ITG-driven turbulence typically dominates core transport (see references cited above), ETG-driven turbulence can become significant when ion-scale modes become suppressed by, e.g., strong $\exb$ shear in the steep-gradient regions of a tokamak (e.g., the pedestal), particularly in spherical or low-aspect-ratio configurations \citep[see][and references therein]{roach05,roach09,guttenfelder13,guttenfelder21,ren17}. As such, developing a better understanding of the turbulent transport driven by both ETG and ITG is a key ingredient in the successful design of future reactors. Despite their similarity in terms of linear physics, previous theoretical \citep{barnes11,adkins22,nies26} and experimental \citep{ghim13} studies suggest that the nonlinear saturation of ETG and ITG can differ significantly, meaning that the resulting transport levels and their scaling with equilibrium parameters certainly cannot be inferred from linear theory alone. Understanding these distinct saturation pathways is essential for constructing reduced, predictive models of turbulent transport.

In this paper, we develop an asymptotic scaling theory for the dependence of both ETG- and ITG-driven turbulent transport on parameters of the plasma equilibrium and the magnetic geometry when driven far above marginality. The final scaling estimates arise from a simple competition between three fundamental timescales associated with gyrokinetic fluctuations: the rates of linear free energy injection, nonlinear decorrelation, and parallel propagation (along the magnetic field). This balance reduces the problem to determining two key quantities: the parallel system scale $\Lpar$ and the perpendicular aspect ratio of the most energetic (`outer-scale') fluctuations $\aspecto$ (defined here as the ratio of their characteristic radial wavenumber to their binormal one). The former is effectively a measure of the parallel correlation length of the turbulence, and is typically determined by the geometry of the confining field. Crucially, the outer-scale aspect ratio is set by the nature of the adiabatic response of the non-driving species, leading to fundamentally different predictions for ETG- and ITG-driven turbulence. In particular, we show that while the unmodified adiabatic ion response on electron scales leads to $\aspecto$ being independent of equilibrium parameters for the ETG case, the modified adiabatic electron response on ion scales causes $\aspecto$ to scale linearly with the ion-temperature gradient. 

For this latter scaling, we draw heavily on a recent theory of ITG-driven turbulence developed by \cite{nies26} for axisymmetric equilibria. In that work, the authors identified a nonlinear saturation mechanism of the turbulence mediated by the so-called `toroidal secondary mode', a novel propagating zonal mode \cite{nies26tsm}. They showed that above some critical fluctuation amplitude, small-scale zonal modes become unstable due to this toroidal secondary and shear apart large-scale turbulent eddies. The resulting balance between the nonlinear decorrelation rate at the outer scale and the radial magnetic-drift timescale imposes a constraint on the radial extent of the turbulence, yielding the aforementioned linear scaling of the outer-scale aspect ratio with the ion temperature gradient. The present work adopts this physical picture as the appropriate saturation mechanism for ITG-driven turbulence, embedding it within a broader asymptotic scaling framework that treats it simultaneously with its ETG-driven counterpart. In this context, the constraint on the turbulence imposed by the toroidal secondary mode naturally enters as a condition on the outer-scale aspect ratio in the ITG case; this condition is absent in the ETG one due to the different (unmodified) adiabatic response.

These differences in the scaling of the outer-scale aspect ratio result in ETG-driven turbulence displaying significantly stiffer transport \cite{wolf03} than its ITG counterpart, the turbulent heat flux scaling cubically with the temperature gradient, as opposed to linearly for ITG (as was also shown by \citep{nies26}). All predictions of our scaling theory are verified against extensive nonlinear gyrokinetic simulations in straight, axisymmetric, and non-axisymmetric magnetic-field geometries, showing remarkable agreement with the numerical results, which also represent the first confirmation of the cubic ETG scaling with the temperature gradient, as predicted by previous theoretical studies \cite{adkins22,adkins23thesis,adkins23}. By distilling the nonlinear saturation physics of ITG and ETG turbulence down to the two parameters $\Lpar$ and $\aspecto$, our approach unifies both regimes within a single theoretical framework. The result is a reduced but interpretable theory that captures the correct asymptotic scalings, providing a physics-based foundation for predictive models. Such models are particularly valuable in reactor design and stellarator optimisation, where scanning across large spaces of magnetic geometries and equilibrium profiles demands fast yet reliable tools informed by first-principles theory.

The remainder of this paper is organised as follows. In \cref{sec:asymptotic_scaling_theory}, we present the asymptotic scaling theory that is the focus of this paper, deriving predictions for the scaling of the heat flux in ETG- and ITG-driven turbulence from a simple balance of timescales. Its implications for slab (\cref{sec:slab_geometry}), axisymmetric (\cref{sec:axisymmetric_geometry}), and non-axisymmetric (\cref{sec:non_axisymmetric_geometry}) geometries are then considered, with the predictions tested, successfully, by a vast array of numerical simulations. We summarise our findings in \cref{sec:summary_and_discussion} and discuss various lines of investigation left open by this work in \cref{sec:open_issues}. Technical details of the numerical simulations are given in \cref{app:numerical_details}, alongside supplementary data in \cref{app:supplementary_data}.

\section{Asymptotic scaling theory}
\label{sec:asymptotic_scaling_theory}
In this section, we develop the theoretical framework wherein we derive the asymptotic scaling theory for the dependence of the turbulent transport on equilibrium parameters. After introducing electrostatic gyrokinetics (\cref{sec:electrostatic_gyrokinetics}) and the local flux-tube limit (\cref{sec:local_limit}), we consider the various time scales appearing in the gyrokinetic equation (\cref{sec:timescales}) before demonstrating how straightforward physical arguments about the nature of the competition between these time scales can be leveraged to determine the properties of the saturated state (\crefrange{sec:outer_scale_dynamics}{sec:heatflux_scaling}). Note that throughout \cref{sec:asymptotic_scaling_theory}, we consider turbulence in plasma models containing a single kinetic species (either electrons or ions). The multi-species dynamics that can arise when electron- and ion-scale fluctuations are coupled are discussed further in \cref{sec:cross_scale_interactions}. In \cref{sec:adiabatic_responses}, we also highlight the contrasting adiabatic responses seen on electron and ion scales, since these differences are responsible for the distinct saturation physics of ETG and ITG turbulence.

\subsection{Electrostatic gyrokinetics}
\label{sec:electrostatic_gyrokinetics}
We consider magnetised-plasma turbulence driven by gradients in the equilibrium density $\dens$ and temperature $\Ts$ of species $\s$ in the presence of a mean magnetic field $\vB$. The turbulent fluctuations are assumed to satisfy the usual gyrokinetic ordering \citep{frieman82,abel13,catto19}:
\begin{align}
	\kperp \rhos \sim \kpar L \sim 1, \quad \frac{\omega}{\Omegas} \sim \frac{\rhos}{L} \equiv \rhostars \ll 1,
	\label{eq:gyrokinetic_ordering}
\end{align}
where $\kperp$ and $\kpar$ are characteristic wavenumbers perpendicular and parallel to $\ub = \vB/|\vB|$, respectively, $\rhos$ and $\Omegas$ are the Larmor radius and frequency of species $\s$, $\omega$ is the inverse time scale associated with the fluctuations, and $L$ is some characteristic length scale of variation of the plasma equilibrium (e.g., of $\dens$, $\Ts$, or $\vB$). We assume that there exist well-defined flux-surfaces labelled by $\psi$ (later taken to be either the poloidal or toroidal flux), allowing us to write the mean magnetic field in Clebsch form $\vB = \grad{\alpha} \times \grad \psi$, with $\alpha$ the field-line label. In this case, it can be shown that, in the absence of sonic flows \cite{abel13}, the plasma equilibrium is, to lowest order in $\rhostars$, a function only of $\psi$, viz., $\dens = \dens(\psi)$ and $\Ts = \Ts(\psi)$. Given that these equilibrium quantities evolve on a slow transport time scale $\tau_E \sim {(\rhostars^2 \omega)^{-1}}$, we treat them as constant over the time scales of the turbulent fluctuations. Note that while we will not consider the effects of any equilibrium flows in our analysis (neglecting, in particular, both equilibrium flow shear and finite-Mach-number rotational drifts), interested readers may find a prescription for modifying our findings in the presence of flow shear in, e.g., \cite{kinsey05,waltz07,casson09,barnes11,mcmillan19,ivanov25}.

If the plasma beta $\betas = 8\pi \dens \Ts/B^2$ is sufficiently small for fluctuations in the magnetic field to be neglected --- we refer to this as the electrostatic limit
--- the only electromagnetic field that participates in the dynamics is the perturbed electrostatic potential $\dphipot$. Then, the non-adiabatic distribution function $\hs = \hs (\vRs, \energy,\mus, t)$, related to the perturbed distribution function by $\dfs = - (\qs \dphipot/\Ts)\Fos + \hs$, evolves according to the gyrokinetic equation \citep{frieman82,abel13,catto19}:
\begin{align}
	\frac{\partial}{\partial t}\left( \hs - \frac{\qs \avgRs{\dphipot}}{\Ts} \Fos\right) + ( \vpar \ub + \vds + \avgRs{\vexb}) \cdot \grad \hs + \avgRs{\vexb} \cdot \grad \Fos = \sum_{\s'} \avgRs{C_{\s\s'}},
	\label{eq:gyrokinetic_equation}
\end{align}
in which $\vRs = \vr - \ub \times \vv/\Omegas$ is the gyrocentre position, $\energy = \ms v^2/2$ is the particle kinetic energy, $\mus = \ms \vperp^2 /2B$ is the magnetic moment, $Z_s e$ and $\Fos$ are the charge and equilibrium Maxwellian distribution of species $\s$, \mbox{$\vds = (\ub/2\Omegas) \times (2 \vpar^2 \ub \cdot \grad \ub + \vperp^2 \grad \log B)$} are the magnetic drifts, \mbox{$\vexb = (c/B) \ub \times \grad \dphipot$} is the perturbed $\exb$ drift, and $\avgRsinline{\dots}$ is the standard gyroaverage at constant gyrocentre position. The term on the right-hand side is the operator encoding the effect of collisions between species $\s$ and $\s'$ on $\hs$. Throughout this paper, we will be concerned with collisionless (or nearly collisionless) dynamics, and so we need not quote a specific form of the collision operator here. The electrostatic gyrokinetic equation \cref{eq:gyrokinetic_equation} is closed by the quasineutrality condition
\begin{align}
	\sum_\s \frac{(\qs)^2 \dens}{\Ts} \dphipot = \sum_\s \qs \int \rmd^3 \vec{v} \avgr{\hs},
	\label{eq:quasineutrality}
\end{align}
where $\avgrinline{\dots}$ denotes the gyroaverage at constant particle position $\vr$.

\subsection{Local flux-tube limit}
\label{sec:local_limit}
\begin{figure*}
	\centering
	\includegraphics[width=0.5\textwidth]{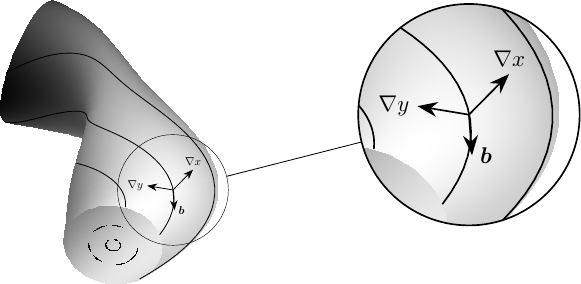}
	
	\caption[]{An illustration of the field-aligned coordinate system $(x, y, z)$ adopted in this work. The flux tube follows a magnetic field line on a given magnetic surface, and is assumed to be infinitesimal in perpendicular extent compared to the equilibrium scales of the plasma and magnetic field. The vectors $\grad x$ and $\grad y$ lie in the plane perpendicular to the magnetic field direction $\ub = \vB/|\vB|$ and are not, in general, orthogonal to one another, i.e., $\grad x \cdot \grad y \neq 0$.}
	\label{fig:flux_tube_geometry}
\end{figure*}

In what follows, we restrict our consideration to the local limit, in that the dynamics supported by \cref{eq:gyrokinetic_equation} will be integrated in a `flux tube': a field-line-following domain of perpendicular size that is infinitesimal in comparison to the length scales associated with the plasma equilibrium $\Fos$ and the equilibrium magnetic field $\vB$. We adopt a set of field-aligned coordinates $(x, y, z)$, in which $z$ is the coordinate along the magnetic field, while $x \propto \psi$ and $y \propto \alpha$ are the radial and binormal coordinates, respectively, with suitable normalisations such that they have units of length and that $\vB = B_0 \grad{x} \times \grad{y}$, with $B_0$ some normalising magnetic field. Note that $x$ and $y$ are not in general orthogonal to one another, viz., $\grad x \cdot \grad y \neq 0$ (see \cref{fig:flux_tube_geometry}). In what follows, we will give the exact definitions of these coordinates in terms of usual toroidal coordinates where appropriate. Perpendicular gradients associated with the equilibrium, such as the density and temperature gradients
\begin{align}
	\frac{1}{\Lns} \equiv - \frac{\rmd \ln \dens}{\rmd x}, \quad \frac{1}{\LTs} \equiv - \frac{\rmd \ln \Ts}{\rmd x}, 
	\label{eq:gradients}
\end{align}
are assumed constant across the domain and equal to their value at the centre of the flux tube \citep[][]{beer95}. Periodic boundary conditions are imposed in the plane perpendicular to the equilibrium magnetic field, allowing any fluctuating quantity to be expressed in terms of its Fourier amplitudes for the perpendicular wavenumber \mbox{$\vkperp = k_x \grad{x} + k_y \grad{y}$}, viz., 
\begin{align}
	\hs (\vRs, \energy,\mus, t) = \sum_{\vkperp} e^{\rmi \vkperp \cdot \vRs} \hskperp(z, \energy,\mus, t) , \quad \dphipot (\vr, t) = \sum_{\vkperp} e^{\rmi \vkperp \cdot \vr} \dphipotkperp(z, t).
	\label{eq:fourier_amplitudes}
\end{align} 

Throughout this paper, the scalar measure of the turbulent transport that will be our focus is the radial turbulent heat (energy) flux of species $\s$, which can be expressed as 
\begin{align}
	Q_\s = \avg{\int \rmd^3 \vec{v} \: \energy \avgr{\hs (\vexb \cdot \grad x)}}{x, y, z}.
	\label{eq:heatflux}
\end{align}
In \cref{eq:heatflux}, the angle brackets $\left< \dots \right>_{x,y,z}$ denote an average over the full spatial domain, with the average over $z$ including a factor of the Jacobian $[(\grad x \times \grad y) \cdot \grad z]^{-1}$ (see, e.g., appendix A of \cite{ivanov25invariant}). While gyrokinetic turbulence can, and often does, support a particle flux $\Gamma_\s$, in the single-kinetic-species simulations that we will consider, $\Gamma_\s$ is identically zero because of the Boltzmann density response of the non-kinetic species.

\subsection{Linear and nonlinear time scales}
\label{sec:timescales}
As in studies of magnetised plasma turbulence in other contexts (see, e.g., \cite{sch09} and references therein), progress towards understanding the turbulence supported by the gyrokinetic equation \cref{eq:gyrokinetic_equation} can be made by considering the various linear and nonlinear time scales contained therein. For our purposes here, there are three important time scales to consider: 

\begin{enumerate}
	\def\itemspacing{\vspace{0.2cm}}
	
	\itemspacing
	\item Linear free energy injection rate: Linear instabilities are the mechanism by which free energy is extracted from the equilibrium gradients in $\dens$ and $\Ts$ at each binormal scale $k_y$ and radial scale $k_x$. In the straight-field-line case, these solutions are parametrised by $\kpar$ because the parallel structure is well-described by a discrete set of Fourier modes. For curved magnetic fields (e.g., in stellarators or tokamaks), such discrete Fourier modes are not appropriate as eigenmodes are in general not Fourier modes. The linear injection rate most relevant for turbulence saturation will be the one maximised over all $k_x$ and the possible parallel structures, which we denote $\omlin$. At this stage, we remain agnostic as to the precise nature of the underlying instability, simply regarding $\omlin$ as the effective rate at which energy is injected into the turbulence.
	
	\itemspacing
	\item Nonlinear decorrelation rate: The time scale of the nonlinear interactions appearing in \cref{eq:gyrokinetic_equation} is simply the rate at which fluctuations are sheared by the perturbed $\exb$ flow, viz., 
	\begin{align}
		\tnl^{-1} \sim \kperp v_E \sim \frac{c }{B} k_x k_y \dphipotamp.
		\label{eq:nonlinear_time_initial}
	\end{align}
	Here, and in what follows, $\dphipotamp$ refers to the characteristic amplitude of the electrostatic potential at the binormal scale $k_y$, rather than to the Fourier transform of the field. Formally, we choose to define it in terms of a summation over the squared Fourier amplitudes at smaller scales, viz.,
	\begin{align}
		\dphipotamp^2 = \sum_{\abs{k_y'}> k_y} \sum_{k_x'} \avg{\left| \dphipot_{\kperp '} \right|^2}{z} .
		\label{eq:amplitude_definition}
	\end{align}
	That said, a precise definition of $\dphipotamp$ is not important since the phenomenological theory that is to follow predicts only scalings; one could alternatively define it via a second-order structure function (see, e.g., \cite{davidson13}). The radial scale $k_x$ appearing in \cref{eq:nonlinear_time_initial} is an implicit function of $k_y$, viz., at each $k_y$, the turbulent fluctuations have a typical radial scale $k_x$ that depends on $k_y$. Following \cite{ivanov25}, we define the \textit{fluctuation aspect ratio} at the scale $k_y$ as
	\begin{align}
		\aspect \equiv \frac{k_x}{k_y},
		\label{eq:aspect_definition}
	\end{align}
	allowing us to rewrite the nonlinear rate \cref{eq:nonlinear_time_initial} as 
	\begin{align}
		\tnl^{-1} \sim \Omegas \aspect (k_y \rhos)^2 {\frac{\qs \dphipotamp}{\Ts}}.
		\label{eq:nonlinear_time}	
	\end{align}
	The presence of $\aspect$ in \cref{eq:nonlinear_time} will turn out to be crucial in determining the difference in scaling of the heat flux between ETG- and ITG-driven turbulence (see \cref{sec:axisymmetric_geometry}). In tokamaks and stellarators, the radial derivative of the fluctuations is highly dependent on the position along the chosen field line due to the presence of magnetic shear. We assume that the radial wavenumber $k_x {\sim \partial \log \dphipotamp/\partial x}$ appearing in $\aspect$ is evaluated at the location in $z$ at which the heat flux is maximal (e.g., the outboard midplane of a tokamak).
	
	\itemspacing
	\item Parallel propagation rate: Though implicit in their definition, neither $\omlin$ nor $\tnl^{-1}$ contains sufficient information to determine the characteristic parallel scales of the turbulence, since both originate from terms in \cref{eq:gyrokinetic_equation} involving only perpendicular gradients of the perturbed fields. This information enters instead through the \textit{parallel propagation rate} $\ompar$, which represents the characteristic time associated with the parallel dynamics along the field lines. As with $\omlin$, we will for now remain agnostic as to the precise form of $\ompar$, but this additional time scale is nevertheless essential for understanding how the parallel dynamics participate in the turbulence. 
	
	\itemspacing
\end{enumerate}

There are, of course, a number of other time scales supported by \cref{eq:gyrokinetic_equation} that we have not listed above, including, for example, the characteristic time scales associated with trapped particles or magnetic drifts (which we discuss further in \cref{sec:axisymmetric_geometry}). For present purposes, however, we will find that the combination of the linear injection rate $\omlin$, the nonlinear decorrelation rate $\tnl^{-1}$, and the parallel propagation rate $\ompar$ is sufficient to formulate a general theory for the asymptotic scaling of the heat flux \cref{eq:heatflux}. 

\subsection{Outer-scale dynamics}
\label{sec:outer_scale_dynamics}
In order to reach a statistical steady state, the free energy extracted from the equilibrium gradients, represented by $\omlin$, must be transferred by the nonlinear term to scales where it can be dissipated. With this in mind, we define the \textit{outer scale} to be the scale at which the rates of nonlinear decorrelation and linear injection balance \citep{barnes11,adkins22,adkins23}:
\begin{align}
	\left(\tnl^{-1}\right)^{\outero} \sim \omlin^{\outero}.
	\label{eq:outer_scale}
\end{align}
Here, and in what follows, the superscript `$\outero$' denotes quantities associated with the outer scale. The inevitability of the balance \cref{eq:outer_scale}, assuming well-developed plasma turbulence, can be seen as follows. Suppose that the nonlinear rate were subdominant to the linear one at all binormal scales $k_y$, so that the system remained effectively linear. Then, amplitudes $\dphipotamp$ would grow exponentially until the nonlinear rate was at least as large as the linear one $\tnl^{-1} \gtrsim \omlin$ at all unstable scales. However, the nonlinear rate cannot be strictly larger than the linear injection rate at all scales, as otherwise there would be no meaningful injection of energy into the turbulence, since the associated rapid decorrelation would disrupt the phase relationships between potential and distribution-function fluctuations required for linear growth. Thus, the existence of an outer scale defined by \cref{eq:outer_scale} is a consequence of saturated turbulence that is homogeneous in time (i.e., not dominated by intermittent behaviour).

To estimate the heat flux $Q_\s$, it will be useful to rewrite its definition \cref{eq:heatflux} explicitly in terms of a sum over the Fourier components of the fluctuations, viz., 
\begin{align}
	Q_\s = \sum_{\vkperp} \avg{Q_{\s \vkperp}}{z}, \quad Q_{\s \vkperp} \equiv \Im \left[\frac{c k_y}{B_0} \int \rmd^3 \vec{v} \: \energy \rmJ_0(\besselarg) \hskperp^* \dphipotkperp \right],
	\label{eq:heatflux_fourier}
\end{align}
where $\rmJ_0(\besselarg)$ is the zeroth-order Bessel function of the first kind, $\besselarg = \kperp \vperp/\Omegas$, and `$^*$' denotes the complex conjugate. Using quasineutrality \cref{eq:quasineutrality} to estimate the non-adiabatic distribution function as $\hs \sim (\qs \dphipot /\Ts) \Fos$, it is clear that the integral in \cref{eq:heatflux_fourier} will be dominated by some outer scale $k_y^\outero \rhos \ll 1$ provided that the amplitudes of the fluctuations decay faster than $k_y^{-1}$ at scales $k_y > k_y^{\outero}$ --- in the so-called `inertial range' --- a condition that is readily satisfied in our simulations (see \cref{fig:sETG_scan_timetraces} or \cref{app:supplementary_data}). Note that we do not provide a detailed treatment of the inertial-range dynamics here, partly because these have been addressed in previous studies \citep{barnes11,adkins22,adkins23}, but primarily because the scaling of the heat flux, the focus of this paper, is determined entirely by the outer-scale dynamics.

Assuming that the phase relationship between $\hs$ and $\dphipot$ does not introduce any nontrivial factors, we estimate an upper bound for $Q_\s$ as follows:
\begin{align}
	Q_\s \sim Q_\s^{\outero} \sim \dens \Ts \vths k_y^{\outero} \rhos \left(\frac{\qs \dphipotampouter}{\Ts}\right)^2.
	\label{eq:heatflux_outer}
\end{align}
Using \cref{eq:nonlinear_time} and \cref{eq:outer_scale}, the dependence of \cref{eq:heatflux_outer} on the outer-scale amplitude can be eliminated, yielding [cf. (3.8) of \cite{ivanov25}]:
\begin{align}
	\frac{Q_\s}{\dens \Ts \vths} \sim \left(\frac{\omlinouter}{\Omegas}\right)^2 \frac{1}{(k_y^{\outero} \rhos)^3 (\aspect^{\outero})^2}.
	\label{eq:heatflux_no_amp}
\end{align}
Up to this point, we have made no assumptions about the specific nature of the unstable modes, beyond requiring that the turbulence is strongly driven and that the temperature fluctuations are comparable to those in the electrostatic potential, viz., $\dTs/\Ts \sim \qs {\dphipot}/\Ts$. To proceed further, we adopt a simple estimate of the growth rate of temperature-gradient-driven instabilities [cf. \cref{eq:gyrokinetic_equation}],
\begin{align}
	\omlinouter \sim \left| \frac{\avgRs{\vexb} \cdot \grad \Fos}{\hs} \right|_{k_y^\outero} \sim \eta_\s \omegasts \equiv \frac{k_y^{\outero} \rhos \vths}{2\LTs},
	\label{eq:omlinouter}
\end{align} 
where, to obtain the final expression, we have once again used quasineutrality \cref{eq:quasineutrality} to estimate $\hs \sim (\qs \dphipot /\Ts) \Fos$. The estimate \cref{eq:omlinouter} is typical of both linear ETG- and ITG-driven electrostatic modes \citep{barnes11,adkins22}, and should be viewed primarily as an estimate for the effective rate of energy injection, rather than the rate at which any individual perturbations grow linearly. Inspection of \cref{eq:heatflux_no_amp} makes it obvious that we then have two remaining unknowns at the outer scale: the binormal wavenumber $k_y^{\outero} \rhos$ and aspect ratio $\aspect^{\outero}$. Let us first focus our attention on the former.

\subsection{Critical balance}
\label{sec:critical_balance}
Given that, at fixed $\aspect^{\outero}$, the heat flux scales inversely with the binormal wavenumber, $Q_\s \propto (k_y^{\outero} \rhos)^{-1}$, it is natural to expect that it will be dominated by the largest accessible binormal scales. In the strongly-driven cases that we are going to consider, this will correspond to $k_y^{\outero} \rhos \ll 1$ \footnote{The fact that the heat flux is dominated by fluctuations at scales $k_y^\outero \rhos \ll 1$ is not always guaranteed. For example, turbulence driven by the short-wavelength ITG \citep{hassam90,guo93,smolyakov02} will have $k_y^\outero \rhos \gtrsim 1$ (see, e.g., \cite{gupta25}), wherein \cref{eq:omlinouter} would also have to be modified. However, such a state requires the ITG to be stable at larger scales, which is unlikely to be achieved except in the presence of strong equilibrium flow shear (see, e.g., \cite{ivanov25}), or for $\eta_\s \sim 1$, regimes that lie outside the scope of the present theory.}. At such scales, the system approaches the drift-kinetic limit of gyrokinetics, in which all gyroaverages in \cref{eq:gyrokinetic_equation} are well-approximated by unity operators to leading order. However, in this limit, there is no intrinsic perpendicular scale: the dynamics are scale-invariant \citep{adkins23}, and so there is no microscopic mechanism that can be used to determine $k_y^{\outero}$ --- other than the sizes $L_x$ and $L_y$ of our flux-tube domain in the radial and binormal directions, respectively. However, recall that the flux-tube approximation is predicated on the assumption that the fluctuations remain microscopic in the perpendicular direction. For this assumption to be valid, as the perpendicular domain size is increased, there must come a point at which the turbulence, and the resultant heat flux, become independent of it; if this were not the case, then the heat flux would diverge as $L_x/\rhos, L_y/\rhos \rightarrow \infty$, implying that the local flux-tube approximation is not a valid model of the plasma dynamics. This divergence is not observed in electrostatic gyrokinetic simulations. Hence, the outer scale cannot be set solely by perpendicular considerations; instead, it must be fixed by parallel dynamics. A natural way to determine this scale is through \textit{critical balance}, viz., the condition that the time scales of parallel propagation and perpendicular nonlinear decorrelation be comparable. The logic is straightforward: two points separated along a field line can only remain correlated with one another if information can propagate between them faster than they are decorrelated by the (perpendicular) nonlinearity.\footnote{Although we here associate $\ompar$ with the rate of parallel propagation, it may equivalently be interpreted as some characteristic parallel dissipation rate. For instance, $\ompar \sim \kpar \vths$ appearing in \cref{eq:critical_balance} represents both the rates of parallel-streaming and (velocity-space) phase mixing associated with Landau damping \cite{sch16}. In the former case, the critical balance \cref{eq:critical_balance} can be interpreted in terms of the causality argument given in the main text; in the latter, the parallel coherence of fluctuations is instead limited (on the small-parallel-scale side) by Landau damping. In either case, however, $\omega_\parallel$ acts to set $\Lpar$.} This is a standard argument originally developed in the context of MHD turbulence \citep{GS95,boldyrev05,sch22}, where the information is carried by Alfv\'en waves. In (collisionless) electrostatic gyrokinetic turbulence, the relevant propagation speed is the parallel streaming rate $\ompar \sim \kpar \vths$, meaning that, at the outer scale, we require, in addition to \cref{eq:outer_scale}, the balance
\begin{align}
	(\tnl^{-1})^{\outero} \sim \ompar^{\outero} \sim \kpar^{\outero} \vths \sim \frac{\vths}{\Lpar}.
	\label{eq:critical_balance}
\end{align}
Here, and throughout, $\Lpar$ denotes the characteristic parallel system scale, which may in principle depend on parameters associated with both the plasma gradients (i.e., on $\LTs$) and the magnetic equilibrium (e.g., on the safety factor $q$). In the cases that we consider in this paper, we find $\Lpar$ to be independent of the plasma gradients (see the numerical simulations of sections \ref{sec:slab_simulations}, \ref{sec:cbc_simulations}, and \ref{sec:stellarator_simulations}), but our general approach does not assume such independence. Combining \cref{eq:outer_scale}, \cref{eq:omlinouter}, and \cref{eq:critical_balance}, we find the following relationship
\begin{align}
	k_y^{\outero} \rhos \sim \frac{\LTs}{\Lpar}.
	\label{eq:binormal_outer_scale}
\end{align}
At first glance, it seems that we have not made any progress, trading the unknown outer-scale binormal wavenumber $k_y^{\outero} \rhos$ for a similarly unknown parallel system scale $\Lpar$. However, the latter is often determined by the magnetic geometry: we will see that, while in a straight magnetic field $\Lpar$ is the parallel box length (\cref{sec:slab_geometry}), and therefore as artificial as $L_x$ or $L_y$, in axisymmetric toroidal systems, it typically corresponds to the connection length (\cref{sec:axisymmetric_geometry}). It is worth emphasising that, while critical balance is usually discussed in the context of inertial-range dynamics, we are only considering dynamics at the outer scale.

\subsection{Heat-flux scaling}
\label{sec:heatflux_scaling}
Substituting \cref{eq:omlinouter} and \cref{eq:binormal_outer_scale} into \cref{eq:heatflux_no_amp}, we arrive at the following expression for the heat flux:
\begin{align}
 	\frac{Q_\s}{\Qgbs}  \sim \frac{\Lpar}{\lref} \left(\frac{\lref}{\LTs}\right)^3 (\aspect^{\outero})^{-2} ,
 	\label{eq:heatflux_final} 
\end{align}
in which $\Qgbs = \dens \Ts \vths (\rhos/\lref)^2$ is the gyro-Bohm heat flux and $\lref$ is some yet-to-be-specified reference length scale for the plasma equilibrium (e.g., the minor radius $a$). Equation \cref{eq:heatflux_final} is the most general scaling that can be obtained for the heat flux in electrostatic gyrokinetic turbulence without committing to a particular equilibrium. To make further progress, one must specify the magnetic geometry and equilibrium profiles in order to determine the parallel system scale $\Lpar$ and outer-scale aspect ratio $\aspect^{\outero}$. Note that $\aspecto$ can (and indeed, does) depend on $\Lpar/\lref$ and/or $\lref/\LTs$. In the following sections, we examine the predictive power of \cref{eq:heatflux_final} for slab (\cref{sec:slab_geometry}), axisymmetric (\cref{sec:axisymmetric_geometry}), and non-axisymmetric (\cref{sec:non_axisymmetric_geometry}) geometries. 

Before doing so, however, it is worth pausing to review the assumptions that led us here. We began by arguing that in order for there to be a statistical steady state, the energy injection due to linear instabilities must be balanced by the nonlinear term. This motivated the definition of the outer scale \cref{eq:outer_scale}, determined by the balance between the rates of nonlinear decorrelation \cref{eq:nonlinear_time} and linear injection due to the temperature-gradient-driven modes \cref{eq:omlinouter}. The resulting expression for the heat flux \cref{eq:heatflux_no_amp} then depended only on the outer-scale binormal wavenumber $k_y^{\outero} \rhos$ and aspect ratio $\aspect^{\outero}$, yet to be determined. However, we could not appeal to any intrinsic perpendicular scale to determine the former owing to the fact that it (generally) lies at large binormal scales, i.e., in the scale-invariant drift-kinetic regime. Instead, we assumed that the dynamics organised themselves so that the rates of parallel propagation and nonlinear decorrelation were comparable at the outer scale, i.e., were in the critical balance \cref{eq:critical_balance}, leading directly to \cref{eq:heatflux_final}. As promised, this series of steps relied only on simple physical arguments about the competition between the time scales introduced in \cref{sec:timescales}. The resulting expression for the electrostatic heat flux \cref{eq:heatflux_final} should therefore be viewed as the natural and quite general consequence of this competition, rather than the product of detailed or restrictive assumptions.  

\subsection{Adiabatic responses}
\label{sec:adiabatic_responses}
As anticipated at the start of this section, we have been considering only the turbulence driven by a single species, either electrons or ions. Such an assumption naturally restricts our consideration to either scales much smaller than the ion Larmor radius, $\kperp \rhoi \gg 1$, in the case of ETG-driven turbulence, or scales on the order of or larger than it, $\kperp \rhoi \lesssim 1$, in the case of ITG-driven turbulence. For $\kperp \rhoi \gg 1$, the ions can easily respond to the electric field fluctuations because the ion gyro-orbits are much larger than the characteristic length of the fluctuations, and the ion gyrofrequency has been assumed to be much larger than the fluctuation frequency [see the ordering \cref{eq:gyrokinetic_ordering}]. Thus, their density response $\dni$ is approximately Boltzmann,
\begin{align}
	\frac{\dni}{\dens[i]} = - \frac{Z_i e \dphipot}{\Ti}.
	\label{eq:adiabatic_ions}
\end{align}
The electron density in the $\kperp \rhoi \lesssim 1$ limit does not have quite the same property. In this limit, we assume the system is dominated by dynamics occurring on ion timescales, meaning that the balances \cref{eq:omlinouter} and \cref{eq:critical_balance} become $\omlinouter \lesssim \vthi/\LTi$ and $(\tnl^\outero)^{-1} \sim \ompar \sim \vthi/\Lpar$, respectively. Then, the parallel streaming of electrons, with characteristic frequency $\vthe/\Lpar$, is sufficiently fast that their motion along the equilibrium magnetic field lines allows them effectively to cover the entire flux-surface. As a result, they respond almost instantaneously. In the radial direction, however, electrons barely move, and so the electron response to a flux-surface-averaged (i.e., zonal) potential is weak. To reflect this anisotropy, one typically adopts the modified adiabatic electron response \citep{dorland93,hammett93,abel13b} when considering ion-scale turbulence, with the zonal component of the potential excused from the Boltzmann response, viz., 
\begin{align}
	\frac{\dne}{\dens[e]} = \frac{e}{\Te} \left(\dphipot - \avg{\dphipot}{y, z}\right),
	\label{eq:adiabatic_electrons}
\end{align}
where $\avg{\dots}{y, z}$ denotes the zonal average (see \cref{sec:cross_scale_interactions} for a discussion on the role of trapped electrons). Unlike the adiabatic ion response \cref{eq:adiabatic_ions}, this explicitly privileges zonal perturbations, making them dynamically distinct. We will see that it is precisely this distinction that gives rise to the differences in saturation between ETG- and ITG-driven turbulence, by affecting the outer-scale aspect ratio $\aspect^{\outero}$. 

\section{Slab geometry}
\label{sec:slab_geometry}
Let us first consider a homogeneous plasma in a uniform magnetic field oriented along the $z$ direction, the so-called `slab' geometry. While this case is not of direct relevance to fusion turbulence, it provides the simplest setting in which \cref{eq:heatflux_final} can be tested, and will allow a reader unfamiliar with these arguments to develop intuition that will prove useful in later sections. The reader more committed to real fusion applications may skip ahead to \cref{sec:axisymmetric_geometry}.  

\subsection{Slab scalings}
\label{sec:slab_scalings}
We focus on turbulence driven by only the temperature gradient $\LTs$ of the non-adiabatic species. In the `slab ETG' (sETG) case, the ion density response is taken to be adiabatic according to \cref{eq:adiabatic_ions}, while in the `slab ITG' (sITG) case, the electron response is taken to be the modified-adiabatic one \cref{eq:adiabatic_electrons}. Let us take a moment to discuss the implications that this has for the outer-scale aspect ratio $\aspecto$. 

The outer scale is, by definition, the scale at which both linear and nonlinear effects are comparable [see \cref{eq:outer_scale}]. This balance must then determine the anisotropy of the fluctuations at that scale. In numerical simulations, we observe that sETG fluctuations have an aspect ratio that does not depend strongly on the gradients, $\aspecto \sim 1$. This is despite the fact that (linear) energy injection is often associated with so-called `streamers': radially-elongated eddies with a vanishingly small radial wavenumber. A possible resolution is provided by the `secondary-instability' picture of saturation \citep{rogers00,rogers05,ivanov20}, which posits that such streamers will go unstable to zonal perturbations that will shear them apart, in so doing generating eddies with $k_x^\outero \sim k_y^\outero$. It is important to emphasise that $\aspecto \sim 1$ does not imply that the turbulence is isotropic at the outer scale, only that whatever outer-scale anisotropy that may exist does not vary with equilibrium quantities. Indeed, we will find that ETG-driven turbulence can have $\aspecto$ both smaller and larger than unity while satisfying this scaling.

For ITG-driven turbulence, however, such a picture of saturation is less obvious. The modified adiabatic electron response \cref{eq:adiabatic_electrons} explicitly privileges the role of zonal perturbations since, unlike their non-zonal counterparts, they receive no contribution from the electron dynamics. This makes zonal flows easier to excite \citep{hammett93} and suggests that they could play a more dominant role, potentially breaking the simple $\aspecto \sim 1$ expectation. Indeed, we find that simulations of sITG turbulence, for all parameter values that were considered, develop very large zonal flows that significantly suppress the turbulence (see \cref{app:sitg_turbulence}). This is known as the Dimits regime \cite{dimits00}. These slab simulations suggest that the presence of curvature-driven ITG is required in order to support strongly driven turbulence (see, e.g., \cite{ivanov20,ivanov22}). Given that the dynamics associated with this Dimits regime lie outside the scope of our theory, we focus on ETG-driven turbulence in the remainder of this section, returning to ITG-driven turbulence in \cref{sec:axisymmetric_geometry} and \cref{sec:non_axisymmetric_geometry}.

Our resulting predictions for the heat flux \cref{eq:heatflux_final}, binormal outer scale \cref{eq:binormal_outer_scale}, and aspect ratio for sETG turbulence are then
\begin{align}
	\frac{Q_e}{\Qgbs[e]} \sim \frac{\Lpar}{\LTe}, \quad k_y^{\outero} \rhoe \sim \frac{\LTe}{\Lpar}, \quad \aspect^{\outero} \sim 1,
	\label{eq:predictions_setg}
\end{align}
with $\Qgbs[e] = \dens[e] \Te \vthe (\rhoe/\LTe)^2$. Note that we have chosen the reference length scale to be the electron-temperature gradient, viz., $\lref = \LTe$, given that the latter is the only non-trivial length scale associated with the plasma equilibrium. This means that the first expression in \cref{eq:predictions_setg} recovers the cubic scaling with the electron-temperature gradient anticipated by previous theoretical studies \cite{adkins22,adkins23thesis,adkins23}. $\Lpar$ should therein be interpreted as the parallel extent of the simulation domain, since in slab geometry there is no other intrinsic parallel length in the absence of finite magnetic shear. Equivalently, the longest distance a turbulent eddy can extend along the field line is the box length itself and so this must set the parallel system scale. Hence, $\Lpar/\LTe$ becomes the sole physical input parameter to the system. 

\subsection{sETG simulations}
\label{sec:slab_simulations}
To test the predictions \cref{eq:predictions_setg}, we solve the electrostatic gyrokinetic system of equations \cref{eq:gyrokinetic_equation} and \cref{eq:quasineutrality} in slab geometry using \texttt{GX} \citep{mandell24}, the code employed for all simulations in this paper. The homogeneous nature of the magnetic field allows periodic boundary conditions along $z$ to be used. We vary $\Lpar/\LTe$ by a factor of two, increasing the parallel resolution accordingly to maintain the same maximum parallel wavenumber $\kpar \LTe$ in the box. We did not perform simulations at $\Lpar/\LTe < 300$ as these would not be sufficiently strongly driven to realise the scalings \cref{eq:predictions_setg}. The perpendicular box sizes and corresponding resolution are held constant. Further details about the numerical setup can be found in \cref{app:numerical_details} (see the row labelled `sETG' in \cref{tab:simulation_parameters}).

\begin{figure*}
	\centering
	\includegraphics[width=\figscale\textwidth]{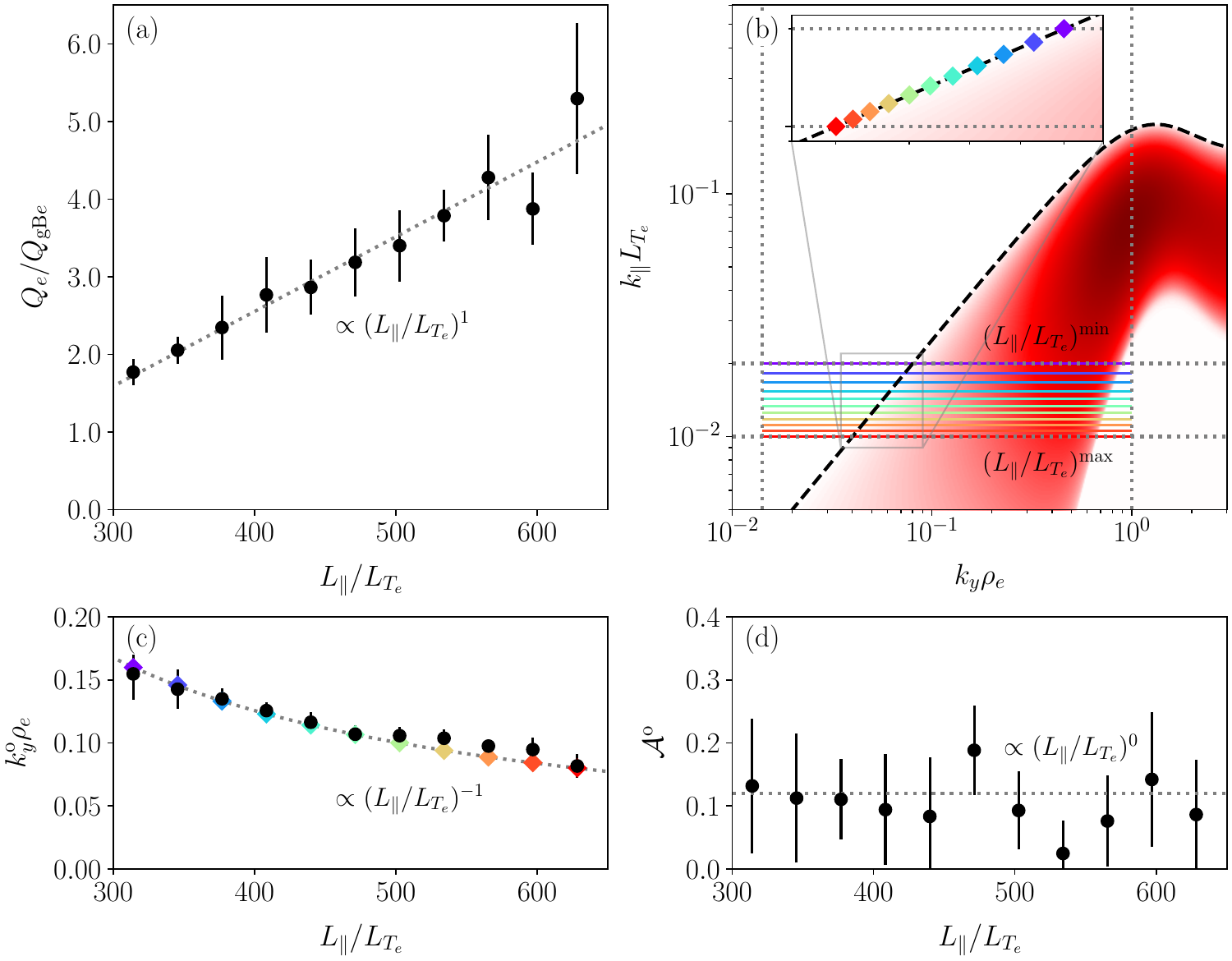}
	
	\caption[]{Scaling results from the sETG simulations. (a) Time-averaged gyro-Bohm-normalised heat flux $Q_e/\Qgbs[e]$ [here $\Qgbs[e]= \dens[e] \Te \vthe (\rhoe/\LTe)^2$] as a function of $\Lpar/\LTe$ (black circles), with the error bars showing the standard deviation over time. The dotted line shows the expected scaling [first expression in \cref{eq:predictions_setg}]. (b) Contour plot of the unstable solutions to the sETG dispersion relation \cref{eq:setg_dispersion_relation} in the $(k_y, \kpar)$ plane for $k_x=0$, $Z_i=1$, and $\Ti = \Te$, with the dashed line showing the associated stability boundary \cref{eq:setg_stability_boundary}. The vertical dotted lines indicate the minimum and maximum $k_y \rhoe$ in the simulation domain. The horizontal dotted lines bound the range of parallel wavenumbers corresponding to the different values of $\Lpar/\LTe$ in panel (a), each indicated by one of the horizontal coloured lines. The coloured diamonds in the inset denote the locations of the intersections of these lines with the stability boundary. The maximum value of $\kpar \LTe = 0.6$ for all simulations corresponds to the top edge of the plot. (c) Time-averaged value of the binormal outer scale $k_y^{\outero} \rhoe$ (black circles), with the error bars showing the standard deviation over time. The coloured diamonds are the values of $k_y \rhoe$ in the inset of panel (b) multiplied by a factor of two [see the discussion following \cref{eq:spectrum_binormal}]. The dotted line shows the expected scaling [the second expression in \cref{eq:predictions_setg}]. (d) Time-averaged outer-scale aspect ratio $\aspect^{\outero}$ (black circles), with the error bars showing the standard deviation over time. The dotted line shows the expected scaling [the third expression in \cref{eq:predictions_setg}].}
	\label{fig:sETG_scan}
\end{figure*}

\begin{figure*}
	\centering
	\includegraphics[width=\figscale\textwidth]{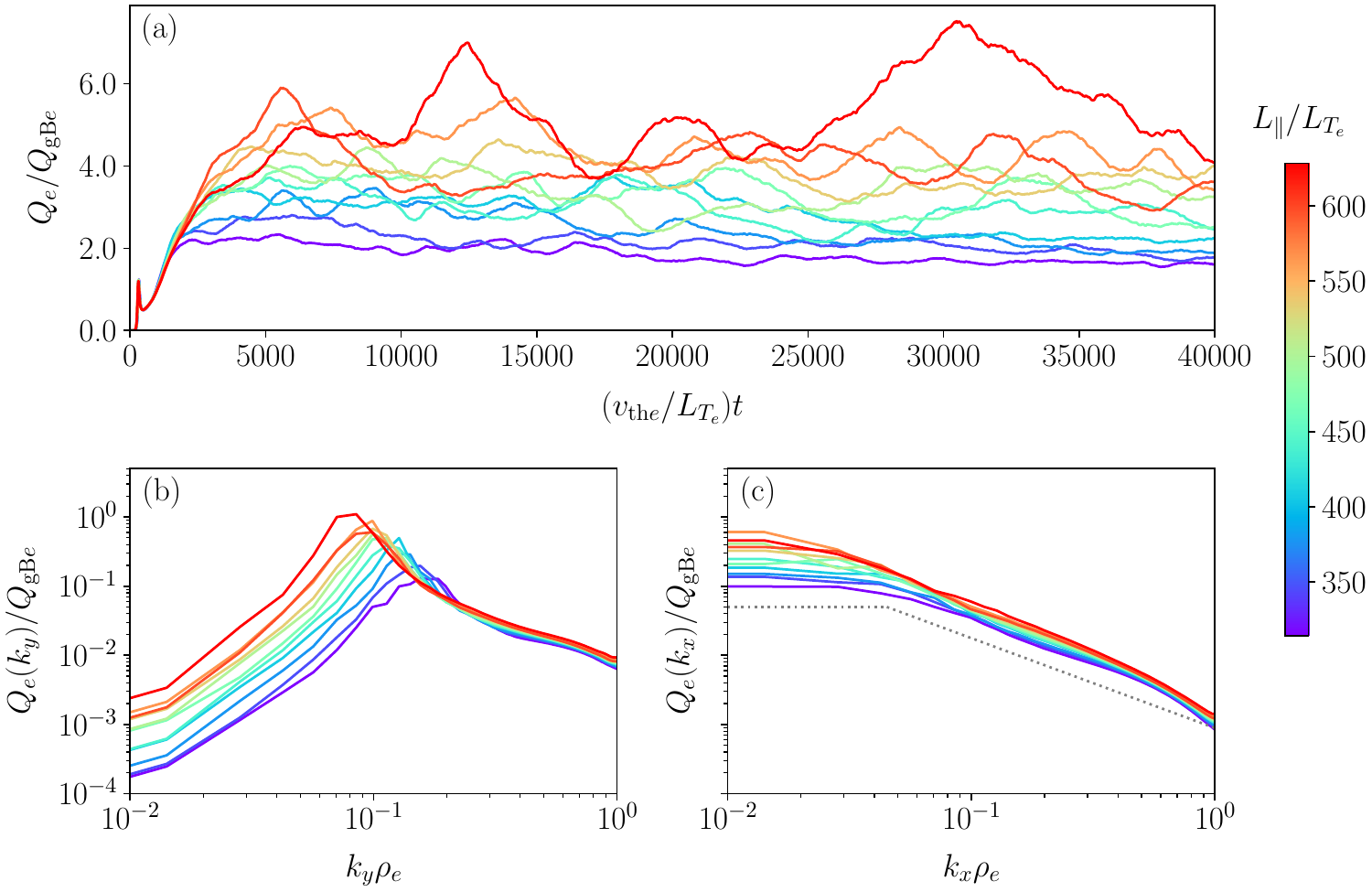}
	
	\caption[]{Gyro-Bohm-normalised heat-flux time traces and spectra from the sETG simulations, with the colours indicating the value of $\Lpar/\LTe$ for a given simulation: (a) heat flux as a function of time; (b) time-averaged binormal spectrum of the heat flux \cref{eq:spectrum_binormal}; (c) time-averaged radial spectrum of the heat flux \cref{eq:spectrum_radial}. The dotted line in panel (c) provides an illustration of the type of broken power law used to calculate the radial outer scale.}
	\label{fig:sETG_scan_timetraces}
\end{figure*}

The results of this scan are shown in \cref{fig:sETG_scan}. In panel (a), we plot the turbulent electron heat flux $Q_e/\Qgbs[e]$ averaged over the last 80\% of each simulation's running time [see \cref{fig:sETG_scan_timetraces}(a)], with the error bars showing the standard deviation over the same period. The first prediction of \cref{eq:predictions_setg} --- that the gyro-Bohm-normalised heat flux should scale linearly with $\Lpar/\LTe$ --- is clearly borne out by the data. At larger $\Lpar/\LTe$, the error bars become noticeably larger, reflecting the presence of stronger oscillations on longer time scales in the heat flux [see \cref{fig:sETG_scan_timetraces}(a)]. These oscillations are a direct consequence of the underlying scale invariance of drift kinetics (see \cite{adkins23}). 

To see why this scaling of the heat flux is, in a sense, inevitable in this simple slab case, it will be instructive to examine the linear dispersion relation for slab ETG modes \citep{liu71,lee87} for $\eta_e \gg 1$:
\begin{align}
	1 + \frac{{Z_i} \Te}{\Ti} + \left\{\zeta \rmZed - \zetast \left[\zeta + \left(\zeta^2 - \frac{1}{2}\right)\rmZed \right]    \right\} \UGamma_0 + \zetast b_e \UGamma_1 \rmZed = 0,
	\label{eq:setg_dispersion_relation}
\end{align}
where $\zeta = \omega/|\kpar| \vthe$ and $\zetast = \eta_e \omega_{*e}/|\kpar| \vthe$ are the normalised frequencies, $\rmZed = \rmZed(\zeta)$ is the standard plasma dispersion function \citep{faddeeva54,fried61}, $\UGamma_0(b_e) = \rmI_0(b_e)\mathrm{e}^{-b_e}$, $\UGamma_1(b_e) = [\rmI_0(b_e) - \rmI_1(b_e)]\mathrm{e}^{-b_e}$, with $b_e = \kperp^2 \rhoe^2 /2$, while $\rmI_0$ and $\rmI_1$ are the modified Bessel functions of the first kind \citep{abramowitz72}. The associated stability boundary  --- the parallel wavenumber above which solutions are no longer unstable --- is
\begin{align}
	\abs{\kpar}\LTe = \frac{k_y \rhoe}{2}\left[\frac{\left(\UGamma_0/2 + b_e \UGamma_1\right)\UGamma_0 }{\left(1 + {Z_i}\Te/\Ti \right)\left(1 + {Z_i}\Te/\Ti - \UGamma_0\right) }\right]^{1/2}.
	\label{eq:setg_stability_boundary}
\end{align}
This is plotted in the $(k_y, \kpar)$ plane for $k_x = 0$ in \cref{fig:sETG_scan}(b), alongside the unstable solutions to \cref{eq:setg_dispersion_relation}.

Now consider a simulation with a given value of $\Lpar/\LTe$, with which we can associate a minimum parallel wavenumber $\kparmin = 2\pi/\Lpar$ accessible by the fluctuations (set by the parallel box size). As we argued at the beginning of \cref{sec:critical_balance}, there is no intrinsic perpendicular scale at $\kperp \rhoe \ll 1$ to set $k_y^{\outero}$. Consequently, at any given $\kpar$, the potential location of the outer scale is pushed to the largest binormal scale that remains unstable; equivalently, $k_y^{\outero} \rhoe$ must lie on, or close to, the linear stability boundary \cref{eq:setg_stability_boundary}. Then, according to the critical-balance arguments of \cref{sec:critical_balance}, the finite parallel extent of the domain pins the outer scale to the intersection of the stability boundary and the line $\kpar = \kparmin$ [viz., the solutions to \cref{eq:setg_stability_boundary} with $\kpar = \kparmin$], which are marked by coloured diamonds in \cref{fig:sETG_scan}(b),(c). It follows that the outer scale observed in nonlinear simulations should coincide with the coloured diamonds along the stability boundary plotted in the inset of panel (b). Panel (c) shows that this expectation is correct: the measured outer-scale wavenumbers, defined as those that maximise the binormal heat-flux spectrum [for $\s = e$; see \cref{eq:heatflux_fourier}]
\begin{align}
	Q_\s(k_y) = \sum_{k_x} \avg{Q_{\s \vkperp}(k_x, k_y, z)}{z},
	\label{eq:spectrum_binormal}
\end{align}
closely track both the order-of-magnitude estimate \cref{eq:predictions_setg} and the values obtained directly from the linear stability boundary. Remarkably, the stability boundary in $k_y \rhoe$ multiplied by a factor of two almost completely overlaps with the measured values. Plotting the binormal spectrum itself in \cref{fig:sETG_scan_timetraces}(b), it is obvious that the peak moves towards larger scales as $\Lpar/\LTe$ is increased. This provides further confirmation of the validity of the simple physical arguments underpinning the results of \cref{sec:asymptotic_scaling_theory}. 

Finally, to examine the scaling of the outer-scale aspect ratio $\aspect^{\outero}$, we need to measure the outer-scale radial wavenumber $k_x^{\outero} \rhoe$. This is less straightforward than for the binormal outer scale, owing to the fact that the radial spectrum of the heat flux [for $\s = e$; see \cref{eq:heatflux_fourier}]
\begin{align}
	Q_\s(k_x) = \sum_{k_y} \avg{Q_{\s \vkperp}(k_x, k_y, z)}{z},
	\label{eq:spectrum_radial}
\end{align} 
is typically observed to peak at $k_x = 0$ in fusion-relevant studies of gyrokinetic turbulence, with the present case being no exception [see \cref{fig:sETG_scan_timetraces}(c)]. We exploit the fact that in such cases the radial spectrum generally exhibits a spectral break at some finite value of $k_x$: at scales larger than the break, the slope of the spectrum is either shallow or flat, while at scales smaller than it, it exhibits a finite slope. We identify the wavenumber location of this break to be the radial outer scale $k_x^{\outero} \rhoe$, which we extract by fitting to the spectrum, at each instant in time, two power-law segments separated by a single break point. For each candidate break location, we measure the combined error from fitting the power laws on either side and select the break location that minimises this error, i.e., the best broken-power-law fit to the spectrum. 

This choice of radial outer scale can be motivated as follows. As we discussed following \cref{eq:heatflux_outer}, our scaling estimate for the heat flux assumes that $\dphipotamp$ decays sufficiently quickly as a function of the binormal wavenumber. Given that we expect the turbulence to be roughly isotropic in the inertial range (see \cite{barnes11,adkins22,adkins23} and appendix A of \cite{ivanov25}), the fluctuations therein will also decay as a function of $k_x$. This means that the break in the spectrum must correspond to the large-scale end of the inertial range, and thus the outer scale, by definition. Note that we have deliberately not attempted to offer a dynamical explanation for the flat spectrum on scales $k_x < k_x^\outero$ because the details thereof are irrelevant for our scaling predictions (and will in all likelihood depend on specifics of the underlying equilibrium in a non-universal way). Indeed, this behaviour of the spectrum could come across as surprising for readers more familiar with studies of isotropic plasma turbulence. For a homogeneous 2D isotropic field, the low-$\kperp$ spectrum can be shown to scale as $\kperp^3$ (see, e.g., appendix A of \cite{sch16}), though a $\kperp^1$ scaling is also possible in some contexts \citep{hosking22forced}. Either of these scalings initially appears inconsistent with the flat or very shallow $k_x$ spectrum at large scales seen here. A possible resolution is that the turbulence above the outer scale is manifestly \textit{not} isotropic, with more power contained in binormal fluctuations [compare the large-scale behaviour of the spectra in panels (b) and (c) of \cref{fig:sETG_scan_timetraces}], and so we would not necessarily expect these kinematic scalings to hold in the present context.

Applying the procedure outlined above, we find that the resulting outer-scale aspect ratio $\aspect^{\outero} = k_x^{\outero}/k_y^{\outero}$ is approximately constant over the full range of $\Lpar/\LTe$ [see \cref{fig:sETG_scan}(d)], in agreement with the third prediction of \cref{eq:predictions_setg}. This supports our earlier claim that the adiabatic ion response \cref{eq:adiabatic_ions} does not promote any scaling of $\aspecto$ with equilibrium parameters. We will find that this is also the case for ETG-driven turbulence in the axisymmetric geometry considered in the next section, whereas ITG-driven turbulence displays markedly different behaviour. It is worth noting that the measured value of $\aspecto$ is numerically small [$\aspecto \approx 0.1$ in \cref{fig:sETG_scan}(d)], which can have important consequences in the presence of externally imposed flow shear: theoretical work \cite{zhang92,hatch18,ivanov25} has shown that eddies with $\aspecto \ll 1$ are suppressed more effectively by the shear than those with $\aspecto \gtrsim 1$.

\section{Axisymmetric geometry}
\label{sec:axisymmetric_geometry}
Let us now consider the predictions presented in \cref{sec:asymptotic_scaling_theory} in the context of axisymmetric toroidal geometry, treating both ETG- and ITG-driven turbulence. We take $\psi$ to be the poloidal magnetic flux divided by $2 \pi$ and define the field-line label as $\alpha = \tor - \safety(\psi) \pol$, where $\tor$ is the toroidal angle (the symmetry angle of the axisymmetric equilibrium), $\pol$ is the straight-field-line poloidal angle, and $\safety$ is the safety factor \citep{kruskal58,dhaeseleer91}. The parallel coordinate is chosen as $z = \pol$, with $z=0$ at the outboard midplane of the device. 

\subsection{Tokamak scalings}
\label{sec:tokamak_scalings}
As in the slab case, our task is to determine both the parallel system scale $\Lpar$ and outer-scale aspect ratio $\aspect^{\outero}$, since together they set the heat-flux scaling \cref{eq:heatflux_final}. Following previous studies \citep{barnes11,adkins22,adkins23,nies26}, we conjecture that the former is set by the distance along the mean field from the outboard to inboard sides of the device, i.e., the connection length $\sim q R$, with $R$ the major radius. The reasoning for this is straightforward: electrostatic temperature-gradient-driven instabilities like curvature-driven ETG and ITG are localised in the regions of `bad' (destabilising) curvature, typically peaked at the outboard midplane, and are stabilised once they extend into the regions of `good' (stabilising) curvature. To within factors of order unity, this distance is the connection length. We note that this is not a unique possibility, and indeed may fail for highly-shaped plasmas: e.g., \cite{parisi20,parisi22} found that curvature-driven ETG fluctuations in a JET pedestal could be localised to at higher values ot the poloidal angle, typically assumed to be in the good-curvature region, whereas slab ETG modes were simultaneously active at other poloidal angles. Nevertheless, $\Lpar \sim qR$ represents the best \textit{a priori} guess for core-relevant turbulence and is, as we will see, supported by our simulations. Since ETG and ITG have isomorphic collisionless linear dispersion relations\footnote{Consider the linearised gyrokinetic equation \cref{eq:gyrokinetic_equation} for a given mode with $k_y \neq 0$. When written in terms of an appropriately normalised potential $\varphi = \qs \dphipot/\Ts$, this is identical for electrons ($\s = e$) and ions ($\s=i$). Doing the same for quasineutrality \cref{eq:quasineutrality}, it is straightforward to see that the resultant linear systems of equations are also identical up to the replacement ${Z_i}\Te/\Ti \mapsto ({Z_i}\Te/\Ti)^{-1}$, converting the ITG case into the ETG one.}, we estimate $\Lpar \sim qR$ in both cases.

It follows that any difference between the heat-flux scaling in ETG- and ITG-driven turbulence must originate from the aspect ratio $\aspect^{\outero}$. Motivated by the results of the previous section, we expect the aspect ratio of ETG eddies once again to be independent of both $\Lpar$ and $\LTe$, viz., $\aspect^{\outero} \sim 1$ \footnote{This is not the only possible outcome. Previous studies (see, e.g., \cite{colyer17,tirkas23}) have shown that ETG turbulence can, in some cases, be regulated at long times by strong zonal flows that grow slowly and eventually dominate the saturation process. Should the amplitude of these zonal flows depend on plasma gradients, then this may result in the assumption of $\aspecto \sim 1$ being broken. In our simulations, however, we see no evidence of such long-time growth (possibly due to the strong drive), and so we regard $\aspect^{\outero} \sim 1$ as the most straightforward \textit{a priori} estimate.}. Substituting this into \cref{eq:heatflux_final} and using our expression for the parallel system scale $\Lpar \sim qR$, we obtain the following predictions for saturated ETG turbulence in axisymmetric geometry:
\begin{align}
	\frac{Q_e}{\Qgbs[e]} \sim q \left(\frac{R}{a}\right) \left(\frac{a}{\LTe}\right)^3, \quad k_y^{\outero} \rhoe \sim \frac{\LTe}{qR}, \quad \aspect^{\outero} \sim 1.
	\label{eq:predictions_etg}
\end{align}
where $\Qgbs[e] = \dens[e] \Te \vthe (\rhoe/a)^2$ uses $\lref = a$. Note that these results are in fact identical to those in the sETG case \cref{eq:predictions_setg}, except we have now specified the parallel system scale and set our reference length scale $\lref$ to be the minor radius $a$, so there is no $\LTe$ dependence hiding in the definition of $\Qgbs[e]$.

For ITG turbulence, however, the situation is different. As discussed in \cref{sec:adiabatic_responses}, the modified adiabatic electron response \cref{eq:adiabatic_electrons} explicitly privileges zonal flows by making them easier to drive than the $k_y \neq0$ modes \citep{hammett93}. Naturally, we might expect zonal fluctuations to play a key role in determining the outer scale. Indeed, recent work \citep{nies26,nies26tsm} has shown that this gives rise to a novel zonal mode --- the so-called `toroidal secondary mode' --- that grows and propagates due to the combined effects of zonal-flow shearing, advection by the radial component of the magnetic drifts, and compressibility of zonal flows in toroidal geometry. Above a certain threshold in the fluctuation amplitude, small-scale toroidal secondary modes become unstable and shear apart what would otherwise be large-scale turbulent eddies, forcing the turbulence to remain close to this threshold. The threshold itself is set by a balance between the nonlinear rate and the time scale associated with the radial magnetic drifts, viz., at the outer scale,
\begin{align}
	 	(\tnl^{-1})^{\outero} \sim \omega_{\rmd i, x}^{\outero} \sim \frac{k_x^{\outero} \rhoi \vthi}{2R} \quad \Rightarrow \quad \aspect^{\outero} \sim \frac{R}{\LTi},
	\label{eq:grand_critical_balance}
\end{align}
where we have used \cref{eq:outer_scale} and \cref{eq:omlinouter}. The combination of \cref{eq:outer_scale}, \cref{eq:critical_balance}, and \cref{eq:grand_critical_balance} is referred to in the literature as `grand critical balance', and is supported by experimental measurements of ion-scale fluctuations in MAST \citep{ghim13}. In the present context, the effect of the toroidal secondary mode can therefore be interpreted as setting the outer-scale aspect ratio of ITG turbulence: as the temperature gradient increases, the aspect ratio of the eddies must also increase, driving the system to smaller radial scales in order to avoid becoming unstable to the toroidal secondary mode. Substituting \cref{eq:grand_critical_balance} into \cref{eq:heatflux_final} and once again using our expression for the parallel system scale $\Lpar \sim qR$, our resulting set of predictions for saturated ITG turbulence is
\begin{align}
	\frac{Q_i}{\Qgbs[i]} \sim q \bigg(\frac{a}{R}\bigg) \left(\frac{a}{\LTi}\right), \quad k_y^{\outero} \rhoi \sim \frac{\LTi}{q R}, \quad \aspect^{\outero} \sim \frac{R}{\LTi}.
	\label{eq:predictions_itg}
\end{align}
These latter estimates are different to those originally proposed for ITG turbulence  by \cite{barnes11}, who were unaware of the toroidal secondary mode, but are (unsurprisingly) identical to those proposed and numerically verified in \cite{nies26}.

Comparing \cref{eq:predictions_etg} and \cref{eq:predictions_itg}, it is clear that ETG turbulent heat fluxes are predicted to exhibit a much stiffer scaling with the temperature gradient than ITG ones, a direct consequence of the difference in how the outer-scale aspect ratio is set. This comparison highlights a fundamental asymmetry between ETG- and ITG-driven turbulence in toroidal systems: although they share the same parallel connection length, it is the distinct role of zonal dynamics (or lack thereof) that controls the outer-scale anisotropy and ultimately the scaling of the heat flux. We note that while this asymmetry is also responsible for the emergence of the Dimits regime \citep{dimits00,rogers00,kim02,diamond05,kobayashi12,villard13,villard14,zhu20PRL,ivanov20,ivanov22} in the ITG-driven case, we do not consider such dynamics here due to the assumption of strong linear drive --- our simulations are far from any characteristic linear or nonlinear critical gradients.

\subsection{CBC simulations}
\label{sec:cbc_simulations}
To put the above predictions on a quantitative basis, we perform simulations of both ETG- and ITG-driven turbulence in the well-studied `Cyclone-base-case' (CBC) geometry \citep{lin99,dimits00}, in which flux surfaces have a circular cross section with aspect ratio $R/a = 2.78$. We consider dynamics on the flux surface located at mid-radius $r/a = 0.5$, with the normalised magnetic shear $\shat = 0.2$ \footnote{This magnetic shear is lower than the more standard $\shat=0.8$ usually employed in CBC studies at $r/a=0.5$. We adopt this reduced value to avoid the well-documented issue of the ETG instability's failure to saturate at higher shear (see, e.g., \cite{dorland00,jenko00}). Since our focus is on the mechanism of ETG saturation when it occurs, rather than on the conditions under which saturation may fail, we leave the latter question aside for the present study. Performing the simulations at this different value of $\shat$ also provides an opportunity to demonstrate that the toroidal-secondary mechanism of \cite{nies26,nies26tsm} is robust to changes in magnetic shear, as could be anticipated from the theory thereof.}. The simulations evolve a single kinetic species, either electrons or ions, employing the appropriate adiabatic closure, viz., \cref{eq:adiabatic_ions} for adiabatic ions or \cref{eq:adiabatic_electrons} for adiabatic electrons. The normalised density gradient is fixed at $a/\Lns = 0.8$ throughout, while both the normalised temperature gradient $a/\LTs$ and safety factor $\safety$ are allowed to vary. The parallel simulation domain is extended for a single poloidal turn, viz., $\pol \in [-\pi, \pi]$. Unless otherwise stated, all data has been time-averaged over the last 80\% of the simulation time (see figures \ref{fig:CBC_timetraces_tprim}, \ref{fig:CBC_timetraces_q_etg}, and \ref{fig:CBC_timetraces_q_itg} in \cref{app:supplementary_data}), with any error bars showing the standard deviation over the same time period. Further details about the numerical setup can be found in \cref{app:numerical_details}.

\begin{figure*}
	\centering
	\includegraphics[width=\figscale\textwidth]{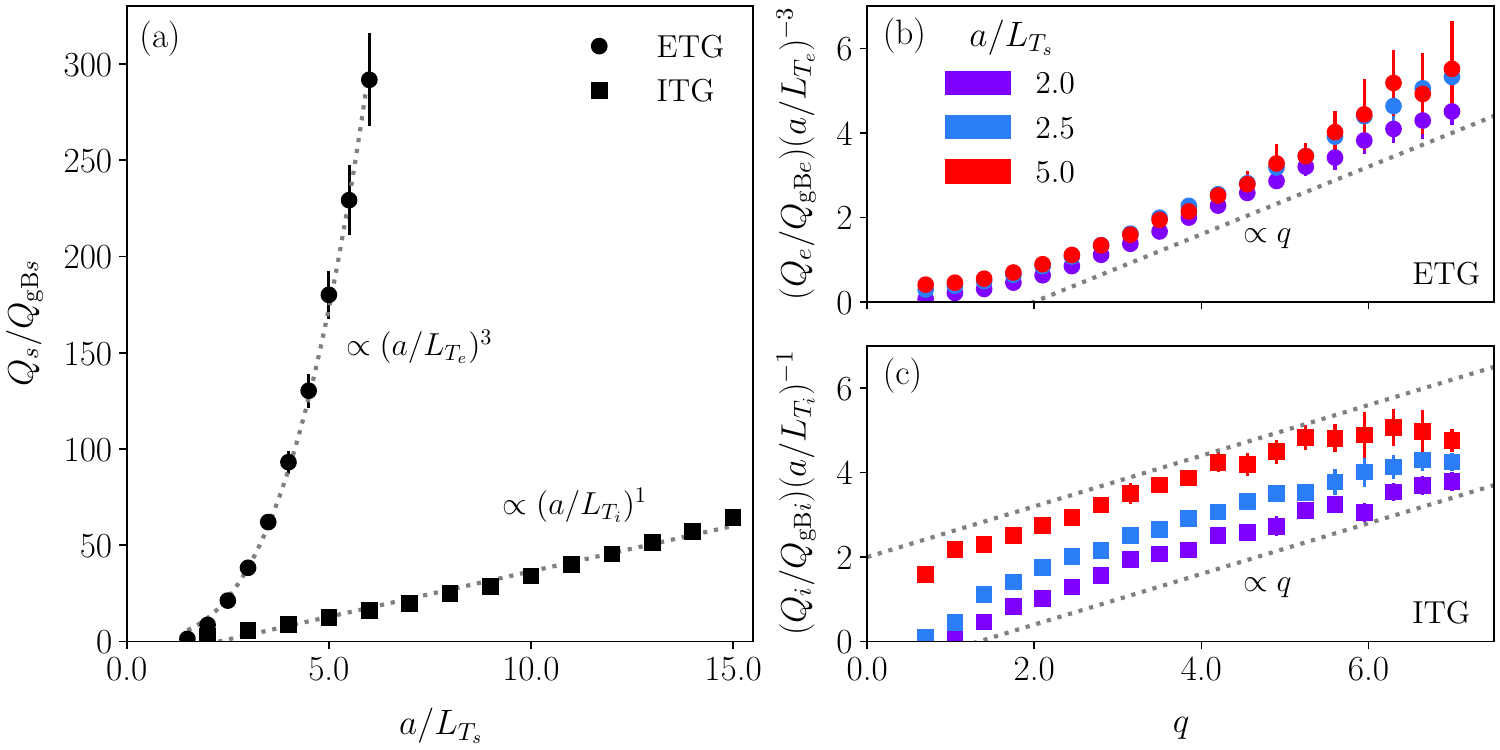}
	
	\caption[]{Heat-flux scalings from the CBC simulations. In all panels, the circles correspond to ETG cases and squares to ITG ones. The error bars show the standard deviation over time. (a) Heat flux as a function of the normalised temperature gradient $a/\LTs$, with the dotted lines showing the predicted scalings [first expressions in \cref{eq:predictions_etg} and \cref{eq:predictions_itg}]. (b) Electron heat flux rescaled by $(a/\LTe)^3$ for three different values of the normalised temperature gradient, indicated by the different colours. The dotted line shows the predicted scaling [first expressions in \cref{eq:predictions_etg} and \cref{eq:predictions_itg}]. (c) Similar to (b) but with the ion heat flux rescaled by $(a/\LTi)^1$.}
	\label{fig:CBC_heatflux_data}
\end{figure*}

\begin{figure*}[!t]
	\centering
	\includegraphics[width=\figscale\textwidth]{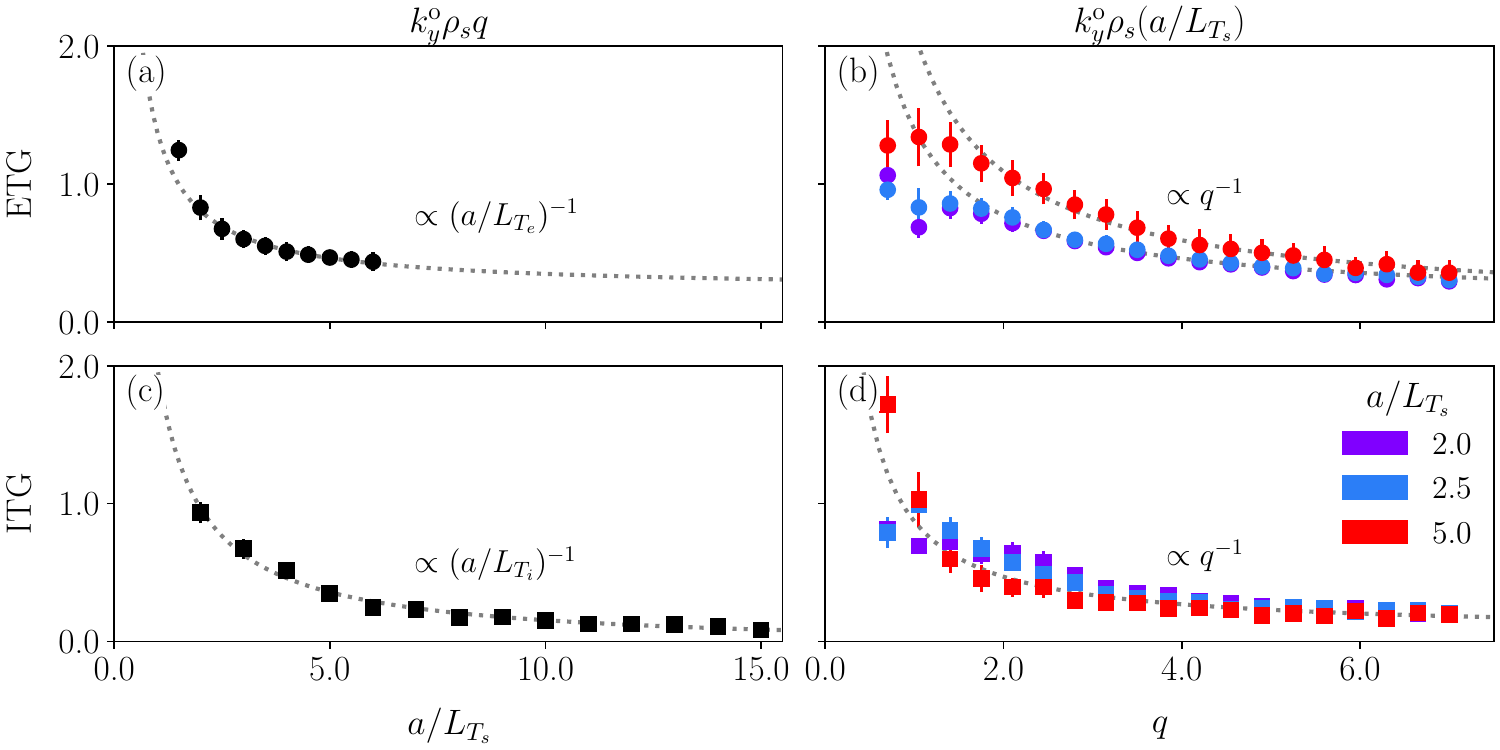}
	
	\caption[]{Outer-scale binormal wavenumber scalings from the same CBC simulations as in \cref{fig:CBC_heatflux_data}. In all panels, the time-averaged values of $k_y^{\outero}\rhos$ are indicated by the points, with circles corresponding to ETG cases and squares to ITG ones. The error bars show the standard deviation over time, while the dotted lines show the predicted scalings [second expressions in \cref{eq:predictions_etg} and \cref{eq:predictions_itg}] at large $a/\LTs$. (a) and (c): $k_y^{\outero} \rhos$ rescaled by the safety factor $q$ as a function of the normalised temperature gradient $a/\LTs$ for ETG and ITG, respectively. (b) and (d): $k_y^{\outero} \rhos$ rescaled by $a/\LTs$ as a function of $q$ at three different values of $a/\LTs$, indicated by the different colours, for ETG and ITG, respectively.}
	\label{fig:CBC_outer_scale_ky}
\end{figure*}

\begin{figure*}[h!t]
	\centering
	\includegraphics[width=\figscale\textwidth]{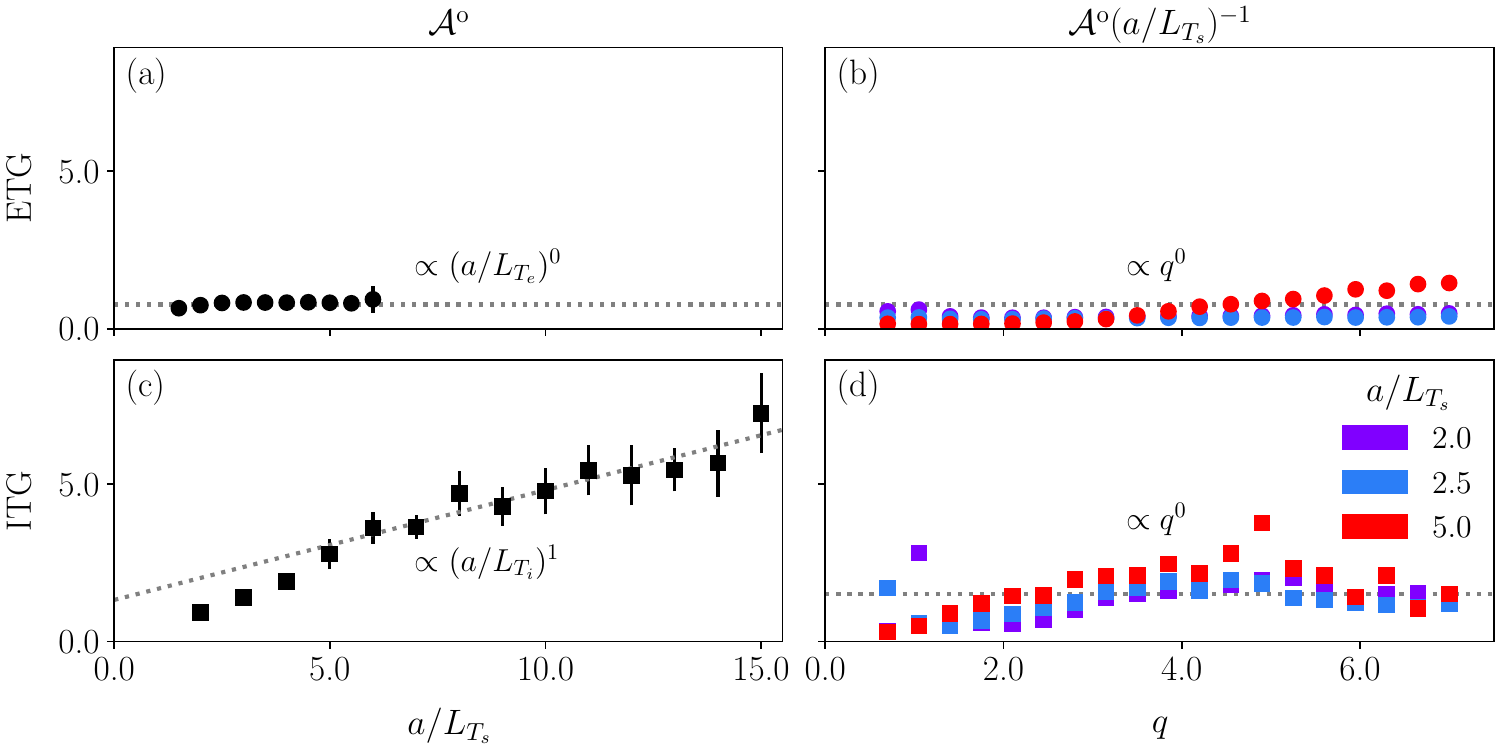}
	
	\caption[]{Outer-scale aspect ratio scalings from the same CBC simulations as in \cref{fig:CBC_heatflux_data}. In all panels, the time-averaged values of $\aspect^{\outero}$ are indicated by the points, with circles corresponding to ETG cases and squares to ITG ones. The error bars show the standard deviation over time, while the dotted lines show the predicted scalings [third expressions in \cref{eq:predictions_etg} and \cref{eq:predictions_itg}]. (a) and (c): $\aspect^{\outero}$ as a function of the normalised temperature gradient $a/\LTs$ for ETG and ITG, respectively. (b) and (d): $\aspect^{\outero}$ rescaled by $a/\LTs$ as a function of $q$ at three different values of $a/\LTs$, indicated by the different colours, for ETG and ITG, respectively.}
	\label{fig:CBC_aspect_ratio}
\end{figure*}

Let us first focus on the predictions for the scaling of the heat fluxes, i.e., the first expressions in \cref{eq:predictions_etg} and \cref{eq:predictions_itg}. \Cref{fig:CBC_heatflux_data}(a) shows the time-averaged gyro-Bohm-normalised heat fluxes from scans in $a/\LTs$ for each species (labelled `CBC~ETG~$\LTe$' and `CBC~ITG~$\LTi$' in \cref{tab:simulation_parameters}). The data displays a striking agreement with the predicted cubic and linear scalings for ETG and ITG, respectively. The ETG simulations could not be extended to higher values of $a/\LTe$ than those shown due to the computational constraints imposed by the saturated large-amplitude fluctuations ($\sim 10^3$ in normalised units) at these values. Panels (b) and (c) show the dependence of the heat fluxes on the safety factor $q$ for three different values of the normalised temperature gradient $a/\LTs$, with the data rescaled by the predicted temperature-gradient scalings (cubic for ETG, linear for ITG). For both ETG and ITG, the expected linear dependence on $q$ --- here acting as a proxy for the parallel system scale $\Lpar \sim qR$ --- is clearly reproduced. These panels also reinforce the temperature-gradient scaling: once rescaled, the datasets from the $q$-scans at different $a/\LTs$ collapse to within order-unity of each other, despite the underlying heat fluxes varying by more than an order of magnitude. The data collapse is somewhat less precise for ITG, which is not unexpected, since the chosen temperature gradients (the same as those in the ETG simulations) correspond to turbulence that is less strongly driven than in the ETG case (due to the presence of the Dimits shift in the former).

\begin{figure*}
	\centering
	\includegraphics[width=\figscale\textwidth]{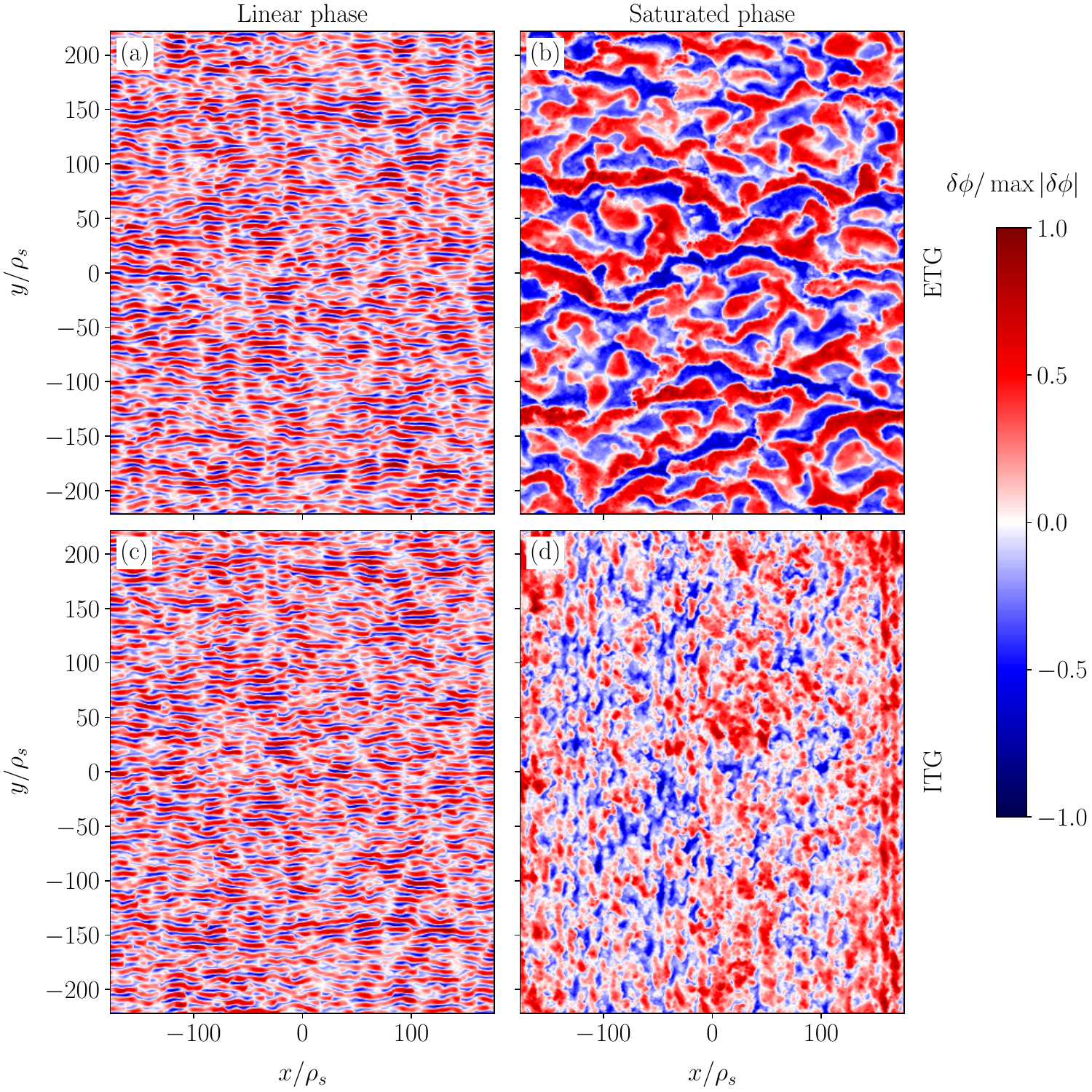}
	
	\caption[]{Snapshots of the perturbed electrostatic potential $\dphipot$ at $\pol=0$ normalised to its instantaneous maximum value from CBC simulations with $a/\LTs = 5.0$ and $q = 2.0$. The left and right columns show the nonlinear simulations during the linear and saturated phases, respectively; the top and bottom rows correspond to ETG and ITG turbulence, respectively. All panels show the entire simulation domain with the appropriate aspect ratio. The difference in the saturated states, and their associated outer-scale aspect ratios, is evident despite the almost identical linear phase.}
	\label{fig:CBC_realspace}
\end{figure*}

Having established that the heat fluxes follow the predicted scalings with both $a/\LTs$ and $q$, we now turn to the outer-scale binormal wavenumber $k_y^{\outero} \rhos$ and the aspect ratio $\aspect^{\outero}$, whose behaviour provides a more detailed test of the asymptotic scaling arguments underlying these results. We measure these in the same way as in \cref{sec:slab_geometry}, viz., $k_y^{\outero} \rhos$ is taken to be the wavenumber that maximises the binormal spectrum of the heat flux, while the $k_x^{\outero} \rhos$ that enters into $\aspect^{\outero}$ is identified with the spectral break in the radial heat-flux spectrum. \Cref{fig:CBC_outer_scale_ky} shows the values of $k_y^{\outero} \rhos$ from the same set of simulations shown in \cref{fig:CBC_heatflux_data}. Panels (a) and (c) are plots of $k_y^{\outero} \rhos$ as a function of the normalised temperature gradients $a/\LTs$, with the results following closely the predicted $(a/\LTs)^{-1}$ scaling from the second expressions in \cref{eq:predictions_etg} and \cref{eq:predictions_itg}. Panels (b) and (d) show $k_y^{\outero}\rhos$ rescaled by $a/\LTs$, now plotted against the safety factor $q$ for three different values of the temperature gradient. Here too the data reproduces the predicted $q^{-1}$ dependence, with the departures at lower values of $q$ likely reflecting the approach to marginal stability, a regime where our asymptotic predictions are not expected to hold. The corresponding outer-scale aspect ratios $\aspect^{\outero}$ are shown in \cref{fig:CBC_aspect_ratio}. Panels (a) and (c) are plots of $\aspect^{\outero}$ as a function of $a/\LTs$. In the ETG case, $\aspect^{\outero}$ remains essentially constant, whereas in the ITG case, it varies by nearly an order of magnitude. Thus, both ETG and ITG are consistent with their respective $(a/\LTe)^0$ and $(a/\LTi)^1$ predictions [the third expressions in \cref{eq:predictions_etg} and \cref{eq:predictions_itg}]. Panels (b) and (d) show $\aspect^{\outero}$ from $q$-scans at fixed $a/\LTs$, confirming the predicted independence on $q$: despite the latter varying by an order of magnitude, the aspect ratios change only by factors of order unity [cf. \cref{fig:sETG_scan}(d)]. To illustrate more vividly this contrast in behaviour between ETG and ITG, in \cref{fig:CBC_realspace} we plot snapshots of the perturbed electrostatic potential $\dphipot$ in CBC simulations with $a/\LTs = 5.0$ and $q = 2.0$ during both the linear and saturated phases of the turbulent evolution. The linear phase [panels (a) and (c)] looks almost identical for both ETG and ITG; this is to be expected given that their linear dispersion relations are isomorphic. Both simulation domains are the same when measured in units of the relevant gyroradius. The saturated phases, however, are manifestly distinct [panels (b) and (d)]. The ETG turbulence exhibits a preference for radially-elongated structures ($\aspect^{\outero} \approx 0.8$), while its ITG counterpart displays very fine-scale structure in the radial direction ($\aspect^{\outero} \approx 2.8$), consistent with it being limited by the toroidal secondary mode \citep{nies26,nies26tsm}. 

The stark visual contrast seen in \cref{fig:CBC_realspace} highlights the fundamentally different ways in which ETG and ITG turbulence are regulated in axisymmetric toroidal geometry. Taken together, figures \ref{fig:CBC_heatflux_data}-\ref{fig:CBC_aspect_ratio} demonstrate that the heat flux, binormal outer scale, and aspect ratio behave precisely as anticipated in \cref{eq:predictions_etg} and \cref{eq:predictions_itg}: ETG turbulence retains a roughly constant outer-scale aspect ratio as a function of equilibrium parameters, viz., $\aspect^{\outero} \sim 1$, while ITG turbulence is strongly shaped by zonal-flow dynamics, leading to systematically larger aspect ratios as the drive increases, $\aspect^{\outero} \sim R/\LTi$. These differences translate directly into the heat-flux scalings: ETG turbulence exhibits a cubic dependence on the normalised temperature gradient, while ITG turbulence is constrained to scale only linearly. Readers concerned that these results are specific to the CBC geometry used here may be reassured by the fact that such behaviour has, at least for ETG, been observed in other contexts: e.g., \cite{chapman22} found a cubic scaling with the temperature gradient in their investigations of nonlinear pedestal turbulence driven by ETG modes [see their equation (1), and the following discussion] --- but of course it will be valuable to build a body of evidence based on many other configurations, standard or otherwise. As a first step in this direction, \cref{app:comparison_to_mackenbach} provides a comparison of the ITG heat flux prediction \cref{eq:predictions_itg} to the database of simulations of ITG turbulence in axisymmetric geometry generated by \cite{mackenbach25}, finding good agreement for the dependence on both the parallel system size and equilibrium temperature gradient.

\section{Non-axisymmetric geometry}
\label{sec:non_axisymmetric_geometry}
Continuing the theme of increasing geometrical complexity, we now consider the predictions of \cref{sec:asymptotic_scaling_theory} in the context of fully three-dimensional toroidal (i.e., stellarator) geometry, once again treating both ETG- and ITG-driven turbulence. We take $\psi$ to be the toroidal magnetic flux divided by $2\pi$ and define the field-line label as $\alpha = \pol - \transform \tor $, in which $\tor$ and~$\pol$ are straight-field-line toroidal and poloidal angles, respectively, and $\transform = 1/\safety$ is the rotational transform. As previously, we adopt $z=\pol$ as the parallel coordinate. The key questions remain how both the parallel system scale $\Lpar$ and the outer-scale anisotropy $\aspect^{\outero}$ are determined in this geometry.

\subsection{Stellarator scalings}
\label{sec:stellarator_scalings}
Determining $\Lpar$ in stellarator geometry is considerably more challenging than in the axisymmetric case. In the latter, all quantities associated with the equilibrium magnetic field sampled along a field line labelled by $\alpha$ are periodic over a single poloidal turn (an interval of $\Delta \pol = 2\pi$). The natural parallel length is therefore just the distance travelled along the field line over one poloidal transit, giving $\Lpar \sim qR$ up to factors of order unity, as we argued in \cref{sec:axisymmetric_geometry}. However, such a simple picture does not obviously carry over to stellarators, where regions of good and bad curvature are distributed along the field lines in a far more complicated way.
	
All stellarator equilibria possess a discrete toroidal symmetry of the magnetic field \cite{dhaeseleer91}
\begin{align}
	B(\pol, \tor) = B(\pol, \tor + \Delta \tor), \quad \Delta \tor = \frac{2\pi}{\nfp},
	\label{eq:toroidal_symmetry}
\end{align}
where $\nfp \geqslant 1$ is the (integer) number of field periods (the number of times the plasma cross-section repeats itself in a single toroidal transit). However, for a given field line labelled by $\alpha$, the poloidal angle $\pol$ shifts by $\Delta \pol = \transform\Delta \tor = 2\pi \transform/\nfp$ in one field period $\Delta \tor = 2\pi/\nfp$. This change in $\pol$ does not in general correspond to a period of the magnetic geometry. This periodicity occurs only when the field line has completed an integer number of both poloidal transits and toroidal field periods, equivalent to the condition that
\begin{align}
	\Delta \pol = 2 \pi M, \quad M = \frac{\transform N}{\nfp},
	\label{eq:delta_theta_definition}
\end{align} 	
for coprime integers $N$ and $M$. While this is only ever precisely satisfied at rational flux surfaces $\iota = M \nfp/N$, we can approximate the effective width $\Delta \pol$ using a near-rational approximation for $N$ by choosing the smallest $M$ for which the toroidal advance differs from an integer number of field periods by less than a prescribed tolerance (here taken to be 1\%). This means that the rescaled poloidal angle $\pol/M$ provides a natural way to compare different magnetic geometries, since one (approximate) repetition of the magnetic geometry sampled along the field line corresponds to the common interval $\Delta(\pol/M)=2\pi$. Note that the axisymmetric case trivially corresponds to $M=1$. 

\begin{figure*}
	\centering
	\includegraphics[width=\figscale\textwidth]{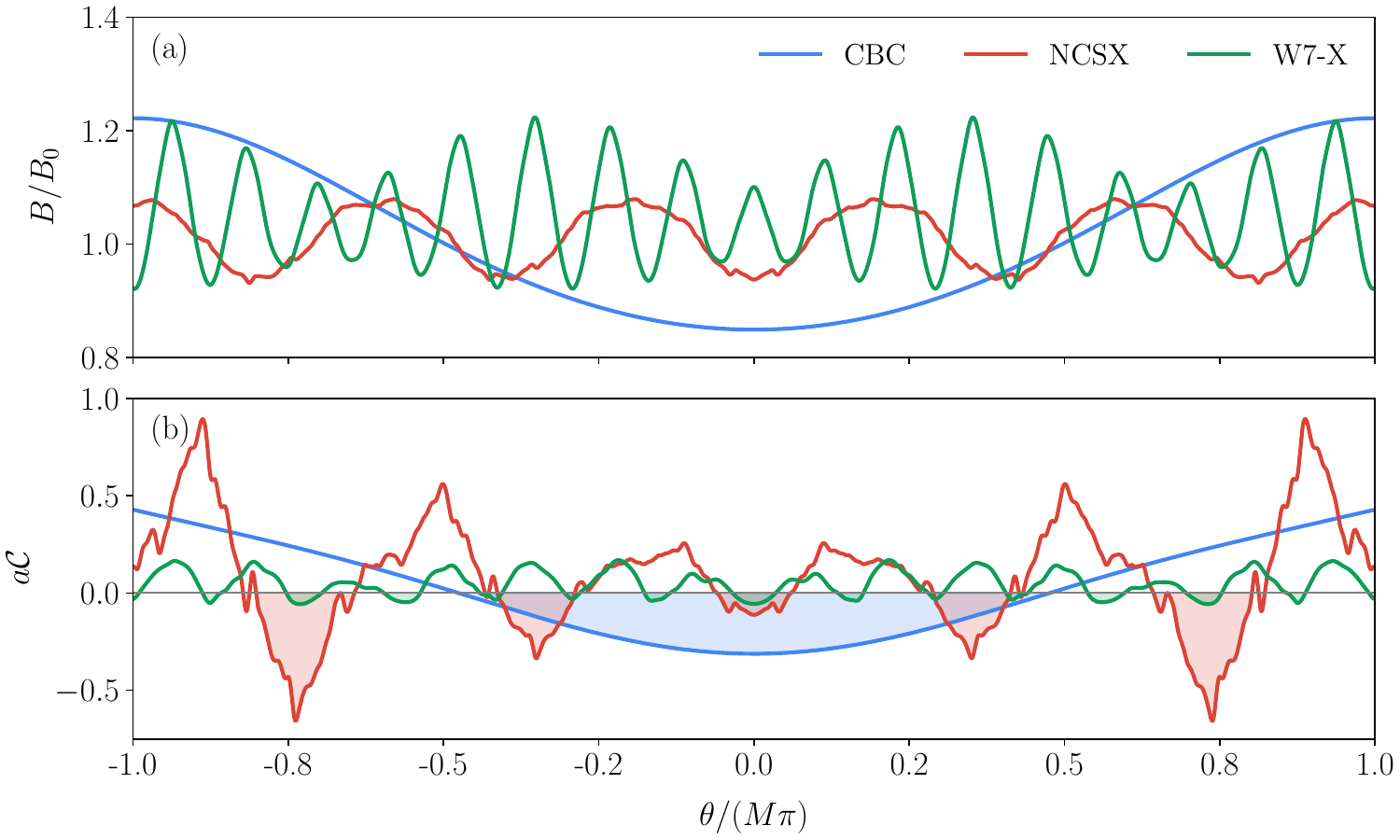}
	
	\caption[]{Magnetic geometry as a function of the rescaled parallel coordinate $\pol/M$ for the $\alpha = 0$ flux tube centred at $\pol = \tor =0$ and $r/a = \sqrt{\psi/\psi_{\mathrm{LCFS}}} = 0.5$. The colours indicate different magnetic equilibria. (a) Magnetic field strength. (b) Local magnetic curvature \cref{eq:curvature}. The shaded regions correspond to regions of bad curvature.}
	\label{fig:comparison_geometry}
\end{figure*}

To illustrate the challenge of determining $\Lpar$ for stellarators, we plot in \cref{fig:comparison_geometry} some quantities associated with the magnetic geometry for the $\alpha = 0$ field line centred at $\pol = \tor =0$ in three different magnetic equilibria: the CBC geometry considered in \cref{sec:cbc_simulations}, the National Compact Stellarator Experiment (NCSX, \cite{zarnstorff01}), and the high-mirror configuration of Wendelstein 7-X (W7-X, \cite{geiger15}), taking the near-rational approximations $M=1, 5, 3$ (corresponding to $\iota = 0.36, 0.49, 0.88$ and $\nfp = 1, 3, 5$), respectively. Panel (a) shows the magnetic field strength, while panel (b) shows the local magnetic curvature for electrostatic modes with $k_x = 0$ (cf. equation (2.30) of \cite{ivanov25invariant}):  
\begin{align}
	\curvature(z) = \frac{B_0}{B} \left(\ub \times \grad{\log B}\right) \cdot \grad y.
	\label{eq:curvature}
\end{align}
Defined in this way, $\curvature>0$ and $\curvature<0$ correspond to regions of good and bad curvature, respectively, meaning that we expect electrostatic modes to be unstable where $\curvature<0$ \citep{plunk14itg,ivanov25invariant}. We emphasise, however, that $\curvature$ is only one aspect of the magnetic geometry that influences stability; here we use it as a simple diagnostic with which to illustrate how the parallel structure of destabilising and stabilising regions differs between axisymmetric and stellarator equilibria. While the CBC case shows a broad, contiguous region of bad curvature, the same is not true for the stellarator cases: both NCSX and W7-X have multiple narrow regions of bad curvature separated by stabilising regions. This means there is no single parallel distance that cleanly separates destabilising from stabilising regions, and so it is not obvious that $\Lpar$ can be straightforwardly identified with a global quantity like the tokamak connection length. Indeed, clarifying what sets $\Lpar$ in stellarator turbulence remains an open problem, with recent work on critical-gradients \citep{robergclark22}, ITG-mode localisation \citep{rodriguez25}, and other optimisation strategies \cite{goodman24,kim24,landreman25} beginning to address this question.

We will therefore proceed by positing that there exists some well-defined $\Lpar$ --- deferring the question of exactly how this is determined for a given equilibrium until \cref{sec:lpar} --- and concentrate instead on the temperature-gradient scaling predicted by our asymptotic theory of \cref{sec:asymptotic_scaling_theory}. We assume, for now, that $\Lpar$ is a geometrical quantity that does not depend on the temperature gradient, an assumption that we will confirm \textit{a posteriori}. Now, the ETG predictions \cref{eq:predictions_etg} relied on axisymmetry only through the identification of $\Lpar$ with the tokamak connection length; the assumption that the aspect ratio remains order unity, $\aspect^{\outero} \sim 1$, is likewise expected to hold in the stellarator case. Substituting this into~\cref{eq:heatflux_final} and choosing $\lref = a$, where $a$ is the minor radius at the last closed flux surface, we obtain
\begin{align}
	\frac{Q_e}{\Qgbs[e]}  \sim \left(\frac{\Lpar}{a}\right) \left(\frac{a}{\LTe}\right)^3, \quad k_y^{\outero} \rhoe \sim \frac{\LTe}{\Lpar}, \quad \aspect^{\outero} \sim 1.
	\label{eq:predictions_etg_stellarator}
\end{align}
This analogy is less straightforward for the ITG case, given that our scaling for $ \aspect^{\outero}$ in the case of tokamaks drew heavily on the toroidal secondary mechanism — a mode that has so far only been demonstrated in axisymmetric systems \citep{nies26,nies26tsm}. Nevertheless, the underlying balance in \cref{eq:grand_critical_balance} --- between the outer-scale values of the ITG growth rate and the radial magnetic drift --- is not inherently specific to tokamaks as the toroidal secondary mode emerges from a competition between zonal-flow shearing and the radial magnetic drifts. Hence, although the specifics of the secondary mode may be modified in more complex stellarator fields, it is reasonable to posit, and test numerically, that a drift-regulated threshold for $\aspecto$ exists also in non-axisymmetric systems. Then our predictions for saturated ITG turbulence are [cf. \cref{eq:predictions_itg}]
\begin{align}
	\frac{Q_i}{\Qgbs[i]} \sim \left(\frac{\Lpar}{a}\right) \bigg(\frac{a}{R}\bigg)^2 \left(\frac{a}{\LTi}\right), \quad k_y^{\outero} \rhoi \sim \frac{\LTi}{\Lpar}, \quad \aspect^{\outero} \sim \frac{R}{\LTi}.
	\label{eq:predictions_itg_stellarator}
\end{align}
It is evident from \cref{eq:predictions_etg_stellarator} and \cref{eq:predictions_itg_stellarator} that the axisymmetric results \cref{eq:predictions_etg} and \cref{eq:predictions_itg} arise as the special case obtained by setting $\Lpar = qR$, as expected. In this sense, \cref{eq:predictions_etg_stellarator} and \cref{eq:predictions_itg_stellarator} represent the most general form of the ETG- and ITG-driven heat-flux scalings applicable to any toroidal geometry.

\subsection{NCSX simulations}
\label{sec:stellarator_simulations}
To test our theory, we perform simulations of both ETG- and ITG-driven turbulence in NCSX geometry. We consider the $\alpha=0$ field line on the flux-surface at mid-radius $r/a = \sqrt{\psi/\psi_{\mathrm{LCFS}}} = 0.5$, where $\psi_{\mathrm{LCFS}}$ denotes the value of $\psi$ at the last closed flux surface and $a$ is the effective minor radius of the toroidally-averaged cross-sectional area. The simulations once again evolve a single kinetic species, either electrons or ions, with the appropriate adiabatic closure applied. The normalised density gradient is fixed at $a/\Lns = 0.5$ throughout — chosen to maximise the heat flux at a given $a/\LTs$ \citep{thienpondt25} — while the normalised temperature gradient $a/\LTs$ is varied. Due to computational cost, the parallel simulation domain is extended for a single poloidal turn, viz., $\pol \in [-\pi, \pi]$. Unless otherwise stated, all data are time-averaged over the last 80\% of the simulation (see \cref{fig:NCSX_timetraces_tprim}), with error bars indicating the standard deviation over the same interval. Further details of the numerical setup are given in \cref{app:numerical_details}.

\begin{figure*}
	\centering
	\includegraphics[width=0.5\textwidth]{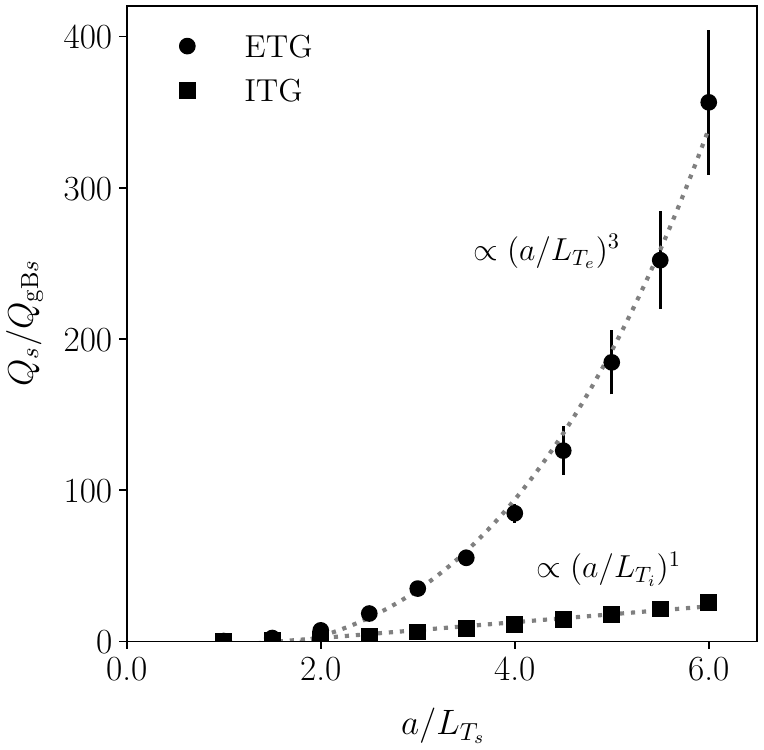}
	
	\caption[]{Time-averaged gyro-Bohm-normalised heat fluxes $Q_\s/\Qgbs$ from the NCSX simulations as functions of the normalised temperature gradients $a/\LTs$, with the dotted lines showing the predicted scalings [first expressions in \cref{eq:predictions_etg} and \cref{eq:predictions_itg}] at large $a/\LTs$. The circles and squares correspond to the ETG and ITG cases, respectively; the error bars show the standard deviation over time.}
	\label{fig:NCSX_heatflux_data}
\end{figure*}

\begin{figure*}
	\centering
	\includegraphics[width=\figscale\textwidth]{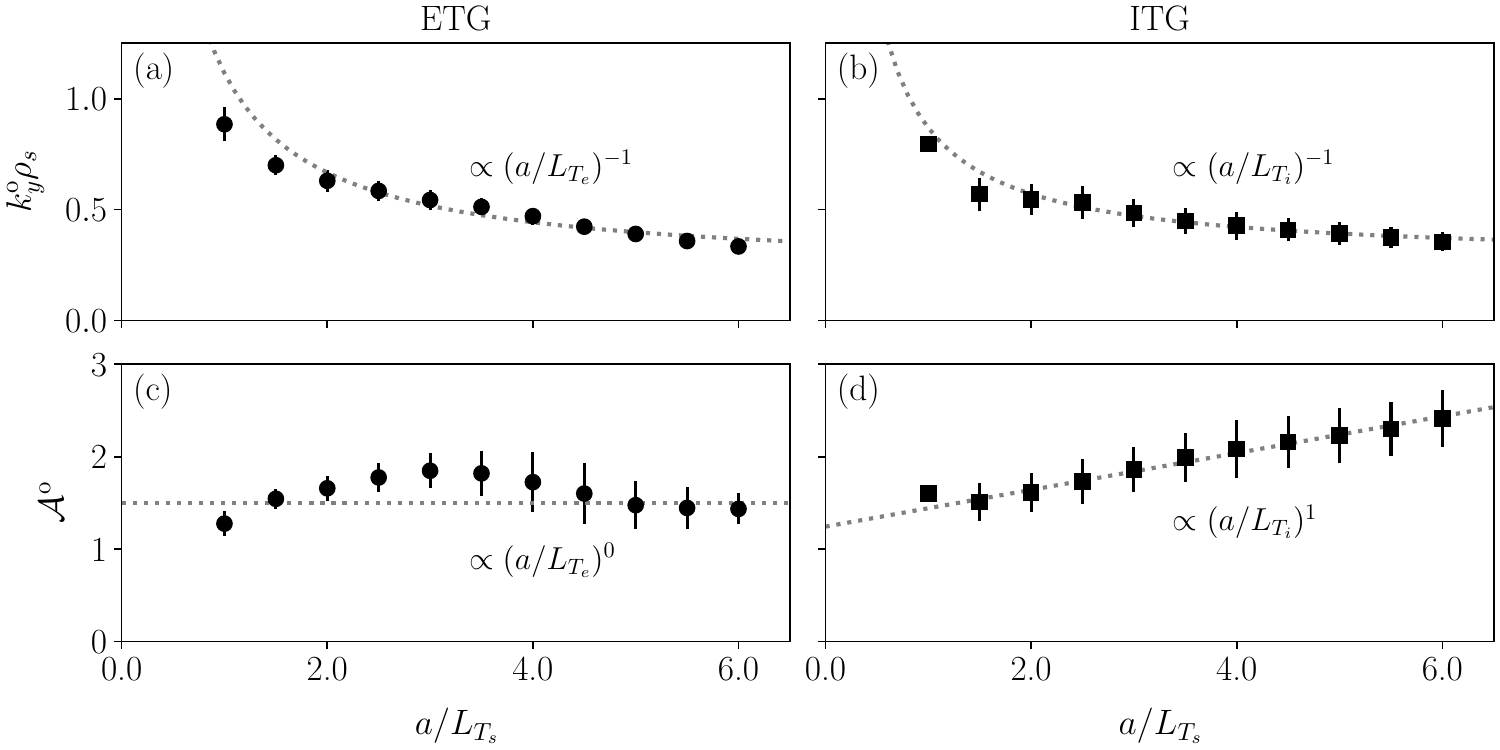}
	
	\caption[]{Outer-scale binormal wavenumber and aspect ratio scalings from the same NCSX simulations as in \cref{fig:NCSX_heatflux_data}. In all panels, the time-averaged values are indicated by the points, with circles corresponding to the ETG cases and squares to the ITG ones. The error bars show the standard deviation over time, while the dotted lines show the predicted scalings [second expressions in \cref{eq:predictions_etg_stellarator} and \cref{eq:predictions_itg_stellarator}]. (a) and (b): $k_y^{\outero} \rhos$ as a function of the normalised temperature gradient $a/\LTs$ for ETG and ITG, respectively. (c) and (d): $\aspect^{\outero}$ as a function of the normalised temperature gradient $a/\LTs$ for ETG and ITG, respectively.}
	\label{fig:NCSX_scales}
\end{figure*} 

The results of these scans in $a/\LTs$ for each species (labelled `NCSX ETG $\LTe$' and `NCSX ITG $\LTi$' in \cref{tab:simulation_parameters}) are shown in figures \ref{fig:NCSX_heatflux_data} and \ref{fig:NCSX_scales}. The heat fluxes (\cref{fig:NCSX_heatflux_data}) closely follow the predicted cubic and linear scalings for ETG and ITG, respectively, while both the binormal outer scale and aspect ratio (\cref{fig:NCSX_scales}) show good overall agreement with theoretical expectations. We were unable to obtain as large a range in $a/\LTs$ for the ITG simulations as in \cref{sec:cbc_simulations}, owing to the fact that more than one poloidal turn was required in order to capture adequately the dynamics at such gradients; that is, we found that the heat flux depended strongly on the parallel domain size for the larger temperature gradients. The computational cost of these larger temperature gradient simulations proved prohibitively expensive. Even within the restricted range of computed temperature gradients, however, the linear scaling of $\aspect^{\outero}$ for ITG is clearly reproduced, as is evident from \cref{fig:NCSX_scales}(d) [though admittedly with less sharp variation than was observed in \cref{fig:CBC_aspect_ratio}(c)]. Taken together, these results show that ETG and ITG turbulence follow the same $a/\LTs$ scalings as in the tokamak case, supporting the predictions \cref{eq:predictions_etg_stellarator} and \cref{eq:predictions_itg_stellarator}. These results also provide tentative confirmation that the toroidal-secondary mechanism of \cite{nies26,nies26tsm} survives the transition to non-axisymmetric geometries, though further study will be required to confirm the robustness of this finding.

\newpage

\begin{figure*}
	\centering
	\includegraphics[width=\figscale\textwidth]{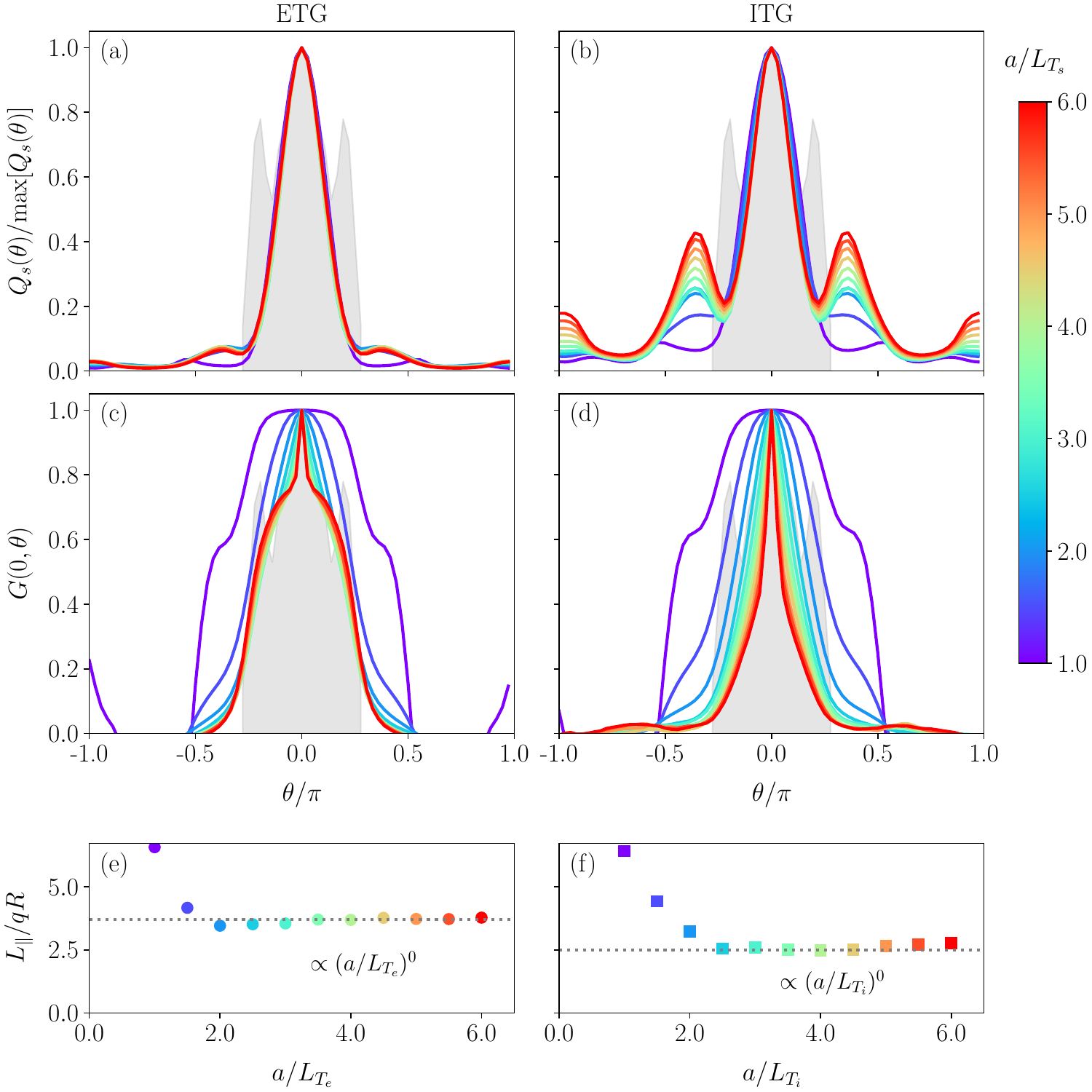}
	
	\caption[]{Parallel structure of the turbulence from the same NCSX simulations as in \cref{fig:NCSX_heatflux_data}, with the left and right columns corresponding to the ETG and ITG cases, respectively. In all panels, the colour indicates the value of the normalised temperature gradient $a/\LTs$. The grey shaded regions in panels (a)-(d) are the areas below the curve given by the negative of the local magnetic curvature \cref{eq:curvature} (plotted in \cref{fig:comparison_geometry}) normalised to its maximum value, indicating the range of $\pol$ in which the plasma is expected to be unstable. (a) and (b): time-averaged parallel spatial dependence of the heat flux \cref{eq:spectrum_parallel} normalised to its maximum value. (c) and (d): parallel correlation function \cref{eq:correlation_function} of the fluctuations at the outboard midplane ($\pol=0$). (e) and (f): the parallel system scale as determined from \cref{eq:lpar_correlation}, viz., $\Lpar = \Lparcorr(0)$, normalised to the connection length $qR$, with circles corresponding to the ETG cases and squares to the ITG ones. The horizontal dotted lines indicate the expected scaling, i.e., that $\Lpar$ is independent of $a/\LTs$.}
	\label{fig:NCSX_lpar}
\end{figure*}

That said, the theory with which we are claiming agreement rests on the assumption that there exists a well-defined $\Lpar$ for the underlying NCSX geometry that is independent of the temperature gradient. To assess the validity of this assumption, we examine directly the parallel structure of the turbulence in the set of simulations shown in figures \ref{fig:NCSX_heatflux_data} and \ref{fig:NCSX_scales}. The results of this are summarised in \cref{fig:NCSX_lpar}. Panels (a) and (b) show the parallel spatial dependence of the heat flux [see \cref{eq:heatflux_fourier}]
\begin{align}
	Q_\s(z) = \sum_{k_x, k_y} Q_{\s \vkperp}(k_x, k_y, z).
	\label{eq:spectrum_parallel}
\end{align}
Recall that we are using $z = \pol$ as the parallel coordinate. For all values of the normalised temperature gradient, the heat flux \cref{eq:spectrum_parallel} has a maximum coinciding with the region of bad curvature localised in the outboard midplane ($\pol =0$), as one would expect from existing theories of the linear ITG modes in stellarators (see, e.g., \cite{rodriguez25}). While the information provided by $Q_\s(z)$ can be useful for, e.g., assessing the correspondence between values of the local magnetic curvature \cref{eq:curvature} and the magnitude of the heat flux at a given location along the field line, it is not necessarily the most appropriate quantity for determining the parallel system scale $\Lpar$. Indeed, the causality argument proposed in \cref{sec:critical_balance}, which led to the binormal outer scale being determined in terms of the parallel system scale, implicitly assumed that $\Lpar$ was the parallel correlation length of the underlying turbulent fluctuations themselves, rather than the parallel length scale associated with the averaged quantity that is the heat flux. The more appropriate quantity from which to determine $\Lpar$ is therefore the two-point parallel correlation function of the non-zonal perturbed electrostatic potential:
\begin{align}
	\corr(z, z') =  \frac{\displaystyle \text{Re}\avg{\sum_{k_x, k_y\neq 0} \dphipotkperp(z, t) \dphipotkperp^*(z', t)}{t}}{\displaystyle\sqrt{\avg{\sum_{k_x, k_y\neq 0} \abs{\dphipotkperp(z, t)}^2}{t}}\sqrt{\avg{\sum_{k_x, k_y\neq 0} \abs{\dphipotkperp(z', t)}^2}{t}}},
	\label{eq:correlation_function}
\end{align}
where $\avg{\dots}{t}$ denotes an average over time. Note that, by definition, $-1 \leqslant \corr(z, z') \leqslant 1$, with the latter equality achieved at $z = z'$. This is plotted in panels (c) and (d) of \cref{fig:NCSX_lpar}. Except at the smallest values of the normalised temperature gradient, the parallel correlation function clearly retains the localisation around the region of bad curvature that we observed with the heat flux, but is significantly less peaked at larger values of $\pol$, particularly in the ITG case. Extracting the effective width of the parallel correlation function
\begin{align}
	\Delta z(z) = \frac{\displaystyle 2\int_{-\infty}^\infty \rmd z' \: |z - z'| \abs{\corr(z, z')}}{\displaystyle \int_{-\infty}^\infty \rmd z' \: \abs{\corr(z, z')}},
	\label{eq:z_width}
\end{align}
one can then estimate the parallel correlation length at a given location $z$:
\begin{align}
	\Lparcorr(z) = \int_{z-\Delta z/2}^{z+\Delta z/2} \frac{\rmd z'}{\abs{\vec{b} \cdot \grad z'}}.
	\label{eq:lpar_correlation}
\end{align}
It is clear from panels (e) and (f) of \cref{fig:NCSX_lpar} that the resultant parallel system scale evaluated at the outboard midplane, $\Lparcorr(0)$, is approximately independent of the normalised temperature gradient at all but the lowest values considered.\footnote{This may seem to contradict our prior statement that we could not perform simulations at higher values of $a/\LTs$ due to more than one poloidal turn being required in order to adequately capture the dynamics at such gradients. The distinction is that $\Lparcorr(0)$ measures the correlation length within the saturated state obtained in a given simulation domain, whereas the limitation at larger gradients arose because the saturated heat flux itself became sensitive to the parallel domain length. This means that a simulation over one poloidal turn may produce an apparently well-defined $\Lparcorr(0)$ while still failing to give a well-converged nonlinear state. We therefore have restricted the scan to the range over which the heat flux itself is converged with respect to the parallel extent of the flux tube.} This validates our \textit{a priori} assumption that $\Lpar$ may be treated as independent of the temperature gradient in the heat-flux scalings \cref{eq:predictions_etg_stellarator} and \cref{eq:predictions_itg_stellarator} and, more importantly, demonstrates that strongly driven electrostatic turbulence in the NCSX equilibrium does indeed organise itself in accordance with the asymptotic scaling theory developed in \cref{sec:asymptotic_scaling_theory}, with the binormal outer scale being determined by the parallel system scale.  

\subsection{Determining the parallel system scale in CBC, NCSX, and W7-X equilibria}
\label{sec:lpar}
As stated previously, what sets the parallel system scale for stellarator equilibria remains an active subject of research. Although we will not attempt to resolve this question definitively, we here offer some discussion aimed at clarifying what ingredients a complete theory will ultimately need to capture, before providing a tentative answer informed by our numerical results.  

We argued in \cref{sec:tokamak_scalings} that for axisymmetric equilibria, the parallel system scale is set by the inhomogeneity scale along the field line, corresponding to the width (in the parallel coordinate $z$) of the bad-curvature region on the outboard side of the device. In stellarator equilibria, however, the complex distribution of good- and bad-curvature regions along a field line suggests that the parallel system scale could be determined in a qualitatively different way. Specifically, the proximity of the bad-curvature regions to one another along the field line [see, e.g., the curve labelled `W7-X' in \cref{fig:comparison_geometry}(b)] raises the possibility that the turbulent fluctuations could remain correlated over several such regions, effectively extending the parallel system scale beyond what would be inferred from any single bad-curvature region, as in the tokamak. If this turns out to be the case, stellarator turbulence should be viewed not as being confined to an isolated unstable region in a given interval of $\pol/M$, but as evolving on a spatially varying sequence of stabilising and destabilising curvature patches in the same interval --- a kind of `corrugated' landscape on which the fluctuations live. In this picture, the effective parallel system scale would then have to reflect how fluctuations sample and respond to these alternating regions, and, in particular, how the turbulence remains correlated between them.

Distinguishing between these two possibilities cannot be done on the basis of our NCSX simulations alone due to the fact that, being a quasi-axisymmetric (QA), tokamak-like configuration, the equilibrium contains only a single bad-curvature region within our simulation domain (a poloidal turn of the outboard midplane). Given that this is the only parallel inhomogeneity that was accessible to the turbulence in our simulations, it is unsurprising that the resulting parallel correlation length was roughly the same as the width of this bad-curvature region [see panels (c) and (d) of \cref{fig:NCSX_lpar}]. We therefore consider a further set of simulations for both ETG and ITG in the CBC, NCSX, and W7-X equilibria where the parallel simulation domain is extended for at least three poloidal turns, viz., $\pol \in [-3\pi, 3\pi]$. The simulations once again evolve a single kinetic species, either electrons or ions, with the appropriate adiabatic closure applied. All simulations are conducted on a surface located at mid-radius $r/a=0.5$, with the stellarator simulations done for the $\alpha = 0$ field line. A normalised temperature gradient of $a/\LTs = 3.0$ is used throughout, while all other parameters such as the normalised density gradient $a/\Lns$, magnetic shear $\shat$, and safety factor $q$ are the same as those used in sections \cref{sec:cbc_simulations} and \cref{sec:stellarator_simulations}, with the same gradients for the W7-X case as for the NCSX one. Unless otherwise stated, all data are time-averaged over the last 80\% of each simulation's running time (see \cref{fig:comparison_timetraces}). Further details of the numerical setup are given in \cref{app:numerical_details}.  

\begin{figure*}
	\centering
	\includegraphics[width=\figscale\textwidth]{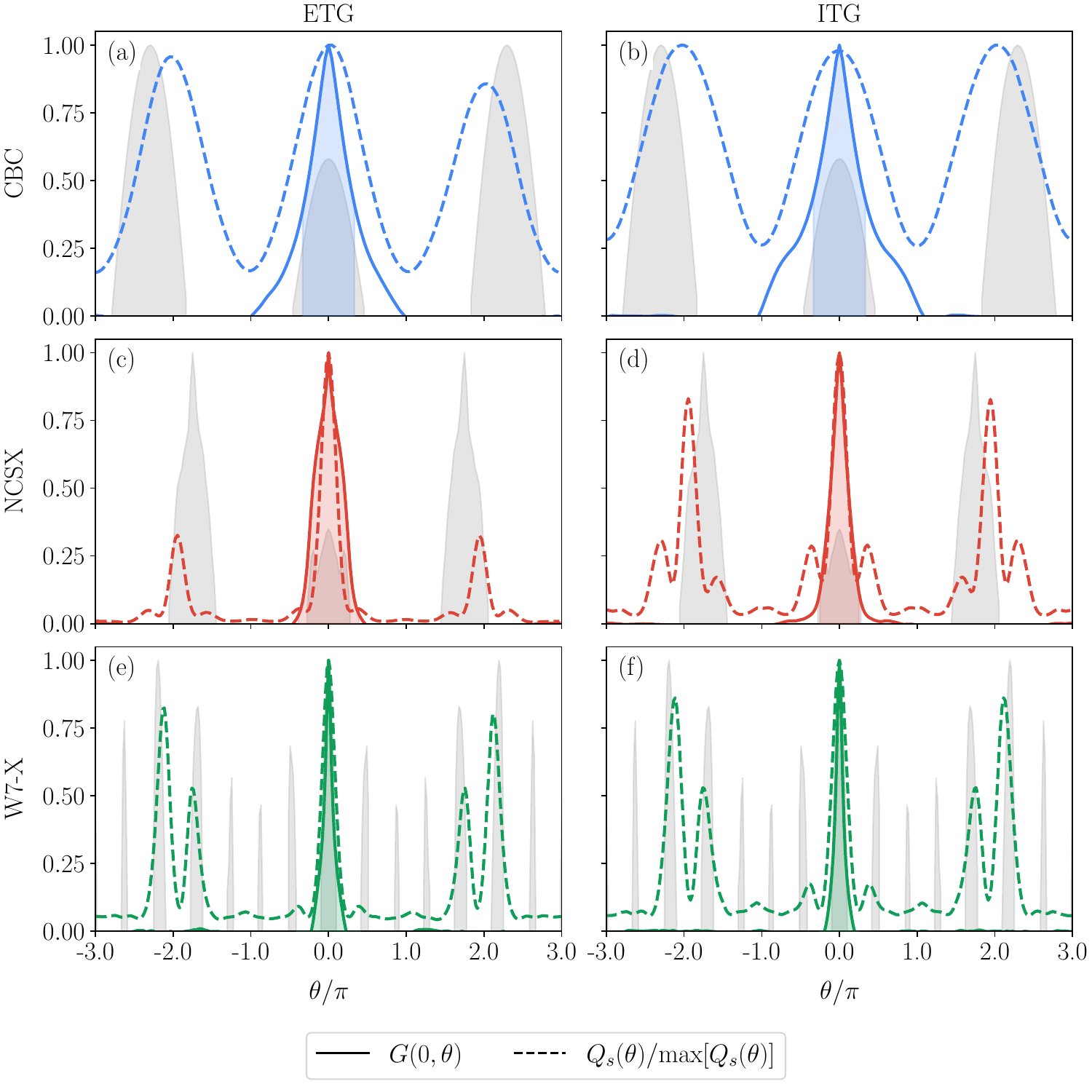}
	
	\caption[]{Parallel structure of the turbulence from the simulations considered in \cref{sec:lpar}. The left and right columns correspond to the ETG and ITG cases, respectively, while the rows correspond to the given particular magnetic equilibrium: CBC (blue), NCSX (red), and W7-X (green), respectively (these colours are the same as those in \cref{fig:comparison_geometry}). In all panels, the solid lines are the parallel correlation functions \cref{eq:correlation_function} of the fluctuations peaked at the outboard midplane ($\pol=0$), while the dashed lines are the time-averaged parallel spatial dependence of the heat flux \cref{eq:spectrum_parallel} normalised to its maximum value. The shaded coloured regions below the solid lines indicate a region of width given by the parallel system scale as determined from \cref{eq:lpar_correlation}, viz., $\Lpar = \Lparcorr(0)$, while the grey shaded regions are the areas below the curve given by the negative of the local magnetic curvature \cref{eq:curvature} normalised to its maximum value, indicating the range of $\pol$ in which the plasma is expected to be unstable. Recall from the discussion following \cref{eq:toroidal_symmetry} that the widths of these structures in $\pol$ cannot be meaningfully compared across the different magnetic equilibria.}
	\label{fig:comparison_lpar}
\end{figure*}

\begin{figure*}
	\centering
	\includegraphics[width=\figscale\textwidth]{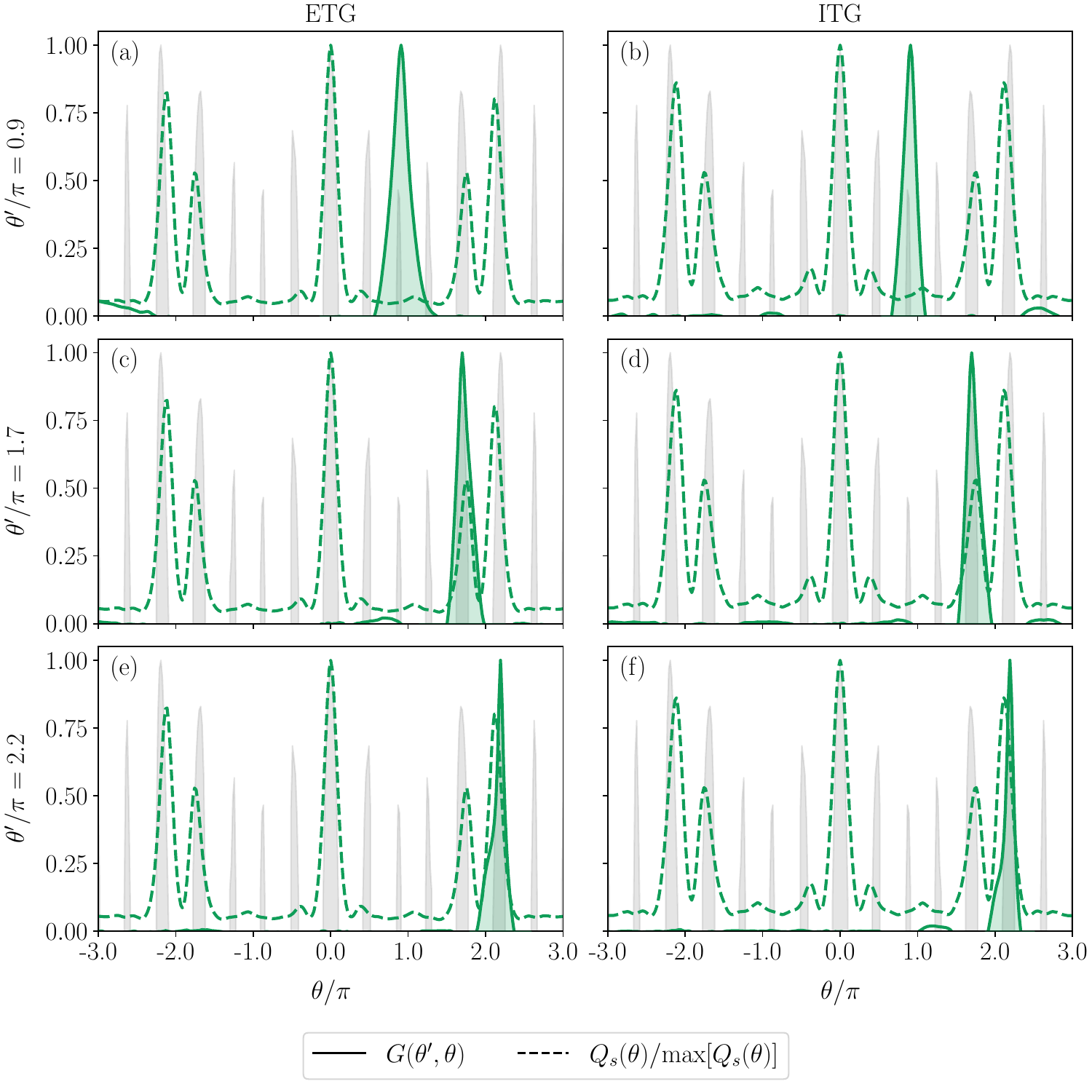}
	
	\caption[]{The same as \cref{fig:comparison_lpar}, except the rows now show the parallel correlation function \cref{eq:correlation_function} for fluctuations at different locations along the field line (solid lines) in the W7-X simulations.}
	\label{fig:comparison_lpar_w7x}
\end{figure*}

The results of these simulations are summarised in \cref{fig:comparison_lpar}. Each panel shows the time-averaged heat flux as a function of $\pol$ \cref{eq:spectrum_parallel} and the parallel correlation function \cref{eq:correlation_function} from an individual simulation, alongside shading indicating the bad-curvature regions [where $\curvature < 0$; see \cref{eq:curvature}]. These quantities are plotted against $\pol$ in order to make visible the parallel structure within each simulation domain. Apparent widths in $\pol$ should therefore be interpreted within each equilibrium separately, rather than as direct quantitative comparisons between them; the appropriate coordinate for comparing full repetitions of the sampled magnetic geometry is instead $\pol/M$. Considering first the CBC cases (top row), it is clear that there are contributions to the heat flux from each of the bad-curvature regions, as one would expect from an axisymmetric equilibrium. Furthermore, as we argued in \cref{sec:tokamak_scalings}, the correlation length of the fluctuations \cref{eq:lpar_correlation} is measured to be comparable to the width of these bad-curvature regions. As one might have guessed from the results of \cref{fig:NCSX_lpar}, the behaviour of the turbulence in the NCSX cases (middle row) is entirely analogous to the CBC cases, with its correlation length again comparable to the width of the central bad-curvature region. Interestingly, and perhaps disappointingly, the turbulence in the W7-X cases (bottom row) displays similar behaviour. Despite there being five distinct bad-curvature regions within a single poloidal turn of the outboard midplane that contribute to the heat flux, the correlation function remains localised around the central bad-curvature region, rather than extending across several such regions, a behaviour similar to the CBC and NCSX cases. The same is also true when considering the correlation function at different locations along the field line, viz., for different values of $\pol'$ in $G(\pol', \pol)$, as can be seen in \cref{fig:comparison_lpar_w7x}. 

These results suggest that the physics that determines the parallel system scale in stellarator turbulence is not qualitatively different to that encountered in axisymmetric geometries: in both cases, the turbulent fluctuations only remain correlated over a small width in the parallel coordinate $z$, corresponding to local bad-curvature regions. This means that the parallel system scale is still set by the scale of the inhomogeneity along the field line. Given, however, that stellarator field lines typically encounter multiple bad-curvature regions of comparable size distributed along $z$, this scale cannot be identified with the width of any single region at a particular location. Instead, it should be understood as some average measure of the length of the bad-curvature regions on a given field line, e.g., the following measure of the effective parallel system scale:
\begin{align}
	\Lpareff = \avg{\int_{z_1}^{z_2}\frac{\rmd z}{\abs{\vec{b} \cdot \grad z}}}{\{(z_1, z_2)\}},
	\label{eq:lpar_geometry}
\end{align}
where $\avg{\dots}{\{(z_1, z_2)\}}$ denotes an average over the set of points $\{(z_1,z_2)\}$ between which $\curvature(z)<0$ within a chosen interval $\Delta z$ of the field-line coordinate. The latter is taken to span one exact or approximate (depending on whether the field line is rational or irrational) repetition of the magnetic geometry when sampled along the field line. For example, when $z=\theta$, this corresponds to choosing $\Delta z = 2\pi M$, with $M$ determined by \cref{eq:delta_theta_definition}. The integral in \cref{eq:lpar_geometry} is therefore the arc length corresponding to a single uninterrupted region of bad curvature, and averaging over all such regions in $\Delta z$ yields a global measure of the parallel system scale that is not tied to a particular local starting point. Our simulation results provide tentative support for this estimate. Using \cref{eq:lpar_geometry} to calculate the right-hand side of the ETG heat-flux prediction \cref{eq:predictions_etg_stellarator} for the CBC, NCSX, and W7-X equilibria, and comparing with the measured heat fluxes, we find that the order-unity (constant) coefficient not fixed by \cref{eq:predictions_etg_stellarator} varies by less than a factor of two. The corresponding coefficient for the ITG heat-flux prediction shows a larger variation, by about a factor of five, perhaps because these ITG simulations are less strongly driven than their ETG counterparts and therefore less robustly in the asymptotic regime assumed by the theory. In any case, determining whether estimates such as \cref{eq:lpar_geometry} are an effective measure of the parallel system scale across a wide array of magnetic equilibria is the subject of ongoing work.

\newpage

An interesting feature of our results is the fact that the parallel correlation function --- and hence the resulting parallel system scale --- differs between ETG- and ITG-driven turbulence (see figures \ref{fig:NCSX_lpar}, \ref{fig:comparison_lpar}, and \ref{fig:comparison_lpar_w7x}). In particular, the parallel correlation length of ETG-driven turbulence is systematically larger than that of its ITG-driven counterpart, with the difference being most pronounced in the NCSX equilibrium [compare, for example, panels (e) and (f) of \cref{fig:comparison_lpar}]. Since the only distinction between the ETG- and ITG-driven simulations considered in this paper is the form of the adiabatic response of the non-driving species, this disparity in correlation lengths is most naturally attributed to differences in zonal-flow dynamics. A more detailed understanding of how zonal flows modify the parallel structure of the turbulence would therefore be required to refine \cref{eq:lpar_geometry}.

A more important limitation to our conclusions here is due to the nature of the local flux-tube approximation. In particular, our nonlinear simulations of NCSX and W7-X were restricted to the $\alpha = 0$ field line on the chosen flux surface, even though the turbulence may differ elsewhere on the same surface. Local studies of stellarator turbulence will sometimes attempt to account for this possible variation by performing nonlinear simulations for a number of different values of $\alpha$ and averaging the resultant turbulent statistics. However, this approach implicitly assumes that the zonal flows determined locally in one flux-tube domain are consistent with those in others --- an assumption that is highly restrictive and almost certainly violated in realistic stellarator equilibria \citep{helander11,alonso17,sanchez20}. A more faithful treatment of the zonal dynamics --- and hence of the saturation level of ITG turbulence, given the role of the toroidal secondary mode in the axisymmetric limit --- may ultimately require simulations that are global in $\alpha$ (while remaining local in $\psi$), an approach currently under active investigation \citep{acton25thesis}.

With these caveats acknowledged, however, our results strongly suggest that the scaling behaviour captured in \cref{eq:predictions_etg_stellarator} and \cref{eq:predictions_itg_stellarator} is already manifest in conventional flux-tube stellarator turbulence. This makes the present work a useful reference point for future poloidally global studies, and a first confirmation that the asymptotic theory developed in \cref{sec:asymptotic_scaling_theory} provides the right framework for interpreting stellarator turbulence and developing a more complete theory for the parallel system scale $\Lpar$.

\section{Summary and discussion}
\label{sec:summary_and_discussion}
Using simple physical arguments about the nature of the competition between the various time scales appearing in temperature-gradient-driven electrostatic gyrokinetic turbulence --- viz., rates of linear free energy injection, nonlinear decorrelation, and parallel propagation --- we have constructed a general theory for the asymptotic scaling of the heat flux with the normalised temperature gradient and other parameters of the magnetic equilibrium (\cref{sec:asymptotic_scaling_theory}). Two quantities emerged as central to this theory: the parallel system scale $\Lpar$ and the outer-scale aspect ratio $\aspect^{\outero} = k_x^{\outero}/k_y^{\outero}$, which together set the scaling of the heat flux.

While the exact value of the parallel system scale depended on the specific magnetic equilibrium --- the parallel size of the box in a uniform magnetic field (\cref{sec:slab_geometry}) or the connection length $\Lpar \sim qR$ in an axisymmetric toroidal one (\cref{sec:axisymmetric_geometry}) --- we argued that the outer-scale aspect ratio was not controlled by the geometry directly. The key factor that determined $\aspect^{\outero}$ was the nature of the response of the adiabatic plasma species. For ETG, the adiabatic ion response \cref{eq:adiabatic_ions} provides no mechanism to enforce a strong anisotropy between the radial and binormal directions, leading to $\aspect^{\outero} \sim 1$ independent of the electron temperature gradient $\LTe$. For ITG, by contrast, the modified adiabatic electron response \cref{eq:adiabatic_electrons} explicitly privileges zonal-flow dynamics, giving rise to the `toroidal secondary mode' identified in recent work \citep{nies26,nies26tsm}. This mode acts to limit the aspect ratio of ITG turbulence by destabilising structures above a certain threshold in aspect ratio, thus causing it to scale with the temperature gradient as $\aspect^{\outero} \sim R/\LTi$, a scaling first derived in axisymmetric geometry by \cite{nies26}. Together, these differences imply that ETG-driven core turbulence has an asymptotic scaling with the temperature gradient that is much stiffer than its ITG-driven counterpart, viz., in a toroidal system with minor radius $a$, 
\begin{align}
	\frac{Q}{\Qgbs} \sim \frac{\Lpar}{a}
	\begin{cases}
		\displaystyle \left(\frac{a}{\LTe}\right)^3,    & \displaystyle \text{ETG,} \\\\
		\displaystyle \left(\frac{a}{\LTi}\right)^1 \bigg(\frac{a}{R}\bigg)^2 ,  & \displaystyle \text{ITG.} 
	\end{cases}
	\label{eq:heatflux_summary}
\end{align}
where $\Qgbs = \dens \Ts \vths (\rhos/a)^2$ is the gyro-Bohm heat flux. For $\Lpar \sim qR$, the latter scaling is that originally proposed by \cite{nies26} in their theory of ITG-driven turbulence in axisymmetric geometries. We have confirmed these scalings using nonlinear gyrokinetic simulations in a slab (for ETG), CBC \citep{lin99,dimits00}, and NCSX \citep{zarnstorff01} magnetic equilibria (\cref{sec:slab_simulations}, \cref{sec:cbc_simulations}, and \cref{sec:stellarator_simulations}, respectively). Notably, this is the first numerical confirmation of the cubic heat-flux scaling with the temperature gradient in ETG-driven turbulence, anticipated by earlier theoretical work \citep{adkins22,adkins23thesis}. In the CBC case, we also verified the scaling of the ETG heat flux with the connection length $\Lpar \sim qR$, and confirmed the same scaling for the ITG heat flux previously demonstrated by \citep{barnes11,nies26}. Although determining a general expression for $\Lpar$ in stellarator magnetic fields remains an active area of research, the results of \cref{sec:non_axisymmetric_geometry} suggest that the physics determining the parallel system scale in stellarator turbulence is not qualitatively different to that encountered in tokamaks, with the turbulent fluctuations only remaining correlated over local bad-curvature regions in both cases. As such, the consistency of the observed scalings across both axisymmetric and non-axisymmetric configurations underscores the generality of the asymptotic scaling theory developed in \cref{sec:asymptotic_scaling_theory}.

A direct implication of \cref{eq:heatflux_summary} is that one would expect electrostatic turbulent transport to be stiffer should it be dominated by the electron channel, which can occur, e.g., when ITG is suppressed by strong $\exb$ shear in steep-gradient regions of a tokamak (e.g., the pedestal), particularly in spherical or low-aspect-ratio configurations (see, e.g., \cite{roach05,roach09,guttenfelder13,guttenfelder21,ren17} and references therein). In such contexts, the cubic scaling of ETG turbulence implies that modest increases in the temperature gradient can trigger disproportionately large increases in transport. Moreover, \cite{parisi22} demonstrated that at gradients well beyond those considered here, ETG transport can become even stiffer, finding empirically that the heat flux scaled quartically with the temperature gradient. This stronger scaling arises from the localisation of ETG modes along the field line away from $\pol=0$ caused by the combined effects of magnetic drifts and finite-Larmor-radius corrections at large drive. In the present context, this can be conceptualised in terms of a modification of $\Lpar$ in such steeper-gradient regions: balancing the diamagnetic and magnetic drifts, one can show that $\Lpar \sim (qR/\shat) (R/\LTe)$ [see equation (41) of \cite{parisi20} and the surrounding discussion], allowing for the quartic scaling to be recovered directly from \cref{eq:heatflux_summary} because $\Lpar$ depends on gradients of the plasma equilibrium. This behaviour has potentially important implications for reactor design and operation: if electron-channel transport dominates, the strong scaling of the ETG heat flux with the electron temperature gradient could make pedestal performance highly sensitive to small changes in gradients, complicating both predictive modelling and experimental control \cite{field23,turica25}.

The scalings \cref{eq:heatflux_summary} also have interesting implications for the relative importance of ETG and ITG transport as a function of the device aspect ratio $R/a$. Estimating the ratio of the ion to electron heat fluxes under the assumptions $\dens[i] \sim \dens[e]$, $\Ti \sim \Te$, and $\LTi \sim \LTe \sim a$, we find that the electron channel dominates once $R/a \gtrsim (m_i/m_e)^{1/4} \approx 7.8$. This threshold lies well above the aspect ratios envisioned for reactor-relevant tokamaks, such as ITER with $R/a = 3.1$ \citep{shimomura01,sips05}, but is directly relevant for recent stellarator designs, which are targeting substantially larger values $R/a \gtrsim 10$ \citep[see, e.g.,][]{hegna25,lion25}. This raises the possibility that in high-aspect-ratio stellarators, confinement could be limited primarily by ETG turbulence rather than ITG. Motivated by comparisons to W7-X experimental data \citep{plunk19,xanthopoulos20}, most stellarator turbulence studies to date have focussed on ITG as the dominant transport channel (note, however, \cite{weir21,wilms24}). If ETG transport becomes comparable to, or even larger than, ITG transport as $R/a$ increases, optimisation strategies may need to shift accordingly, especially if the stellarator configuration is so well optimised to suppress ITG turbulence that ion-scale fluctuations do not significantly shear the ETG-scale turbulence through cross-scale interactions (see \cref{sec:cross_scale_interactions} for further discussion). In particular, greater emphasis would be required on how the parallel system scale $\Lpar$ and field-line geometry shape electron-scale dynamics. These considerations suggest that the role of ETG transport in large-aspect-ratio stellarators is a key open question, with potentially important implications for future optimisation efforts.

To conclude, this work shows that, in the strongly driven regime far from marginality, a remarkably simple asymptotic theory, built only from considering the competition of a few fundamental time scales, can successfully explain the behaviour of both ETG- and ITG-driven electrostatic turbulence. The numerical results presented here therefore do more than confirm a set of asymptotic scalings: they provide a foundation for physics-based reduced models of turbulent transport that can be used to inform reactor design. Existing transport models employed in optimisation studies often rely on linear gyrokinetic theory, which can be a poor predictor of nonlinear performance, or on expensive nonlinear gyrokinetic calculations, which can be difficult to deploy systematically across wide regions of parameter space. By contrast, the asymptotic theory developed here isolates the essential balances that control ETG- and ITG-driven transport, distilling from gyrokinetics a pair of physically transparent parameters: the outer-scale aspect ratio $\aspecto$ and the parallel system scale $\Lpar$. While our results suggest that $\aspecto$ is a quantity insensitive to the details of magnetic geometry, subject to confirmation in a broader range of stellarator configurations, the parallel system scale $\Lpar$ must ultimately be set, at least in part, by that geometry. In tokamaks, this connection is well understood, but in stellarators, its precise determination remains an open problem, making it the key link between geometry and transport. Nevertheless, the fact that our framework reduces turbulent transport to a dependence on just $\Lpar$ and $\aspecto$ means it offers a tractable pathway for building reduced, geometry-aware models that can accelerate stellarator optimisation and guide the design of future reactors.

\section{Open issues}
\label{sec:open_issues}
While the results of this paper show that the theory developed here is broadly general (e.g., being agnostic to the choice of magnetic equilibrium), we necessarily introduced a number of simplifications in order to make analytical progress. Such simplifications obviously come at a cost to general applicability, and so we will here devote some space to a discussion of the most pressing lines of investigation left open by this work.

\subsection{Near-marginal regime}
\label{sec:near_marginal_regime}
One key assumption of \cref{sec:asymptotic_scaling_theory} was that the turbulence was sufficiently strongly driven that it sat far from marginality, i.e., at gradients 
sufficiently far above either the linear critical gradient or, in the case of ITG turbulence, the Dimits threshold \citep{dimits00}. In the latter case, the reason for this is obvious: the dynamics below the Dimits threshold are regulated by strong, long-lived, large-scale zonal flows \citep{rogers00,diamond05,stonge17,zhu20PRL,zhu20JPP,ivanov20}, representing an entirely different type of turbulence than considered here. At first sight, this restriction to large gradients might appear to limit the applicability of our theory, since much of fusion-relevant plasma is expected to sit at or close to the high-transport threshold. However, the results derived under the assumption of strong drive must still connect continuously to the near-threshold state, provided that the heat flux is a continuous function of the equilibrium gradients. For example, transport models developed for ETG turbulence in the tokamak pedestal often parametrise the heat flux as (cf. \cite{chapman22,hatch22,turica25})
\begin{align}
	Q_e \propto \left(\frac{a}{\LTe}\right)^2 \left(\frac{a}{\LTe} - \frac{a}{\LTe^{\mathrm{crit.}}}\right)^\lambda,
	\label{eq:near_marginal}
\end{align} 
where $\LTe^{\mathrm{crit.}}$ is some (linear or nonlinear) critical gradient and $\lambda$ a constant. This reproduces our predicted cubic scaling at gradients much larger than $\LTe^{\mathrm{crit.}}$ if $\lambda = 1$, which is close to the values obtained from simulations based on experimental data \citep{chapman22,turica25}. Models similar to \cref{eq:near_marginal} have also been employed to describe ITG-driven core turbulence (see, e.g., \cite{garbet04,mantica09}). In this sense, our results can be interpreted as the `universal' asymptotic behaviour of electrostatic gyrokinetic turbulence above threshold, to which near-marginal states must connect. Indeed, the predicted scalings were observed already for gradients only a factor of order unity above the linear critical values [see \cref{fig:CBC_heatflux_data}(a) and \cref{fig:NCSX_heatflux_data}], suggesting that extreme drive is not required for the scalings to manifest. This could make the theory particularly valuable for constructing reduced, physics-based transport models: it identifies the fundamental physical scalings that persist from the strongly driven limit into regimes close to marginality.

\subsection{Effect of trapped electrons and cross-scale interactions}
\label{sec:cross_scale_interactions}
The theory developed in \cref{sec:asymptotic_scaling_theory} was constructed for turbulence driven by a single species (either electrons or ions) --- with the other species treated as adiabatic through the appropriate closure --- in order to isolate the fundamental dynamics underpinning the turbulence. This, however, is only an approximation: any realistic fusion plasma necessarily contains at least four kinetic species (deuterium, tritium, electrons, helium), and their coupled dynamics can in principle alter both the drive and saturation mechanisms of the turbulence.

For ion-scale dynamics, the modified adiabatic electron closure \cref{eq:adiabatic_electrons} is predicated on the assumption of (infinitely) fast parallel electron streaming, viz., that the electrons will sample the entire flux surface in a single dynamical time. Since the nonlinear decorrelation rate of the most energetic ion-scale fluctuations is comparable to the ions' parallel-streaming rate [see \cref{eq:critical_balance}], this approximation remains accurate to leading order in the square root of the electron-ion mass ratio $\sqrt{m_e/m_i}$. Nevertheless, retaining kinetic electrons can give rise to additional ion-scale instabilities, such as trapped-electron modes \citep{adam73,adam76,catto78,cheng81}, which are often considered to be an important source of turbulent fluctuations and have been the focus of some schemes for optimising stellarator configurations \citep{proll12,helander13,rodriguez24}. Kinetic electrons may also affect ion-scale saturation indirectly by modifying the zonal-flow response: in stellarators, trapped-particle radial drifts influence the collisionless damping and residual level of zonal flows \citep{sugama05,sugama06,mishchenko08,helander11}, with kinetic electrons providing an additional correction \citep{monreal16,nicolau21}. Since the zonal-flow response is known to influence ITG turbulent transport in stellarator geometry \citep{watanabe08,xanthopoulos11}, these effects may change the quantitative level of transport without necessarily changing the underlying ITG-dominated character of the turbulence. Consistent with this picture, fully kinetic-electron simulations have found quantitative changes to ITG growth rates and transport levels while retaining zonal-flow-regulated ITG saturation \citep{singh22}. In regimes where the nonlinear state remains ITG-driven (rather than TEM-driven), we expect the scalings derived here to remain the relevant leading-order description, though it is up to future work to determine whether this is indeed the case.

In the case of ETG-driven turbulence, the adiabatic ion closure relies on a very well satisfied time scale separation (the ion gyrofrequency must be much larger than the characteristic frequencies of the turbulence), and on the length-scale separation between the ion-Larmor radius and the characteristic wavenumbers of the fluctuations, $\kperp \rhoi \gg 1$. However, given the scaling of the binormal outer scale $k_y^{\outero} \rhoe \sim \LTe/\Lpar$ [see \cref{eq:predictions_etg_stellarator}], the dominant injection scale shifts towards progressively larger values as the temperature gradient increases. At sufficiently large gradients, this binormal scale could approach the ion Larmor radius, at which point the effects of kinetic ions can no longer be neglected. Assuming $\Ti \sim \Te$, this crossover occurs when $\Lpar/\LTe \sim qR/\LTe \sim \sqrt{m_i/m_e}$. While such values are much higher than those typically realised in the core, they are comparable to gradients expected in the edge and pedestal regions (see, e.g., \cite{parisi22}), implying that ETG turbulence in these regimes certainly requires a fully kinetic treatment of both species. 

Numerically expensive gyrokinetic simulations with enough resolution to span both ion and electron scales have shown that ion-scale turbulence can directly suppress electron-scale fluctuations \citep{candy07,waltz07,maeyama15,howard16a}, with experimental results supporting this possibility \citep{howard16b}. This effect is attributed to the ion-scale turbulence providing strong $\exb$ shear flows that break up the turbulent electron-scale eddies, either in the perpendicular plane or via the parallel-to-the-field $\exb$ shearing mechanism proposed by \cite{hardman20}. It is worth emphasising that, unlike the breakdown of the adiabatic-electron approximation described in the previous paragraph, such cross-scale suppression of ETG does not require large gradients to manifest, and so further work will be required to determine whether our results are robust to such considerations, even at the comparatively modest gradients used in our simulations.

\subsection{Electromagnetic fluctuations}
\label{sec:electromagnetic_fluctuations}
We have restricted our analysis to electrostatic turbulence, in which the dynamics are mediated only through the perturbed electrostatic potential $\dphipot$ [see \cref{eq:gyrokinetic_equation} and \cref{eq:quasineutrality}]. Reactor-relevant tokamak and stellarator plasmas, however, will generally operate at finite beta (the ratio of the thermal to magnetic pressures), where magnetic-field fluctuations cannot be ignored. This not only introduces additional linear instabilities — most prominently the microtearing (MTM \cite{drake77,drake80,hassam80a,hassam80b,zocco15,larakers20,larakers21,chandran22,hardman23,patel25}) and kinetic-ballooning (KBM \cite{tang80,snyder99thesis,snyder01,snyder01gf,pueschel08,pueschel10,waltz10,wan12,wan13,guttenfelder13,ishizawa13,ishizawa14,ishizawa19,terry15,aleynikova17,kennedy23,kennedy24,giacomin24}) modes — but can also change the way in which those instabilities localise along field lines \citep{hardman23,ivanov25invariant}. In particular, while KBMs remain tied to bad-curvature regions, MTMs can extend more broadly, meaning that both the critical-balance estimate \cref{eq:critical_balance} and the identification $\Lpar \sim qR$ may need to be revisited in electromagnetic regimes. Moreover, finite fluctuations in the parallel magnetic-vector potential $\dApar$ and magnetic-field strength $\dBpar$ are expected to play a direct role in mediating outer-scale injection. Such electromagnetic fluctuations introduce more nonlinear terms into the gyrokinetic system of equations, meaning that the assumption that the nonlinear decorrelation rate is set by the (electrostatic) $\exb$ shearing rate \cref{eq:nonlinear_time_initial} may have to be modified. Developing a full scaling theory that includes these electromagnetic effects remains an open challenge, with tentative predictions already proposed by \cite{adkins22}. The present results provide a necessary electrostatic baseline from which such extensions can be built.

\section*{Acknowledgements}
The authors would like to thank G.~Acton, S.~C.~Cowley, F.~J.~Escoto, and G.~W.~Hammett for helpful discussions and suggestions at various stages of this project.

\section*{Data availability}
The data that supports the findings of this study are openly available \cite{adkins26data}.

\section*{Funding}
This work was supported in part by the Engineering and Physical Sciences Research Council (EPSRC) [EP/R034737/1 and EP/W006839/1]. The work of T.A. was supported in part by the Laboratory Directed Research and Development (LDRD) Program at the Princeton Plasma Physics Laboratory for the U.S. Department of Energy under Contract No. DE-AC02-09CH11466. The United States Government retains a non-exclusive, paid-up, irrevocable, world-wide license to publish or reproduce the published form of this manuscript, or allow others to do so, for United States Government purposes. The work of S.B. was supported by a grant from the Simons Foundation/SFARI (560651, AB) and the Department of Energy Award No. DE-SC0024548 (until March 31, 2025). The work of I.G.A. was supported by US Department of Energy grants DE-FG0293ER54197, DE-SC0018429, and DE-SC0024425. The work of M.B. and A.A.S. was supported in part by the EPSRC grant EP/R034737/1; the work of A.A.S. was also supported in part by the Simons Foundation via a Simons Investigator award. The work of R.N. was supported by a Leverhulme Trust International Professorship Grant (No. LIP-2020-014). The authors acknowledge the support of the Royal Society Te Ap\=arangi, through Marsden-Fund grant MFP-UOO2221 (T.A. and J.S.) and MFP-U0020 (R.M.). The work of R.M. was also supported in part by NASA grant 80NSSC24K0171. This research used resources of the National Energy Research Scientific Computing Center (NERSC), a Department of Energy User Facility using NERSC awards FES-ERCAP-0032464 and FES-ERCAP-0031087. Some of the simulations presented in this article were performed on computational resources managed and supported by Princeton University’s Research Computing, and on the HPC systems Raven \& Viper at the Max Planck Computing and Data Facility.

\section*{Declaration of interests}
I.G.A. is a co-founder and owner of GridFire Inc., a private fusion company.

\begin{appendix}

\section{Numerical details}
\label{app:numerical_details}
\begin{table}
	
	\centering
	
		\begin{tabular}{l | c c c c c c | c c c c c c | c}
			& $\npol$ & $\lref/\LTs$ & $\lref/\Lns$ & $\safety$ & $L_x/\rhos$ & $L_y/\rhos$   & $\ntheta$ & $N_x$ & $N_y$  & $\nhermite$ & $\nlaguerre$ & $\Dhyper$ & \text{Sims.} \\
			\hline
			sETG & --- & 1.00 & 0.00 & --- & 444 & 444   & (64, 128) & 216 & 216  & 24 & 6 & 0.05 & 11 \\
			\hline
			sITG & --- & 1.00 & 0.00 & --- & 444 & 444   & (64, 128) & 216 & 216  & 24 & 6 & 0.05 & 2 \\
			\hline
			CBC ETG $\LTe$ & 1.0 & (1.50, 6.00) & 0.80 & 2.80 & 354 & 444   & 48 & 486 & 512  & 12 & 6 & 0.10 & 10 \\
			\hline
			CBC ITG $\LTi$ & 1.0 & (2.00, 15.0) & 0.80 & 2.80 & 354 & 444   & 48 & 486 & 512  & 12 & 6 & 0.10 & 14 \\
			\hline
			CBC ETG $q$	& 1.0 & 2.00 & 0.80 & (0.70, 7.00) & 354 & 444   & 64 & 384 & 512  & 12 & 6 & 0.10 & 19 \\
			& 1.0 & 2.50 & 0.80 & (0.70, 7.00) & 354 & 444   & 64 & 384 & 512  & 12 & 6 & 0.10 & 19 \\
			& 1.0 & 5.00 & 0.80 & (0.70, 7.00) & 354 & 444   & 64 & 384 & 512  & 12 & 6 & 0.10 & 19 \\
			\hline
			CBC ITG $q$	& 1.0 & 2.00 & 0.80 & (0.70, 7.00) & 442 & 444   & 16 & 1024 & 432  & 8 & 12 & 0.10 & 19 \\
			& 1.0 & 2.50 & 0.80 & (0.70, 7.00) & 442 & 444   & 16 & 1024 & 432  & 8 & 12 & 0.10 & 19 \\
			& 1.0 & 5.00 & 0.80 & (0.70, 7.00) & 442 & 444   & 16 & 1024 & 432  & 8 & 12 & 0.10 & 19 \\
			\hline
			NCSX ETG $\LTe$ & 1.0 & (1.00, 6.00) & 0.50 & 2.06 & 587 & 444   & 72 & 648 & 512  & 16 & 6 & 0.10 & 11 \\  
			\hline 
			NCSX ITG $\LTi$ & 1.0 & (1.00, 6.00) & 0.50 & 2.06 & 587 & 444   & 72 & 648 & 512  & 16 & 6 & 0.10 & 11 \\  
			\hline
			CBC ETG $\Lpar$ & 3.0 & 3.00 & 0.80 & 2.80 & 237 & 222   & 144 & 324 & 256  & 12 & 6 & 0.10 & 1 \\
			\hline
			CBC ITG $\Lpar$ & 3.0 & 3.00 & 0.80 & 2.80 & 237 & 222   & 144 & 324 & 256  & 12 & 6 & 0.10 & 1 \\
			\hline
			NCSX ETG $\Lpar$ & 3.0 & 3.00 & 0.50 & 2.06 & 244 & 222   & 216 & 324 & 256  & 16 & 6 & 0.10 & 1 \\
			\hline
			NCSX ITG $\Lpar$ & 3.0 & 3.00 & 0.50 & 2.06 & 244 & 222   & 216 & 324 & 256  & 16 & 6 & 0.10 & 1 \\
			\hline
			W7-X ETG $\Lpar$ & 3.2 & 3.00 & 0.50 & 1.14 & 240 & 222   & 324 & 324 & 256  & 12 & 6 & 0.10 & 1 \\
			\hline
			W7-X ITG $\Lpar$ & 3.2 & 3.00 & 0.50 & 1.14 & 240 & 222   & 324 & 324 & 256  & 12 & 6 & 0.10 & 1 \\  
			\hline    
		\end{tabular} 

	\caption{The numerical parameters used for the gyrokinetic simulations considered in this paper. Values in parentheses indicate the minimum and maximum values for the corresponding column, with the final column (`sims') indicating the number of simulations in a given set. A dash in an entry indicates that the given simulation does not take that parameter as an input.}
	\label{tab:simulation_parameters}
	
\end{table}

All simulations in this paper were performed with the gyrokinetic code \texttt{GX} \citep{mandell24}, for which the relevant numerical input parameters are detailed in \cref{tab:simulation_parameters}. The quantities $L_x$ and $L_y$ denote the box sizes in the radial and binormal directions, respectively, and are related to the associated minimum wavenumber by $k^{\mathrm{min}}_{x, y} = 2\pi/L_{x, y}$. The spatial resolution is determined by $N_\parallel$, the number of parallel grid points, as well as $N_x$ and $N_y$, the numbers of Fourier modes in the radial and binormal directions, respectively. The velocity-space resolution is determined by the number of Hermite moments $\nhermite$ (dual to the parallel velocity) and Laguerre moments $\nlaguerre$ (dual to the magnetic moment). Perpendicular hyperviscosity of the form $\propto \kperp^4$ is employed to remove energy at the grid scale in perpendicular-wavenumber space. The strength of this dissipation is controlled by $\Dhyper$. A small amount of physical collisions, $(\lref/\vths)\nu_{\s\s} \approx 0.01$ --- in addition to `hypercollisions` (using the standard implementation in \texttt{GX}, see \cite{mandell24}) --- were included to provide a sufficient sink of energy at small scales in velocity space. Resolution studies were performed in all of these parameters to ensure numerical convergence. The flux-tube simulations shown in sections \ref{sec:cbc_simulations} and \ref{sec:stellarator_simulations} were extended for a single poloidal turn ($\npol =1.0$), while those in \cref{sec:lpar} were extended for three or more poloidal turns ($\npol \geqslant 3.0$). Standard or generalised `twist-and-shift' boundary conditions \citep{beer95,martin18} were used in all cases. All simulations were run for at least 2000 normalised time units, with the sETG/sITG simulations (see \cref{sec:slab_geometry}) having been run for at least 40000 normalised time units in order to have adequate statistics for time-averaging.

\section{Zonal-flow dominance in sITG turbulence}
\label{app:sitg_turbulence}
In this appendix, we show numerical results demonstrating the dominance of zonal flows in sITG turbulence. The numerical setup of the sITG simulations is identical to that used for the sETG simulations (see the discussion at the start of \cref{sec:slab_simulations} and \cref{app:numerical_details}), except with the electron response taken to be the modified-adiabatic one \cref{eq:adiabatic_electrons}. We consider two values of $\Lpar/\LTi$ that correspond to the maximum and minimum values of $\Lpar/\LTe$ used in the sETG scan in \cref{sec:slab_simulations}.

\begin{figure*}
	\centering
	\includegraphics[width=\figscale\textwidth]{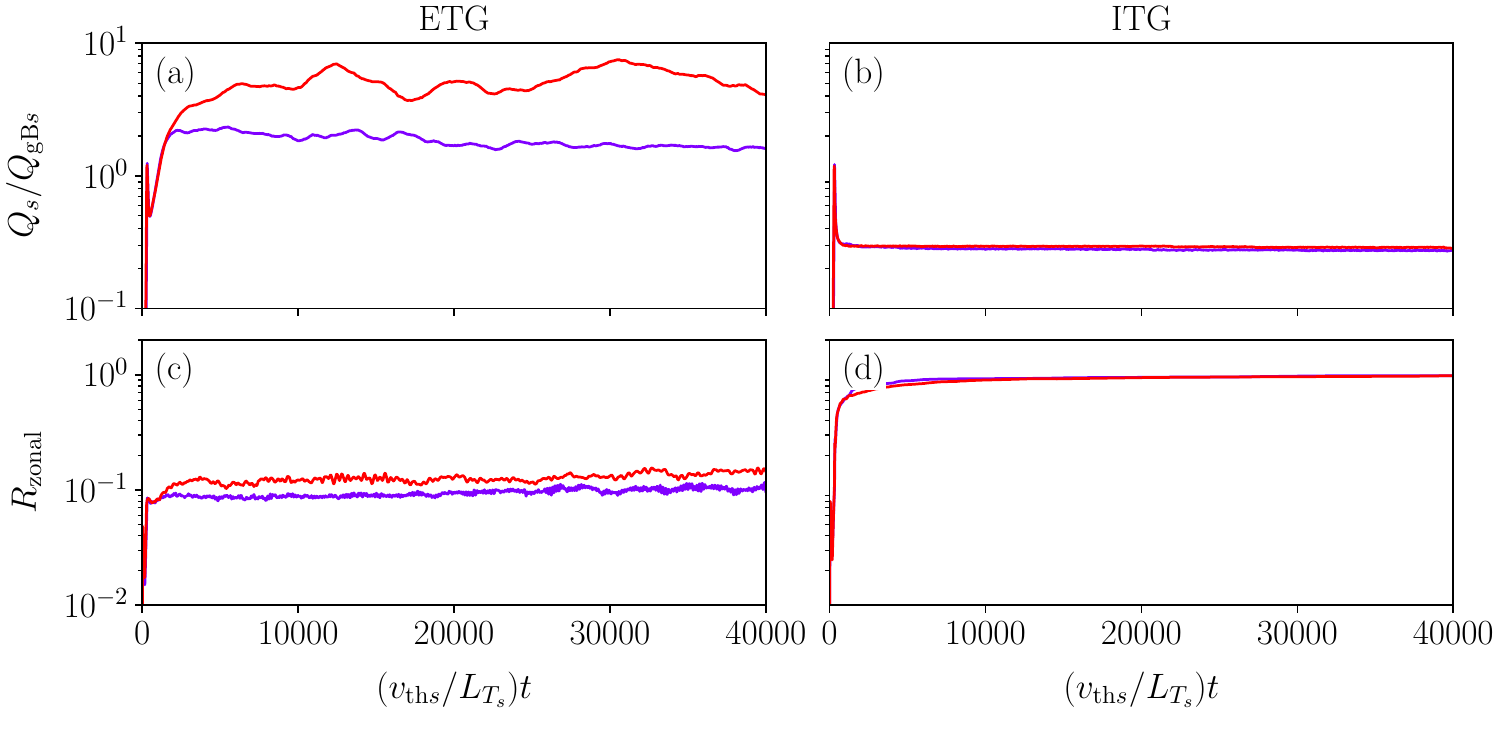}
	
	\caption[]{Time traces comparing the results of sETG and sITG simulations for $\Lpar/\LTs = 10 \pi$ (purple) and $\Lpar /\LTs = 20 \pi$ (red). (a) and (b): gyro-Bohm-normalised heat fluxes $Q_\s/\Qgbs$, where $\Qgbs= \dens \Ts \vths (\rhos/\LTs)^2$. (c) and (d): $R_\text{zonal}$ \cref{eq:zonal_ratio}.}
	\label{fig:slab_comparison_timetraces}
\end{figure*}

\begin{figure*}
	\centering
	\includegraphics[width=\figscale\textwidth]{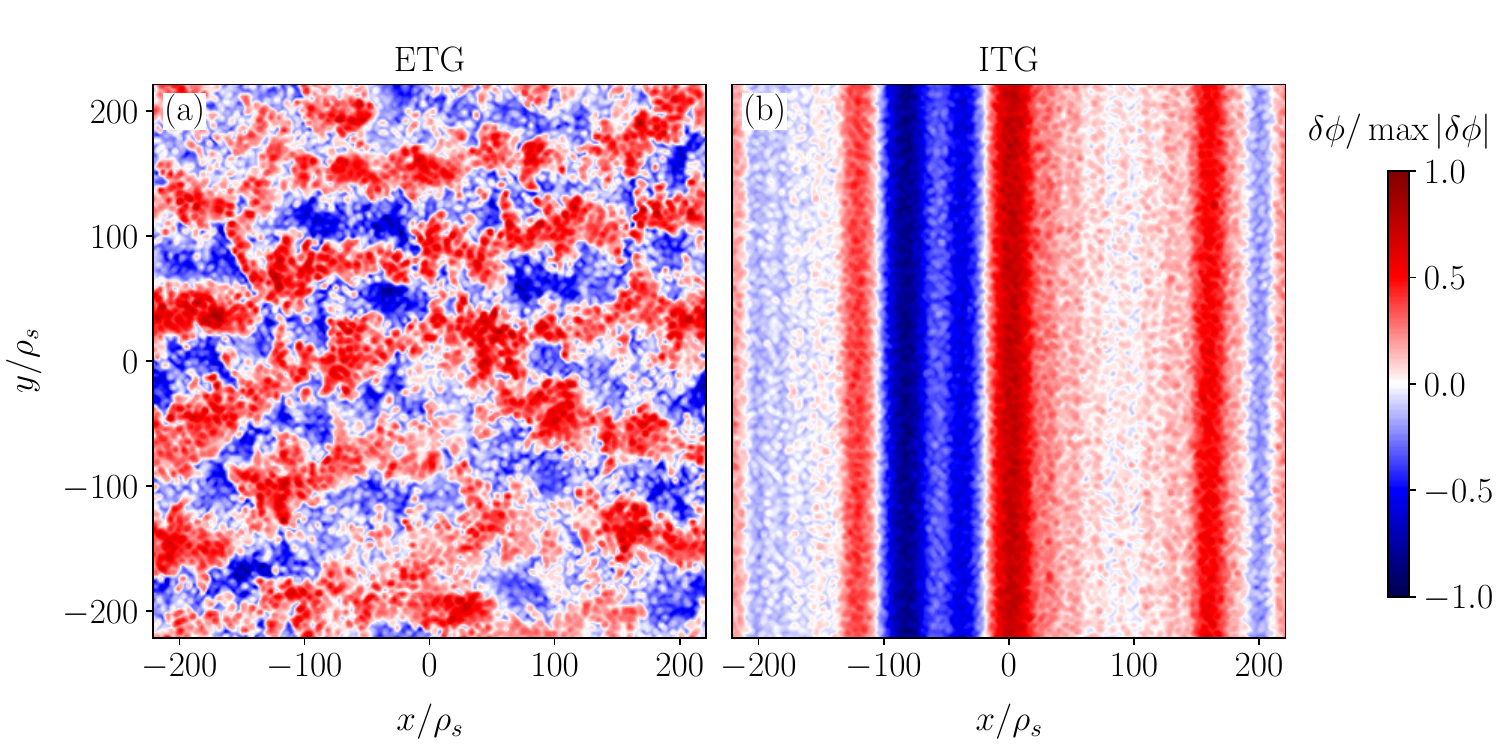}
	
	\caption[]{Snapshots of the perturbed electrostatic potential $\dphipot$ at $z=0$ normalised to its instantaneous maximum value from slab simulations with $\Lpar/\LTs = 20 \pi$. The left and right columns correspond to ETG and ITG turbulence, respectively, and both panels show the entire simulation domain with the appropriate aspect ratio.}
	\label{fig:slab_comparison_realspace}
\end{figure*}

In panels (a) and (b) of \cref{fig:slab_comparison_timetraces}, we plot time traces of the heat fluxes from simulations of both sETG and sITG turbulence at two different values of the parallel domain size, $\Lpar/\LTs = 10 \pi$ and $\Lpar/\LTs = 20 \pi$. While the sETG simulations show a variation in the average heat-flux level consistent with the scalings \cref{eq:predictions_setg} [see panel (a) of \cref{fig:sETG_scan}], the sITG simulations show negligible variation therein. This is due to the dominance of zonal flows in the saturated state; plotting the ratio of the amplitudes in the zonal and total electrostatic potential fluctuations 
\begin{align}
	\Rzonal^2 = \frac{\displaystyle \avg{\sum_{\vkperp, k_y=0}\left| \dphipotkperp \right|^2}{z}}{\displaystyle \avg{\sum_{\vkperp}\left| \dphipotkperp \right|^2}{z}},
	\label{eq:zonal_ratio}
\end{align}
we find that $\Rzonal \ll 1$ for the sETG cases, but $\Rzonal \approx 1$ for the sITG ones. This can also be seen clearly by plotting snapshots of the perturbed electrostatic potential $\dphipot$ in the saturated state of the $\Lpar/\LTs = 20 \pi$ simulations, which are included in \cref{fig:slab_comparison_realspace}. The dominant zonal bands associated with significant zonal-flow activity are evident in the sITG case [panel (b)]. It is worth noting, however, that these sITG simulations have not been run for long enough in order to observe the decay of the zonal flows due to collisions or hyperdissipation, these both being numerically small (see \cref{app:numerical_details}), and so we make no claim to the asymptotic longevity of this zonally-dominated state. 

\section{Comparison with the simulation dataset of \cite{mackenbach25}}
\label{app:comparison_to_mackenbach}
We here provide a comparison of our prediction for the scaling of the ITG-driven heat flux in axisymmetric (tokamak) geometry, given by the first expression in \cref{eq:predictions_itg}, against the database of \texttt{GX} simulations generated by \cite{mackenbach25} [see their appendix D and \url{https://zenodo.org/records/15799186}]. All simulations use a Miller-Turnbull magnetic equilibrium \cite{turnbull99} (the generalisation of the usual Miller equilibrium to include flux-surface `squareness'), for which the geometric coefficients (and their derivatives) have been randomly sampled over some prescribed range. The equilibrium plasma gradients of both density and temperature are also randomly sampled from a normal distribution. Further details about the database generation, including numerical convergence of the simulations, can be found in \cite{mackenbach25}. In the following analysis, we exclude simulations with time-averaged heat fluxes satisfying $Q_i/\Qgbs[i] \leqslant 0.1$ and $Q_i/\Qgbs[i] \geqslant 1000$, regarding the former as `stable' and the latter as insufficiently well-resolved given the numerical setup adopted by \cite{mackenbach25}.

\begin{figure*}
	\centering
	\includegraphics[width=\figscale\textwidth]{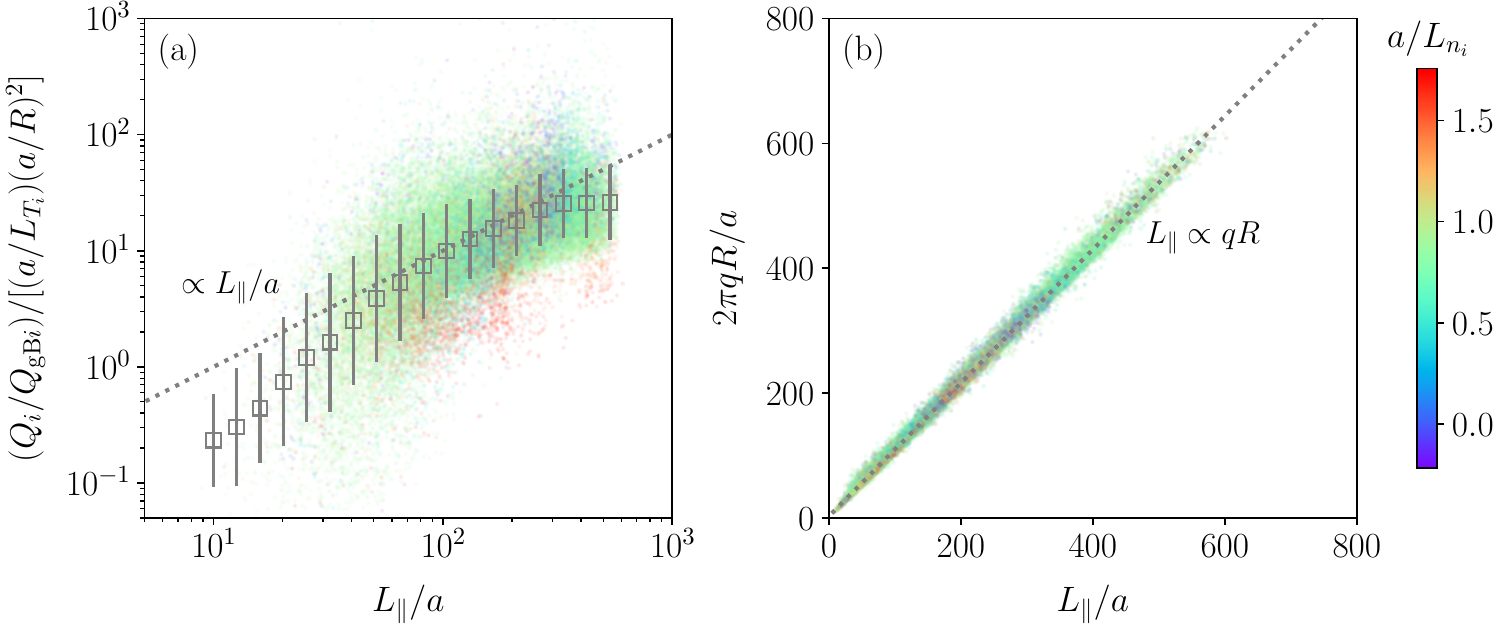}
	
	\caption[]{Heat flux scalings with the parallel system size $\Lpar$ in the ITG-driven simulations of \cite{mackenbach25}. In both panels, the colour indicates the value of the normalised density gradient $a/\Lni$. (a) The time-averaged ion heat flux $Q_i/\Qgbs[i]$ normalised to isolate the dependence on $\Lpar$ [cf. \cref{eq:predictions_itg}], given here by the connection length \cref{eq:connection_length}. The dotted line shows the predicted scaling, while markers show the geometric mean of the heat flux values in (logarithmic) bins in $\Lpar/a$, with the error bars showing the (logarithmic) standard deviation. (b) A comparison between the estimate for the connection length $2\pi \safety R$ and the computed value \cref{eq:connection_length}.}
	\label{fig:mackenbach_data_lpar}
\end{figure*}

\begin{figure*}
	\centering
	\includegraphics[width=\figscale\textwidth]{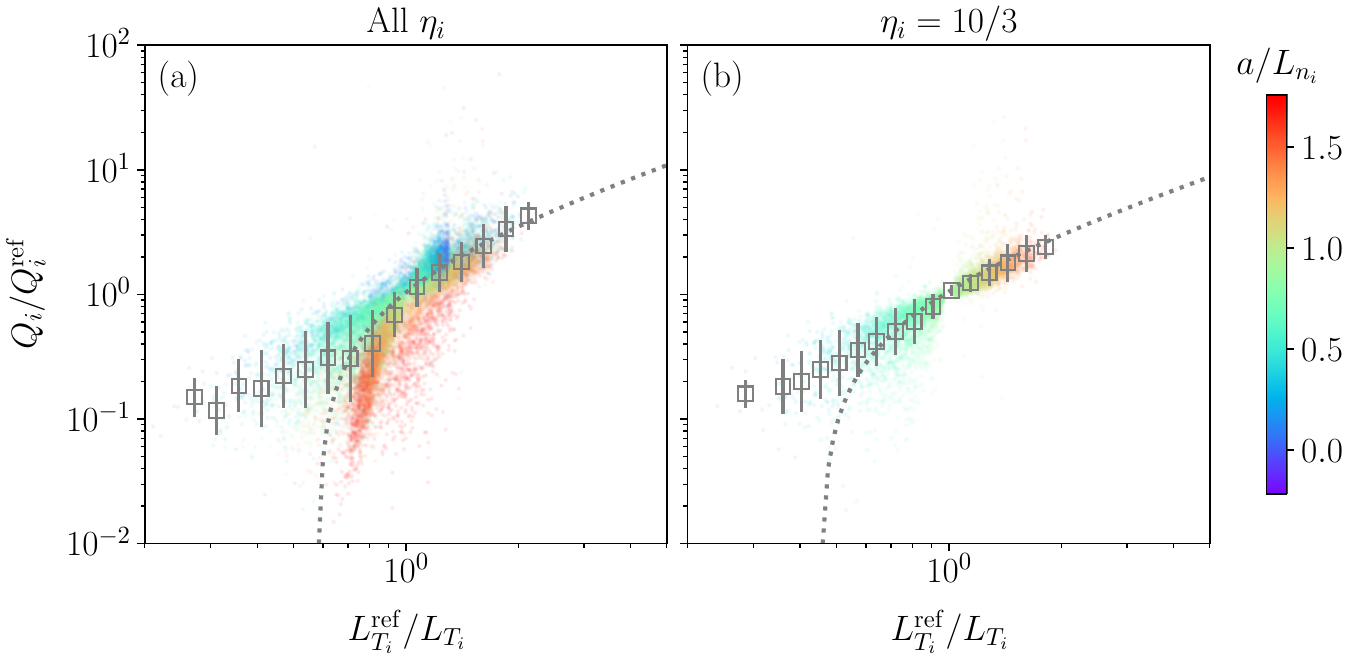}
	
	\caption[]{Heat flux scalings with the normalised temperature gradient in the ITG-driven simulations of \cite{mackenbach25}. For each magnetic equilibrium considered, the time-averaged ion heat flux $Q_i$ is normalised to some reference value $Q_i^{\mathrm{ref}}$ corresponding to normalised gradients $a/\LTi = 3.0$ and $a/\Lni = 0.9$. In both panels, the colour indicates the value of the normalised density gradient $a/\Lni$, and the markers are the same as those in \cref{fig:mackenbach_data_lpar}. The dotted line shows an offset-linear scaling of the form $Q_i \propto a/\LTi - \left(a/\LTi\right)^{\mathrm{crit.}}$, for some critical normalised temperature gradient $\left(a/\LTi\right)^{\mathrm{crit.}}$. (a) Data from all combinations of gradients $a/\LTi$ and $a/\Lni$. (b) Data from simulations with $\eta_i = 10/3$.}
	\label{fig:mackenbach_data_grad}
\end{figure*}

In \cref{fig:mackenbach_data_lpar}(a), we plot the time-averaged ion heat flux as a function of the parallel system size $\Lpar$, taken here to be the connection length, viz., 
\begin{align}
	\Lpar = \int_{-\pi}^{\pi} \frac{\rmd z}{\ub \cdot \grad z}.
	\label{eq:connection_length}
\end{align}
The data shows an approximately linear scaling with \cref{eq:connection_length}, particularly for the higher values of the normalised gradients at which you would expect the asymptotic scaling theory of \cref{sec:asymptotic_scaling_theory} to be most applicable. Panel (b) also shows that, unsurprisingly, the estimate of the connection length as $\Lpar \sim \safety R$ that we adopted in \cref{sec:tokamak_scalings} remains valid across a wide variety of equilibria. In \cref{fig:mackenbach_data_grad}, we plot the time-averaged ion heat flux as a function of the normalised temperature gradient $a/\LTi$. To isolate the effects of changes in the plasma gradients from those of the magnetic equilibrium, the heat flux data from each distinct magnetic equilibrium is normalised to some reference value corresponding to the gradients $a/\LTi = 3.0$ and $a/\Lns = 0.9$. The data shows good agreement with the linear prediction from \cref{eq:predictions_itg} at higher values of $a/\LTi$ --- particularly for the constant $\eta_i = 10/3$ dataset shown in panel (b) --- and is generally consistent with an offset-linear scaling of the form $Q_i \propto a/\LTi - \left(a/\LTi\right)^{\mathrm{crit.}}$, for some critical normalised temperature gradient $\left(a/\LTi\right)^{\mathrm{crit.}}$. This is to be expected, given that the asymptotic scaling theory of \cref{sec:asymptotic_scaling_theory} is only valid for values of $a/\LTi$ sufficiently far above marginality. Overall, it appears that the ITG heat flux scaling \cref{eq:predictions_itg} is consistent with the simulation database, providing further evidence that the predictions are not specific to the CBC geometry in which they were tested in \cref{sec:cbc_simulations}.

\section{Supplementary simulation data}
\label{app:supplementary_data}
\begin{figure*}
	\centering
	\includegraphics[width=\figscale\textwidth]{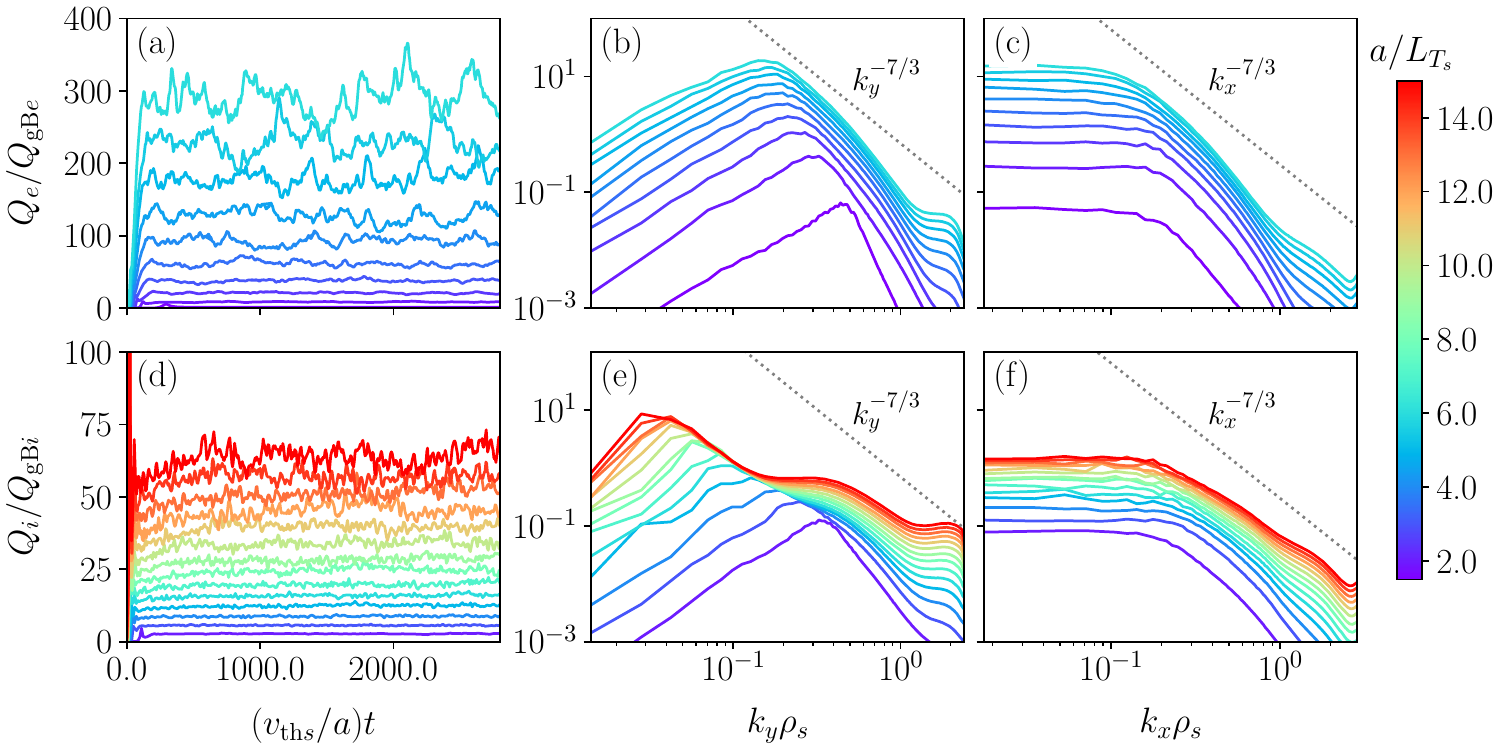}
	
	\caption[]{Data from the simulations labelled `CBC ETG $\LTe$' and `CBC ITG $\LTi$' in \cref{tab:simulation_parameters}, shown in the top and bottom rows, respectively. The colours indicate the value of $a/\LTs$ for the given simulation, and all data shown is normalised to $\Qgbs = \dens \Ts \vths (\rhos/a)^2$. First column: heat fluxes as functions of time. Second column: time-averaged one-dimensional binormal heat-flux spectra \cref{eq:spectrum_binormal}. Third column: time-averaged one-dimensional radial heat-flux spectra \cref{eq:spectrum_radial}.}
	\label{fig:CBC_timetraces_tprim}
\end{figure*}

\begin{figure*}
	\centering
	\includegraphics[width=\figscale\textwidth]{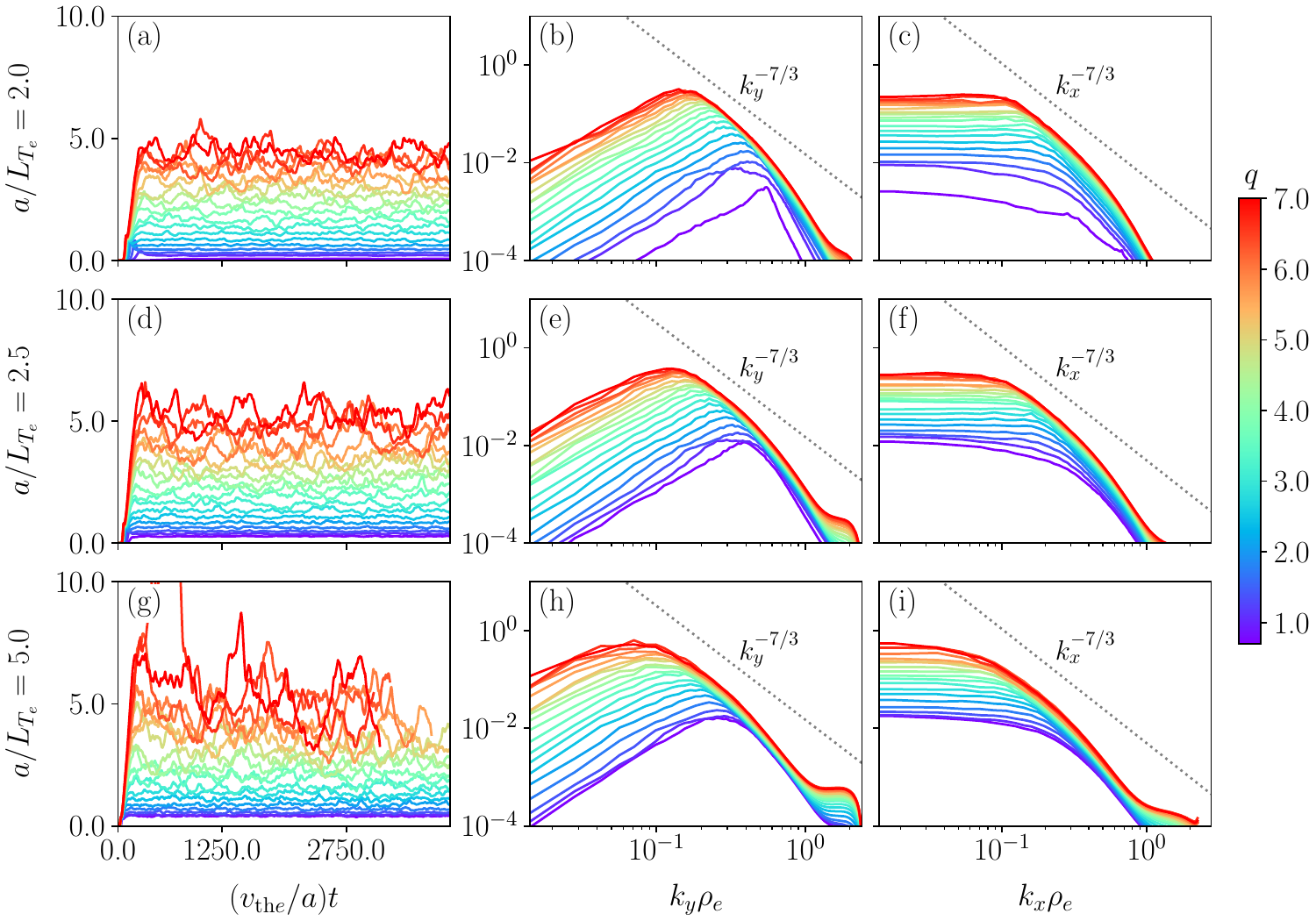}
	
	\caption[]{Data from the simulations labelled `CBC ETG $q$' in \cref{tab:simulation_parameters}, with each row corresponding to a different value of $a/\LTe$. The colours indicate the value of $q$ for a given simulation, where all data shown is normalised to $\Qgbs[e] = \dens[e]\Te \vthe (\rhoe/a)^2$ and rescaled by $(a/\LTe)^3$ [see \cref{eq:predictions_etg}]. First column: heat fluxes as functions of time. Second column: time-averaged one-dimensional binormal heat-flux spectra \cref{eq:spectrum_binormal}. Third column: time-averaged one-dimensional radial heat-flux spectra \cref{eq:spectrum_radial}.}
	\label{fig:CBC_timetraces_q_etg}
\end{figure*}

\begin{figure*}
	\centering
	\includegraphics[width=\figscale\textwidth]{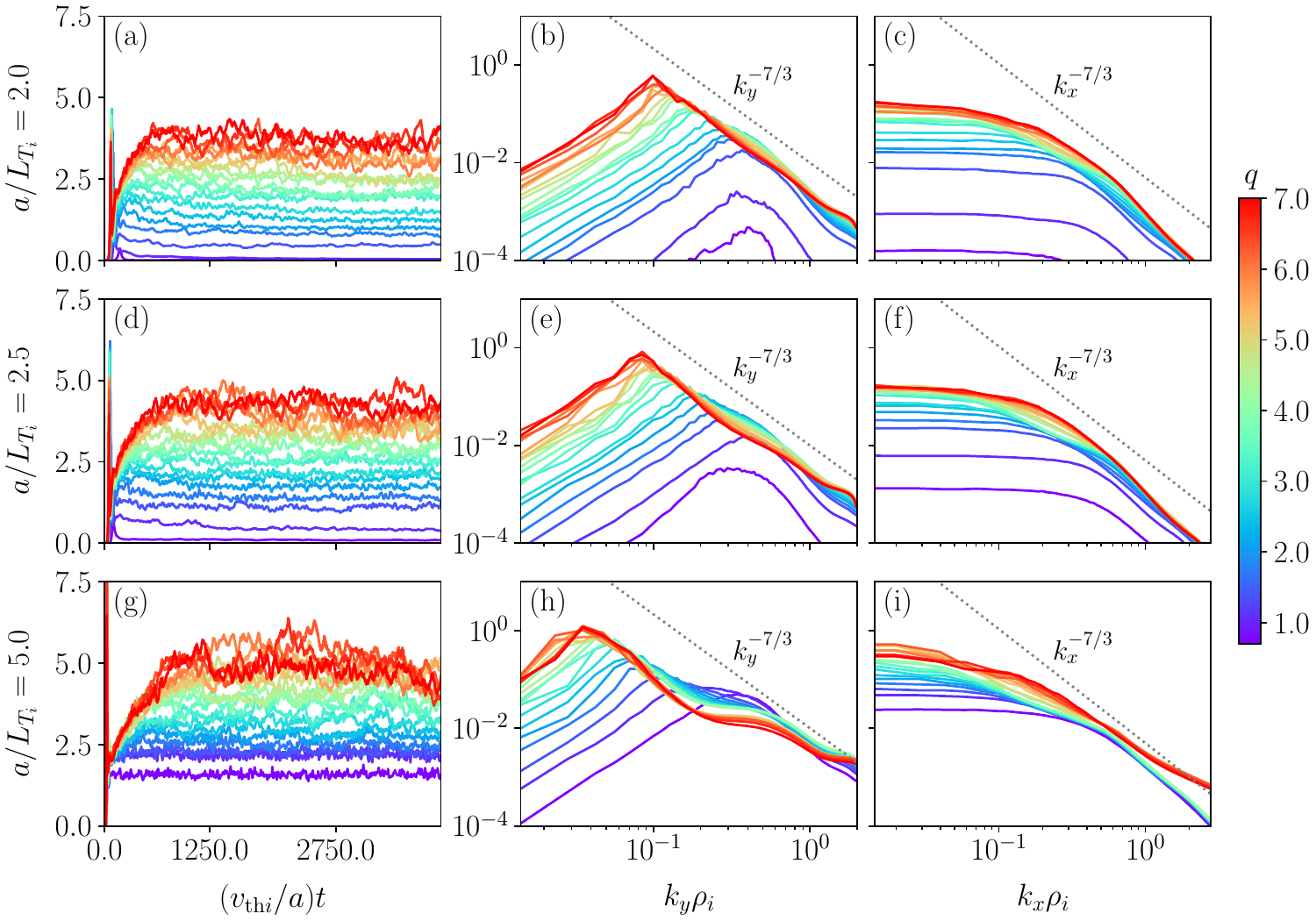}
	
	\caption[]{Data from the simulations labelled `CBC ITG $q$' in \cref{tab:simulation_parameters}, with each row corresponding to a different value of $a/\LTi$. The colours indicate the value of $q$ for a given simulation, and all data shown is normalised to $\Qgbs[i] = \dens[i]\Ti \vthi (\rhoi/a)^2$ and rescaled by $(a/\LTi)^1$ [see \cref{eq:predictions_itg}]. First column: heat fluxes as functions of time. Second column: time-averaged one-dimensional binormal heat-flux spectra \cref{eq:spectrum_binormal}. Third column: time-averaged one-dimensional radial heat-flux spectra \cref{eq:spectrum_radial}.}
	\label{fig:CBC_timetraces_q_itg}
\end{figure*}

\begin{figure*}
	\centering
	\includegraphics[width=\figscale\textwidth]{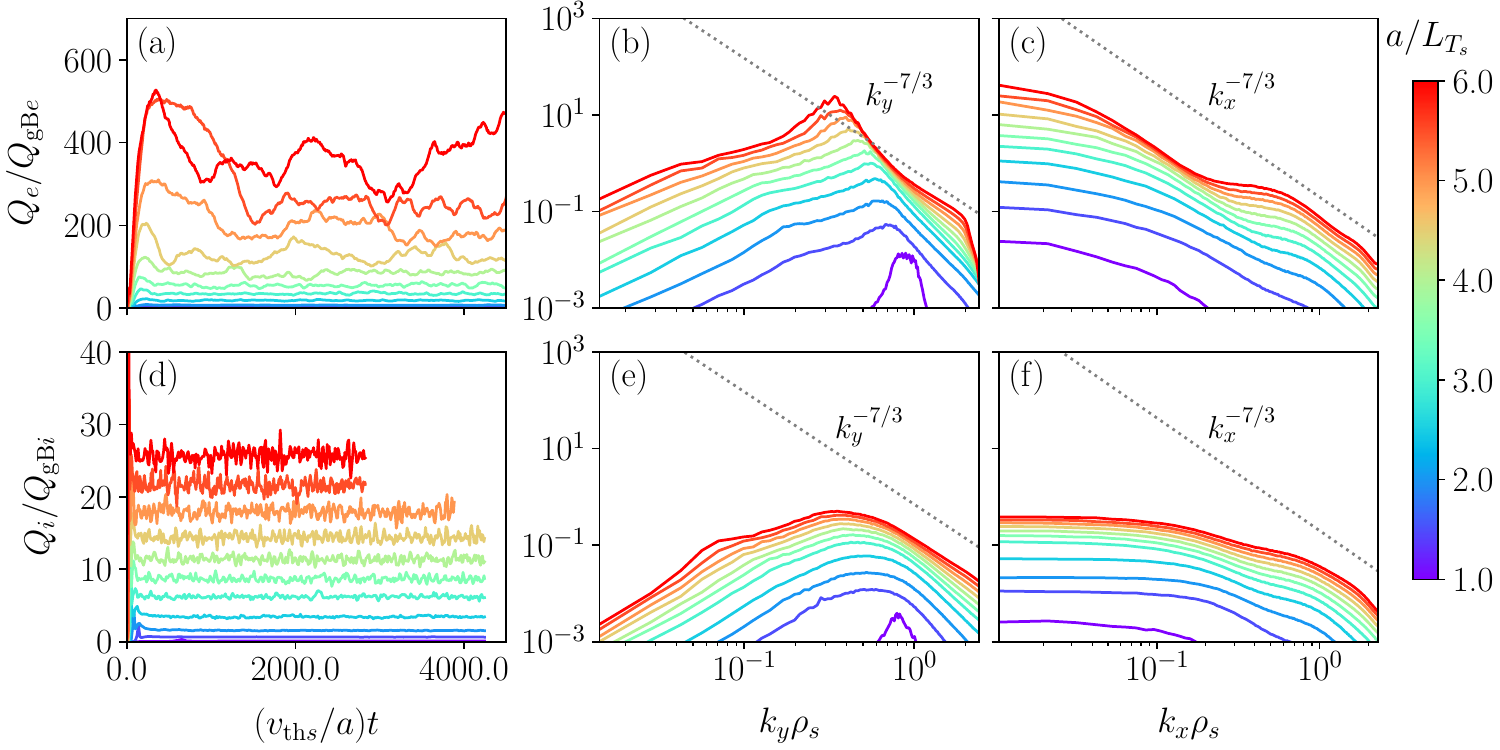}
	
	\caption[]{Data from the simulations labelled `NCSX ETG $\LTe$' and `NCSX ITG $\LTi$' in \cref{tab:simulation_parameters}, shown in the top and bottom rows, respectively. The colours indicate the value of $a/\LTs$ for the given simulation, and all data shown is normalised to $\Qgbs = \dens \Ts \vths (\rhos/a)^2$. First column: heat fluxes as functions of time. Second column: time-averaged one-dimensional binormal heat-flux spectra \cref{eq:spectrum_binormal}. Third column: time-averaged one-dimensional radial heat-flux spectra \cref{eq:spectrum_radial}.}
	\label{fig:NCSX_timetraces_tprim}
\end{figure*}

\begin{figure*}
	\centering
	\includegraphics[width=0.852\textwidth]{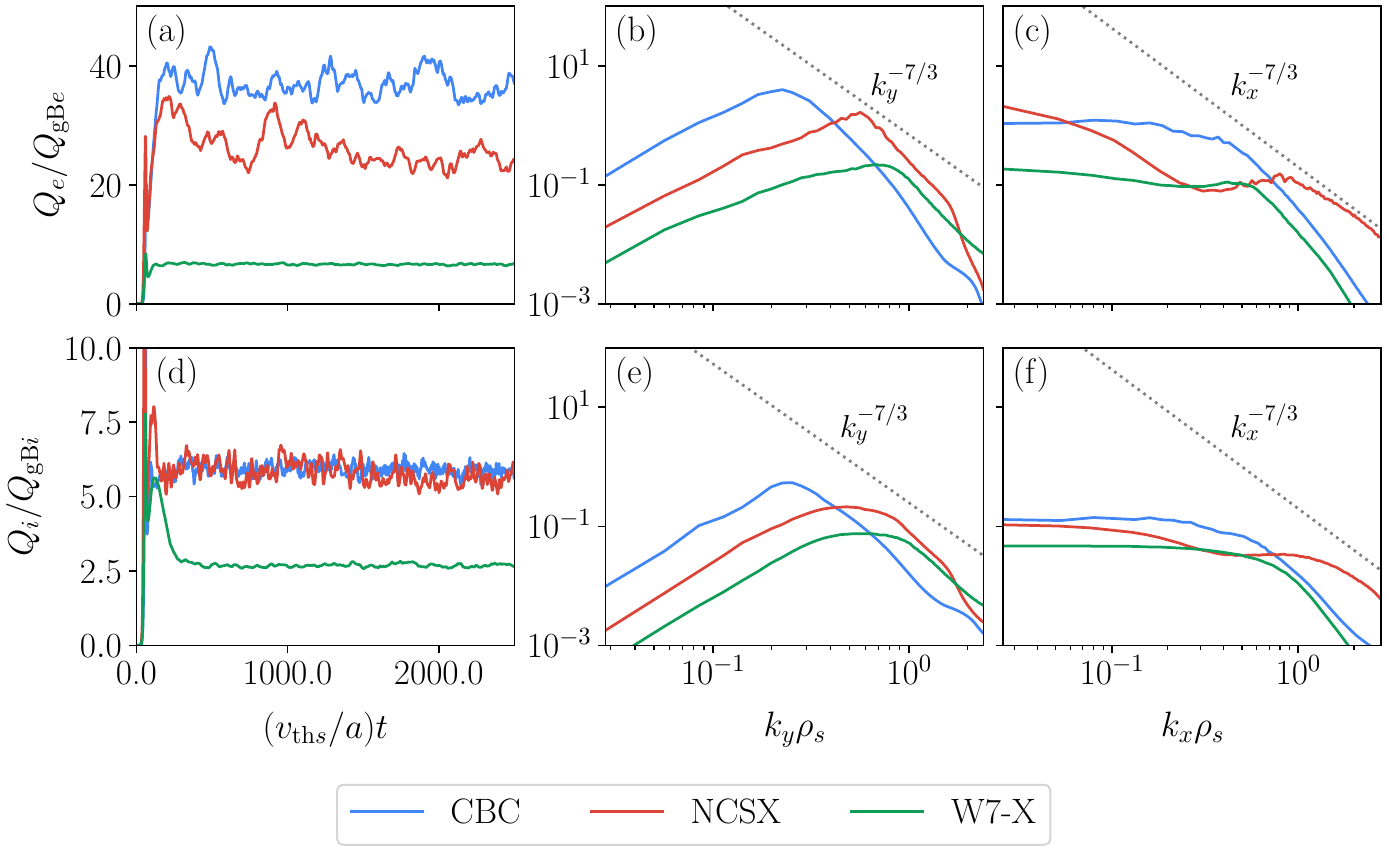}
	
	\caption[]{Data from the simulations labelled `CBC ETG $\Lpar$', `CBC ITG $\Lpar$', `NCSX ETG $\Lpar$', `NCSX ITG $\Lpar$', `W7-X ETG $\Lpar$', and `W7-X ITG $\Lpar$' in \cref{tab:simulation_parameters}. The colours indicate different magnetic equilibria, and all data shown is normalised to $\Qgbs = \dens \Ts \vths (\rhos/a)^2$. First column: heat fluxes as functions of time. Second column: time-averaged one-dimensional binormal heat-flux spectra \cref{eq:spectrum_binormal}. Third column: time-averaged one-dimensional radial heat-flux spectra \cref{eq:spectrum_radial}.}
	\label{fig:comparison_timetraces}
\end{figure*}

In this appendix, we include supplementary data from the simulations presented in \cref{sec:cbc_simulations} (\cref{fig:CBC_timetraces_tprim,fig:CBC_timetraces_q_etg,fig:CBC_timetraces_q_itg}), \cref{sec:stellarator_simulations} (\cref{fig:NCSX_timetraces_tprim}), and \cref{sec:lpar} (\cref{fig:comparison_timetraces}). These results are included for completeness and to aid readers interested in the data underlying the time-averaged values shown in the main text. Where shown, dotted lines indicate the scalings predicted by previous theories of the inertial range \citep{barnes11,adkins22,adkins23}. We emphasise, however, that the asymptotic scaling theory developed in \cref{sec:asymptotic_scaling_theory} does not rely on agreement with these predictions, remaining agnostic to whether or not they are realised in practice. 

\end{appendix}
	
\bibliography{bibliography.bib}

\end{document}

%% file: packages.tex
\usepackage[makeroom]{cancel}
\usepackage{scrextend} 
\usepackage{bold-extra} 
\usepackage{color}
\usepackage[T1]{fontenc} 
\usepackage{fancyhdr}
\usepackage{ulem}
\usepackage{xcolor}
\usepackage{relsize}
\usepackage{textcomp}
\usepackage{textgreek}
\usepackage{lipsum}
\usepackage{setspace}
\usepackage{empheq}

\usepackage{float} 
\usepackage{epstopdf, epsfig}
\usepackage{graphicx} 
\usepackage{graphics} 
\usepackage{tabu} 
\usepackage{tabulary} 
\usepackage{tabularx} 
\usepackage{booktabs}

\usepackage{framed} 
\usepackage{listings} 

\usepackage{amsmath} 
\usepackage{amssymb} 
\usepackage{mathrsfs}

\usepackage{algorithm} 
\usepackage{algpseudocode} 

\usepackage{tensor}
\usepackage{upgreek}
\usepackage{physics}

\usepackage[toc, xindy, acronym]{glossaries}
\usepackage{natbib}

\usepackage[title]{appendix}

\usepackage[setpagesize=false]{hyperref} 
\hypersetup{
	colorlinks,
	linkcolor={blue!80!black},
	citecolor={blue!80!black},
	urlcolor={blue!80!black}
}

\usepackage{enumitem}
\usepackage{zref-savepos}
\usepackage{cleveref}
\crefformat{section}{\S#2#1#3} 
\crefformat{subsection}{\S#2#1#3}
\crefformat{subsubsection}{\S#2#1#3}
\crefformat{figure}{figure~#2#1#3}
\crefformat{appendix}{appendix~#2#1#3}
\crefformat{equation}{(#2#1#3)}

\usepackage{tikz}
\usepackage{circuitikz}
\usetikzlibrary{calc}
\usetikzlibrary{patterns}
\usetikzlibrary{arrows}
\usetikzlibrary{shapes}
\usetikzlibrary{plotmarks}
\usetikzlibrary{decorations.markings}

\usepackage{pgfplots}
\usepgfplotslibrary{polar}
\usepgflibrary{shapes.geometric}
\usepackage[]{pstricks}

\tikzset{
	set arrow inside/.code={\pgfqkeys{/tikz/arrow inside}{#1}},
	set arrow inside={end/.initial=>, opt/.initial=},
	/pgf/decoration/Mark/.style={
		mark/.expanded=at position #1 with
		{
			\noexpand\arrow[\pgfkeysvalueof{/tikz/arrow inside/opt}]{\pgfkeysvalueof{/tikz/arrow inside/end}}
		}
	},
	arrow inside/.style 2 args={
		set arrow inside={#1},
		postaction={
			decorate,decoration={
				markings,Mark/.list={#2}
			}
		}
	},
}

%% file: commands.tex
\newcommand{\closesymbol}{\!}
\newcommand{\fourier}[1]{\tilde{#1}}  
\renewcommand{\fourier}[1]{#1}  

\newcommand{\rmd}{\text{d}}
\newcommand{\rmi}{\text{i}}

\ifdefined\rmJ
\renewcommand{\rmJ}{\text{J}}
\else
\newcommand{\rmJ}{\text{J}}
\fi

\ifdefined\rmI
\renewcommand{\rmI}{\text{I}}
\else
\newcommand{\rmI}{\text{I}}
\fi

\newcommand{\rmZed}{\text{Z}}
\newcommand{\UGamma}{\mathord{\text{\textGamma}}}

\newcommand{\besselarg}[1][\s]{a_{#1}}

\ifdefined\partd
\renewcommand{\partd}[3][]{\frac{{\partial^{#1} #2}}{{\partial #3}^{#1}}}
\else
\newcommand{\partd}[3][]{\frac{{\partial^{#1} #2}}{{\partial #3}^{#1}}}
\fi

\renewcommand{\vec}[1]{\boldsymbol{#1}}
\ifdefined\vec  
\renewcommand{\vec}[1]{\boldsymbol{#1}}
\else
\newcommand{\vec}[1]{\boldsymbol{#1}}
\fi

\ifdefined\div
\renewcommand{\div}{{\vec{\nabla} \cdot}}
\else
\newcommand{\div}{{\vec{\nabla} \cdot}}
\fi

\ifdefined\vr
\renewcommand{\vr}{{\vec{r}}}
\else
\newcommand{\vr}{{\vec{r}}}
\fi

\newcommand{\vk}{{\vec{k}}}
\newcommand{\vv}{{\vec{v}}}
\newcommand{\vkperp}{{\vk_{\closesymbol\perp}}}
\newcommand{\kperp}{k_\perp}
\newcommand{\kpar}{k_\parallel}
\newcommand{\vperp}{v_\perp}

\newcommand{\vpar}{v_\parallel}

\newcommand{\s}{s}

\newcommand{\vths}[1][\s]{v_{\text{th}{#1}}}
\newcommand{\vthi}{\vths[i]}
\newcommand{\vthe}{\vths[e]}
\newcommand{\qs}[1][\s]{q_{#1}}

\newcommand{\ms}[1][\s]{m_{#1}}

\newcommand{\rhoi}{\rho_i}
\newcommand{\rhoe}{\rho_e}
\newcommand{\rhos}{\rho_\s}

\newcommand{\betas}[1][\s]{\beta_{#1}}

\newcommand{\dens}[1][\s]{n_{#1}}

\newcommand{\dns}[1][\s]{\delta n_{#1}}
\newcommand{\dne}{\dns[e]}
\newcommand{\dni}{\dns[i]}

\newcommand{\Ts}[1][\s]{T_{#1}}
\newcommand{\Te}{\Ts[e]}
\newcommand{\Ti}{\Ts[i]}

\newcommand{\dTs}[1][\s]{\delta T_{#1}}

\newcommand{\Qgbs}[1][\s]{Q_{\mathrm{gB}#1}}

\newcommand{\Fos}{F_{0\s}}
\newcommand{\dfs}{\delta \closesymbol f_{\s}}

\newcommand{\energy}{\varepsilon_\s}
\newcommand{\mus}{\mu_\s}

\newcommand{\rhostars}{\rho_{*\s}}

\newcommand{\vRs}[1][\s]{{\vec{R}_{#1}}}

\newcommand{\hs}[1][\s]{h_{#1}}

\newcommand{\hskperp}[1][\s]{\fourier{h}_{{#1}\vkperp}}

\newcommand{\phipot}{\phi}  

\newcommand{\dphipot}{\delta \phipot}  

\newcommand{\dphipotkperp}{{\fourier{\dphipot}}_{\vkperp}}

\newcommand{\dApar}{\delta \closesymbol  A_\parallel}

\newcommand{\dBpar}{{\delta \closesymbol  B_\parallel}}

\newcommand{\vE}{\vec{E}}
\newcommand{\vB}{\vec{B}}
\newcommand{\ub}{\vec{b}}

\newcommand{\exb}{\vE\times\vB}

\newcommand{\vds}{\vec{v}_{\rmd \s}}

\newcommand{\vexb}{\vec{v}_E}

\newcommand{\avgRs}[1]{\left\langle #1 \right\rangle_{\vec{R}_\s}}

\newcommand{\avgRsinline}[1]{{\langle #1 \rangle_{\vec{R}_\s}}}
\newcommand{\avgr}[1]{\left\langle #1 \right\rangle_{\vec{r}}}
\newcommand{\avgrinline}[1]{{\langle #1 \rangle_{\vec{r}}}}
\newcommand{\avg}[2]{\left< #1 \right>_{#2}}

\newcommand{\Ln}[1][]{L_{n_{#1}}}
\newcommand{\LT}[1][]{L_{T_{#1}}}
\newcommand{\LTs}{\LT[\s]}
\newcommand{\LTe}{\LT[e]}
\newcommand{\LTi}{\LT[i]}
\newcommand{\Lns}{\Ln[\s]}

\newcommand{\Lni}{\Ln[i]}

\newcommand{\shat}{\hat{s}}
\newcommand{\safety}{q}
\newcommand{\transform}{\iota}

\newcommand{\tor}{\zeta}
\newcommand{\pol}{\theta}

\newcommand{\Omegas}[1][\s]{\Omega_{#1}}

\newcommand{\omegasts}[1][\s]{\omega_{*{#1}}}

\newcommand{\zetasts}[1][\s]{\zeta_{*#1}}

\newcommand{\zetast}{\zetasts[\relax]}

%% file: bibliography.bib
@string{AA       = {Astron.\ Astrophys.}}

@string{AN       = {Astron.\ Nachr.}}

@string{ApJ      = {Astrophys.~J.}}

@string{ApJS     = {Astrophys.\ J.\ Suppl.}}

@string{FED      = {Fus.\ Eng.\ Des.}}

@string{JCP      = {J.\ Comp.\ Phys.}}

@string{JPP      = {J.\ Plasma Phys.}}

@string{MHD      = {Magnetohydrodynamics}}

@string{NF       = {Nucl.\ Fusion}}

@string{NJP      = {New\ J.\ Phys.}}

@string{PFB      = {Phys.\ Fluids B}}

@string{PoF      = {Phys.\ Fluids}}

@string{PoP      = {Phys.\ Plasmas}}

@string{PPCF     = {Plasma Phys.\ Control.\ Fusion}}

@string{PRL      = {Phys.\ Rev.\ Lett.}}

@string{PRR      = {Phys.\ Rev.\ Res.}}

@string{PRXE     = {Phys.\ Rev.\ X Energy}}

@string{PTRS     = {Philos.\ Trans.\ R.\ Soc.\ London~A}}

@string{ROPP     = {Rep.\ Prog.\ Phys.}}

@article{abel13b,
	title        = {{Multiscale gyrokinetics for rotating tokamak plasmas: II. Reduced models for electron dynamics}},
	author       = {{Abel}, I.~G. and {Cowley}, S.~C.},
	year         = 2013,
	journal      = NJP,
	volume       = 15,
	pages        = {023041},
	doi          = {10.1088/1367-2630/15/2/023041}
}

@article{abel13,
	title        = {{Multiscale gyrokinetics for rotating tokamak plasmas: fluctuations, transport and energy flows}},
	author       = {{Abel}, I.~G. and {Plunk}, G.~G. and {Wang}, E. and {Barnes}, M. and {Cowley}, S.~C. and {Dorland}, W. and {Schekochihin}, A.~A.},
	year         = 2013,
	journal      = ROPP,
	volume       = 76,
	pages        = 116201,
	doi          = {10.1088/0034-4885/76/11/116201}
}

@book{abramowitz72,
	title        = {{Handbook of Mathematical Functions}},
	author       = {{Abramowitz}, M. and {Stegun}, I.~A.},
	year         = 1972,
	adsurl       = {https://ui.adsabs.harvard.edu/abs/1972hmfw.book.....A}
}

@article{adkins22,
	title        = {Electromagnetic instabilities and plasma turbulence driven by electron-temperature gradient},
	author       = {Adkins, T. and Schekochihin, A.~A. and Ivanov, P.~G. and Roach, C.~M.},
	year         = 2022,
	journal      = JPP,
	publisher    = {Cambridge University Press},
	volume       = 88,
	pages        = 905880410,
	doi          = {10.1017/S0022377822000654}
}

@phdthesis{adkins23thesis,
	title        = {Electromagnetic instabilities and plasma turbulence driven by the electron-temperature gradient},
	author       = {{Adkins}, T. G.},
	year         = 2023,
	school       = {University of Oxford (URL: \url{https://ora.ox.ac.uk/objects/uuid:337d61a1-7997-435c-975e-4928cc3902ea)}}
}

@article{adkins23,
	title        = {{Scale invariance and critical balance in electrostatic drift-kinetic turbulence}},
	author       = {{Adkins}, T. and {Ivanov}, P.~G. and {Schekochihin}, A.~A.},
	year         = 2023,
	journal      = JPP,
	volume       = 89,
	pages        = 905890406,
	doi          = {10.1017/S0022377823000600}
}

@article{aleynikova17,
	title        = {{Quantitative study of kinetic ballooning mode theory in simple geometry}},
	author       = {{Aleynikova}, K. and {Zocco}, A.},
	year         = 2017,
	journal      = PoP,
	volume       = 24,
	pages        = {092106},
	doi          = {10.1017/jfm.2019.394}
}

@article{barnes11,
	title        = {{Critically balanced ion temperature gradient turbulence in fusion plasmas}},
	author       = {{Barnes}, M. and {Parra}, F.~I. and {Schekochihin}, A.~A.},
	year         = 2011,
	journal      = PRL,
	volume       = 107,
	pages        = 115003,
	doi          = {10.1103/PhysRevLett.107.115003}
}

@article{boldyrev05,
	title        = {{On the spectrum of magnetohydrodynamic turbulence}},
	author       = {{Boldyrev}, S.},
	year         = 2005,
	journal      = ApJ,
	volume       = 626,
	pages        = {L37},
	doi          = {10.1086/431649}
}

@article{candy16,
	title        = {{A high-accuracy Eulerian gyrokinetic solver for collisional plasmas}},
	author       = {{Candy}, J. and {Belli}, E.~A. and {Bravenec}, R.~V.},
	year         = 2016,
	month        = nov,
	journal      = JCP,
	volume       = 324,
	pages        = 73,
	doi          = {10.1016/j.jcp.2016.07.039}
}

@article{candy07,
	title        = {{The effect of ion-scale dynamics on electron-temperature-gradient turbulence}},
	author       = {{Candy}, J. and {Waltz}, R.~E. and {Fahey}, M.~R. and {Holland}, C.},
	year         = 2007,
	journal      = PPCF,
	volume       = 49,
	pages        = 1209,
	doi          = {10.1088/0741-3335/49/8/008}
}

@article{catto78,
	title        = {{Linearized gyro-kinetics}},
	author       = {{Catto}, P.~J.},
	year         = 1978,
	month        = jul,
	journal      = {Plasma Physics},
	volume       = 20,
	pages        = {719},
	doi          = {10.1088/0032-1028/20/7/011}
}

@article{chapman22,
	title        = {The role of {ETG} modes in {JET}{\textendash}{ILW} pedestals with varying levels of power and fuelling},
	author       = {{Chapman-Oplopoiou}, B. and {Hatch}, D.~R. and {Field}, A.~R. and {Frassinetti}, L. and {Hillesheim}, J. and {Horvath}, L. and {Maggi}, C.~F. and {Parisi}, J. and {Roach}, C.~M. and {Saarelma}, S. and {Walker}, J.},
	year         = 2022,
	journal      = NF,
	volume       = 62,
	pages        = {086028},
	doi          = {10.1088/1741-4326/ac7476}
}

@article{colyer17,
	title        = {{Collisionality scaling of the electron heat flux in ETG turbulence}},
	author       = {{Colyer}, G.~J. and {Schekochihin}, A.~A. and {Parra}, F.~I. and {Roach}, C.~M. and {Barnes}, M.~A. and {Ghim}, Y.-c. and {Dorland}, W.},
	year         = 2017,
	journal      = PPCF,
	volume       = 59,
	pages        = {055002},
	doi          = {10.1088/1361-6587/aa5f75}
}

@article{cowley91,
	title        = {{Considerations of ion-temperature-gradient-driven turbulence}},
	author       = {{Cowley}, S.~C. and {Kulsrud}, R.~M. and {Sudan}, R.},
	year         = 1991,
	journal      = PFB,
	volume       = 3,
	pages        = 2767,
	doi          = {10.1063/1.859913}
}

@book{davidson13,
	title        = {{Turbulence in Rotating, Stratified and Electrically Conducting Fluids}},
	author       = {{Davidson}, P.~A.},
	year         = 2013,
	publisher    = {Cambridge University Press},
	address      = {Cambridge}
}

@book{dhaeseleer91,
	title        = {{Flux Coordinates and Magnetic Field Structure}},
	author       = {{D'Haeseleer}, W.~D. and {Hitchon}, W.~N.~G. and {Callen}, J.~D. and {Shohet}, J.~L.},
	year         = 1991,
	publisher    = {Springer Berlin},
	address      = {Heidelberg}
}

@article{diamond05,
	title        = {{Zonal flows in plasma---a review}},
	author       = {{Diamond}, P.~H. and {Itoh}, S.-I. and {Itoh}, K. and {Hahm}, T.~S.},
	year         = 2005,
	journal      = PPCF,
	volume       = 47,
	pages        = R35,
	doi          = {10.1088/0741-3335/47/5/R01}
}

@article{dimits00,
	title        = {{Comparisons and physics basis of tokamak transport models and turbulence simulations}},
	author       = {{Dimits}, A.~M. and {Bateman}, G. and {Beer}, M.~A. and {Cohen}, B.~I. and {Dorland}, W. and {Hammett}, G.~W. and {Kim}, C. and {Kinsey}, J.~E. and {Kotschenreuther}, M. and {Kritz}, A.~H. and {Lao}, L.~L. and {Mandrekas}, J. and {Nevins}, W.~M. and {Parker}, S.~E. and {Redd}, A.~J. and {Shumaker}, D.~E. and {Sydora}, R. and {Weiland}, J.},
	year         = 2000,
	journal      = PoP,
	volume       = 7,
	pages        = 969,
	doi          = {10.1063/1.873896}
}

@article{dorland93,
	title        = {{Gyrofluid turbulence models with kinetic effects}},
	author       = {{Dorland}, W. and {Hammett}, G.~W.},
	year         = 1993,
	journal      = PFB,
	volume       = 5,
	pages        = 812,
	doi          = {10.1063/1.860934}
}

@article{dorland00,
	title        = {{Electron temperature gradient turbulence}},
	author       = {{Dorland}, W. and {Jenko}, F. and {Kotschenreuther}, M. and {Rogers}, B.~N.},
	year         = 2000,
	journal      = PRL,
	volume       = 85,
	pages        = 5579,
	doi          = {10.1103/PhysRevLett.85.5579}
}

@article{drake80,
	title        = {{Microtearing modes and anomalous transport in tokamaks}},
	author       = {{Drake}, J.~F. and {Gladd}, N.~T. and {Liu}, C.~S. and {Chang}, C.~L.},
	year         = 1980,
	journal      = PRL,
	volume       = 44,
	pages        = 994,
	doi 	     = {10.1103/PhysRevLett.44.994}
}

@article{drake77,
	title        = {{Kinetic theory of tearing instabilities}},
	author       = {{Drake}, J.~F. and {Lee}, Y.~C.},
	year         = 1977,
	journal      = PoF,
	volume       = 20,
	pages        = 1341,
	doi          = {10.1063/1.862017}
}

@book{faddeeva54,
	title        = {Tables of Values of the Function $w(z)=\exp(-z^2)(1+2i/\sqrt{\pi}\int_0^z \exp(t^2) \rmd t)$ for Complex Argument},
	author       = {{Faddeeva}, V.~N. and {Terent'ev}, N.~M.},
	year         = 1954,
	publisher    = {Moscow: Gostekhizdat, English translation: New York: Pergamon Press, 1961}
}

@book{fried61,
	title        = {{The Plasma Dispersion Function}},
	author       = {{Fried}, B.~D. and {Conte}, S.~D.},
	year         = 1961,
	publisher    = {Academic Press},
	address      = {New York}
}

@article{frieman82,
	title        = {{Nonlinear gyrokinetic equations for low-frequency electromagnetic waves in general plasma equilibria}},
	author       = {{Frieman}, E.~A. and {Chen}, L.},
	year         = 1982,
	journal      = PoF,
	volume       = 25,
	pages        = 502,
	doi          = {10.1063/1.863762}
}

@article{ghim13,
	title        = {{Experimental signatures of critically balanced turbulence in MAST}},
	author       = {{Ghim}, Y.-c. and {Schekochihin}, A.~A. and {Field}, A.~R. and {Abel}, I.~G. and {Barnes}, M. and {Colyer}, G. and {Cowley}, S.~C. and {Parra}, F.~I. and {Dunai}, D. and {Zoletnik}, S.},
	year         = 2013,
	journal      = PRL,
	volume       = 110,
	pages        = 145002,
	doi          = {10.1103/PhysRevLett.110.145002}
}

@article{GS95,
	title        = {{Toward a theory of interstellar turbulence. 2: Strong Alfv\'enic turbulence}},
	author       = {{Goldreich}, P. and {Sridhar}, S.},
	year         = 1995,
	journal      = ApJ,
	volume       = 438,
	pages        = 763,
	doi          = {10.1086/175121}
}

@article{guttenfelder21,
	title        = {{Testing predictions of electron scale turbulent pedestal transport in two DIII-D ELMy H-modes}},
	author       = {{Guttenfelder}, W. and {Groebner}, R.~J. and {Canik}, J.~M. and {Grierson}, B.~A. and {Belli}, E.~A. and {Candy}, J.},
	year         = 2021,
	journal      = NF,
	volume       = 61,
	pages        = {056005},
	doi          = {10.1088/1741-4326/abecc7}
}

@article{guttenfelder13,
	title        = {{Progress in simulating turbulent electron thermal transport in NSTX}},
	author       = {{Guttenfelder}, W. and {Peterson}, J.~L. and {Candy}, J. and {Kaye}, S.~M. and {Ren}, Y. and {Bell}, R.~E. and {Hammett}, G.~W. and {LeBlanc}, B.~P. and {Mikkelsen}, D.~R. and {Nevins}, W.~M. and {Yuh}, H.},
	year         = 2013,
	journal      = NF,
	volume       = 53,
	pages        = {093022},
	doi          = {10.1088/0029-5515/53/9/093022}
}

@article{hammett93,
	title        = {{Developments in the gyrofluid approach to tokamak turbulence simulations}},
	author       = {{Hammett}, G.~W. and {Beer}, M.~A. and {Dorland}, W. and {Cowley}, S.~C. and {Smith}, S.~A.},
	year         = 1993,
	journal      = PPCF,
	volume       = 35,
	pages        = 973,
	doi          = {10.1088/0741-3335/35/8/006}
}

@article{hardman20,
	title        = {Stabilisation of short-wavelength instabilities by parallel-to-the-field shear in long-wavelength {ExB} flows},
	author       = {{Hardman}, M. R. and {Barnes}, M. and {Roach}, C. M.},
	year         = 2020,
	journal      = JPP,
	volume       = 86,
	pages        = 905860601,
	doi          = {10.1017/S0022377820001294}
}

@article{hardman23,
	title        = {{New linear stability parameter to describe low-{\ensuremath{\beta}} electromagnetic microinstabilities driven by passing electrons in axisymmetric toroidal geometry}},
	author       = {{Hardman}, M.~R. and {Parra}, F.~I. and {Patel}, B.~S. and {Roach}, C.~M. and {Ruiz Ruiz}, J. and {Barnes}, M. and {Dickinson}, D. and {Dorland}, W. and {Parisi}, J.~F. and {St-Onge}, D. and {Wilson}, H.},
	year         = 2023,
	journal      = PPCF,
	volume       = 65,
	pages        = {045011},
	doi          = {10.1088/1361-6587/acb9ba}
}

@article{hassam80a,
	title        = {{Fluid theory of tearing instabilities}},
	author       = {{Hassam}, A.~B.},
	year         = 1980,
	journal      = PoF,
	volume       = 23,
	pages        = 2493,
	doi          = {10.1063/1.862950}
}

@article{hassam80b,
	title        = {{Higher-order Chapman-Enskog theory for electrons}},
	author       = {{Hassam}, A.~B.},
	year         = 1980,
	journal      = PoF,
	volume       = 23,
	pages        = 38,
	doi          = {10.1063/1.862860}
}

@article{helander11,
	title        = {{Oscillations of zonal flows in stellarators}},
	author       = {{Helander}, P. and {Mishchenko}, A. and {Kleiber}, R. and {Xanthopoulos}, P.},
	year         = 2011,
	journal      = PPCF,
	publisher    = {{IOP} Publishing},
	volume       = 53,
	pages        = {054006},
	doi          = {10.1088/0741-3335/53/5/054006}
}

@article{hosking22forced,
	title        = {{Emergence of long-range correlations and thermal spectra in forced turbulence}},
	author       = {{Hosking}, D.~N. and {Schekochihin}, A.~A.},
	year         = 2022,
	journal      = {arXiv e-prints},
	volume       = {2202.00462}
}

@article{howard16a,
	title        = {{Multi-scale gyrokinetic simulation of tokamak plasmas: enhanced heat loss due to cross-scale coupling of plasma turbulence}},
	author       = {{Howard}, N.~T. and {Holland}, C. and {White}, A.~E. and {Greenwald}, M. and {Candy}, J.},
	year         = 2016,
	month        = jan,
	journal      = {Nuclear Fusion},
	volume       = 56,
	pages        = {014004},
	doi          = {10.1088/0029-5515/56/1/014004}
}

@article{howard16b,
	title        = {{Multi-scale gyrokinetic simulations: Comparison with experiment and implications for predicting turbulence and transport}},
	author       = {{Howard}, N.~T. and {Holland}, C. and {White}, A.~E. and {Greenwald}, M. and {Candy}, J. and {Creely}, A.~J.},
	year         = 2016,
	month        = may,
	journal      = {Physics of Plasmas},
	volume       = 23,
	pages        = {056109},
	doi          = {10.1063/1.4946028}
}

@article{ishizawa13,
	title        = {{Gyrokinetic turbulence simulations of high-beta tokamak and helical plasmas with full-kinetic and hybrid models}},
	author       = {{Ishizawa}, A. and {Maeyama}, S. and {Watanabe}, T. -H. and {Sugama}, H. and {Nakajima}, N.},
	year         = 2013,
	journal      = NF,
	volume       = 53,
	pages        = {053007},
	doi          = {10.1088/0029-5515/53/5/053007}
}

@article{ishizawa19,
	title        = {{Persistence of Ion Temperature Gradient Turbulent Transport at Finite Normalized Pressure}},
	author       = {{Ishizawa}, A. and {Urano}, D. and {Nakamura}, Y. and {Maeyama}, S. and {Watanabe}, T. -H.},
	year         = 2019,
	journal      = PRL,
	volume       = 123,
	pages        = {025003},
	doi          = {10.1103/PhysRevLett.123.025003}
}

@article{ishizawa14,
	title        = {Electromagnetic gyrokinetic turbulence in finite-beta helical plasmas},
	author       = {{Ishizawa}, A. and {Watanabe}, T.-H. and {Sugama}, H. and {Maeyama}, S. and {Nakajima}, N.},
	year         = 2014,
	journal      = PoP,
	volume       = 21,
	pages        = {055905},
	doi          = {10.1063/1.4876960}
}

@article{ivanov20,
	title        = {{Zonally dominated dynamics and Dimits threshold in curvature-driven ITG turbulence}},
	author       = {{Ivanov}, P.~G. and {Schekochihin}, A.~A. and {Dorland}, W. and {Field}, A.~R. and {Parra}, F.~I.},
	year         = 2020,
	journal      = JPP,
	volume       = 86,
	pages        = 855860502,
	doi          = {10.1017/S0022377820000938}
}

@article{ivanov22, 
	title        = {Dimits transition in three-dimensional ion-temperature-gradient turbulence},
	author       = {Ivanov, P.~G. and Schekochihin, A.~A. and Dorland, W.},
	year         = {2022},
	journal      = JPP,
	volume       = 88,
	pages        = 905880506,
	doi          = {10.1017/S002237782200071X}
}

@article{jenko00GENE,
	title        = {{Massively parallel Vlasov simulation of electromagnetic drift-wave turbulence}},
	author       = {{Jenko}, F.},
	year         = 2000,
	month        = mar,
	journal      = {Comp. Phys. Comms.},
	volume       = 125,
	pages        = 196,
	doi          = {10.1016/S0010-4655(99)00489-0}
}

@article{jenko00,
	title        = {{Electron temperature gradient driven turbulence}},
	author       = {{Jenko}, F. and {Dorland}, W. and {Kotschenreuther}, M. and {Rogers}, B.~N.},
	year         = 2000,
	journal      = PoP,
	volume       = 7,
	pages        = 1904,
	doi          = {10.1063/1.874014}
}

@article{kotschenreuther95,
	title        = {{Quantitative predictions of tokamak energy confinement from first-principles simulations with kinetic effects}},
	author       = {{Kotschenreuther}, M. and {Dorland}, W. and {Beer}, M.~A. and {Hammett}, G.~W.},
	year         = 1995,
	journal      = PoP,
	volume       = 2,
	pages        = 2381,
	doi          = {10.1063/1.871261}
}

@article{kotschenreuther95GS2,
	title        = {{Comparison of initial value and eigenvalue codes for kinetic toroidal plasma instabilities}},
	author       = {{Kotschenreuther}, M. and {Rewoldt}, G. and {Tang}, W.~M.},
	year         = 1995,
	journal      = {Comp. Phys. Comms.},
	volume       = 88,
	pages        = 128,
	doi          = {10.1016/0010-4655(95)00035-E}
}

@article{kruskal58,
	title        = {{Equilibrium of a Magnetically Confined Plasma in a Toroid}},
	author       = {{Kruskal}, M.~D. and {Kulsrud}, R.~M.},
	year         = 1958,
	month        = jul,
	journal      = {Physics of Fluids},
	volume       = 1,
	pages        = {265--274},
	doi          = {10.1063/1.1705884}
}

@article{larakers20,
	title        = {A comprehensive conductivity model for drift and micro-tearing modes},
	author       = {{Larakers}, J.~L. and {Hazeltine}, R.~D. and {Mahajan}, S.~M.},
	year         = 2020,
	journal      = PoP,
	volume       = 27,
	pages        = {062503},
	doi          = {10.1063/5.0006215}
}

@article{larakers21,
	title        = {Global Theory of Microtearing Modes in the Tokamak Pedestal},
	author       = {{Larakers}, J.~L. and {Curie}, M. and {Hatch}, D.~R. and {Hazeltine}, R.~D. and {Mahajan}, S.~M.},
	year         = 2021,
	journal      = PRL,
	volume       = 126,
	pages        = 225001,
	doi          = {10.1103/PhysRevLett.126.225001}
}

@article{lee87,
	title        = {Collisionless electron temperature gradient instability},
	author       = {Lee,Y. C. and Dong,J. Q. and Guzdar,P. N. and Liu,C. S.},
	year         = 1987,
	journal      = {The Physics of Fluids},
	volume       = 30,
	pages        = 1331,
	doi          = {10.1063/1.866248}
}

@article{liu71,
	title        = {Instabilities in a Magnetoplasma with Skin Current},
	author       = {Liu, C.~S.},
	year         = 1971,
	journal      = {Phys. Rev. Lett.},
	volume       = 27,
	pages        = 1637,
	issue        = 24,
	doi          = {10.1103/PhysRevLett.27.1637}
}

@article{maeyama15,
	title        = {{Cross-Scale Interactions between Electron and Ion Scale Turbulence in a Tokamak Plasma}},
	author       = {{Maeyama}, S. and {Idomura}, Y. and {Watanabe}, T. -H. and {Nakata}, M. and {Yagi}, M. and {Miyato}, N. and {Ishizawa}, A. and {Nunami}, M.},
	year         = 2015,
	month        = jun,
	journal      = PRL,
	volume       = 114,
	pages        = 255002,
	doi          = {10.1103/PhysRevLett.114.255002}
}

@article{mandell24,
	title        = {{GX: a GPU-native gyrokinetic turbulence code for tokamak and stellarator design}},
	author       = {{Mandell}, N.~R. and {Dorland}, W. and {Abel}, I. and {Gaur}, R. and {Kim}, P. and {Martin}, M. and {Qian}, T.},
	year         = 2024,
	journal      = JPP,
	volume       = 90,
	pages        = 905900402,
	doi          = {10.1017/S0022377824000631}
}

@article{parisi20,
	title        = {Toroidal and slab {ETG} instability dominance in the linear spectrum of {JET}-{ILW} pedestals},
	author       = {{Parisi}, J.~F. and {Parra}, F.~I. and {Roach}, C.~M. and {Giroud}, C. and {Dorland}, W. and {Hatch}, D.~R. and {Barnes}, M. and {Hillesheim}, J.~C. and {Aiba}, N. and {Ball}, J. and {Ivanov}, P.~G. and JET contributors},
	year         = 2020,
	journal      = NF,
	publisher    = {{IOP} Publishing},
	volume       = 60,
	pages        = 126045,
	doi          = {10.1088/1741-4326/abb891}
}

@article{parisi22,
	title        = {Three-dimensional inhomogeneity of electron-temperature-gradient turbulence in the edge of tokamak plasmas},
	author       = {J.~F. {Parisi} and F.~I. {Parra} and C.~M. {Roach} and M.~R. {Hardman} and A.~A. {Schekochihin} and I.~G. {Abel} and N. {Aiba} and J. {Ball} and M. {Barnes} and B. {Chapman-Oplopoiou} and D. {Dickinson} and W. {Dorland} and C. {Giroud} and D.~R. {Hatch} and J.~C. {Hillesheim} and J. {Ruiz Ruiz} and S. {Saarelma} and D. {St-Onge} and JET Contributors},
	year         = 2022,
	journal      = NF,
	publisher    = {{IOP} Publishing},
	volume       = 62,
	pages        = {086045},
	doi          = {10.1088/1741-4326/ac786b}
}

@article{pueschel10,
	title        = {Transport properties of finite-$\beta$ microturbulence},
	author       = {{Pueschel}, M.~J. and {Jenko}, ~F.},
	year         = 2010,
	journal      = PoP,
	volume       = 17,
	pages        = {062307},
	doi          = {10.1063/1.3435280}
}

@article{pueschel08,
	title        = {Gyrokinetic turbulence simulations at high plasma beta},
	author       = {{Pueschel}, M.~J. and {Kammerer}, M. and {Jenko}, F.},
	year         = 2008,
	journal      = PoP,
	volume       = 15,
	pages        = 102310,
	doi          = {10.1063/1.3005380}
}

@article{ren17,
	title        = {{Recent progress in understanding electron thermal transport in NSTX}},
	author       = {{Ren}, Y. and {Belova}, E. and {Gorelenkov}, N. and {Guttenfelder}, W. and {Kaye}, S.~M. and {Mazzucato}, E. and {Peterson}, J.~L. and {Smith}, D.~R. and {Stutman}, D. and {Tritz}, K. and {Wang}, W.~X. and {Yuh}, H. and {Bell}, R.~E. and {Domier}, C.~W. and {LeBlanc}, B.~P.},
	year         = 2017,
	journal      = NF,
	volume       = 57,
	pages        = {072002},
	doi          = {10.1088/1741-4326/aa4fba}
}

@article{roach09,
	title        = {{Gyrokinetic simulations of spherical tokamaks}},
	author       = {{Roach}, C.~M. and {Abel}, I.~G. and {Akers}, R.~J. and {Arter}, W. and {Barnes}, M. and {Camenen}, Y. and {Casson}, F.~J. and {Colyer}, G. and {Connor}, J.~W. and {Cowley}, S.~C. and {Dickinson}, D. and {Dorland}, W. and {Field}, A.~R. and {Guttenfelder}, W. and {Hammett}, G.~W. and {Hastie}, R.~J. and {Highcock}, E. and {Loureiro}, N.~F. and {Peeters}, A.~G. and {Reshko}, M. and {Saarelma}, S. and {Schekochihin}, A.~A. and {Valovic}, M. and {Wilson}, H.~R.},
	year         = 2009,
	journal      = PPCF,
	volume       = 51,
	pages        = 124020,
	doi          = {10.1088/0741-3335/51/12/124020}
}

@article{roach05,
	title        = {{Microstability physics as illuminated in the spherical tokamak}},
	author       = {{Roach}, C.~M. and {Applegate}, D.~J. and {Connor}, J.~W. and {Cowley}, S.~C. and {Dorland}, W.~D. and {Hastie}, R.~J. and {Joiner}, N. and {Saarelma}, S. and {Schekochihin}, A.~A. and {Akers}, R.~J. and {Brickley}, C. and {Field}, A.~R. and {Valovic}, M. and {MAST Team}},
	year         = 2005,
	journal      = PPCF,
	volume       = 47,
	pages        = {B323},
	doi          = {10.1088/0741-3335/47/12B/S23}
}

@article{rogers00,
	title        = {{Generation and stability of zonal flows in ion-temperature-gradient mode turbulence}},
	author       = {{Rogers}, B.~N. and {Dorland}, W. and {Kotschenreuther}, M.},
	year         = 2000,
	journal      = PRL,
	volume       = 85,
	pages        = 5336,
	doi          = {10.1103/PhysRevLett.85.5336}
}

@article{sch09,
	title        = {{Astrophysical gyrokinetics: kinetic and fluid turbulent cascades in magnetized weakly collisional plasmas}},
	author       = {{Schekochihin}, A.~A. and {Cowley}, S.~C. and {Dorland}, W. and {Hammett}, G.~W. and {Howes}, G.~G. and {Quataert}, E. and {Tatsuno}, T.},
	year         = 2009,
	journal      = ApJS,
	volume       = 182,
	pages        = 310,
	doi          = {10.1088/0067-0049/182/1/310}
}

@article{sch16,
	title        = {{Phase mixing versus nonlinear advection in drift-kinetic plasma turbulence}},
	author       = {{Schekochihin}, A.~A. and {Parker}, J.~T. and {Highcock}, E.~G. and {Dellar}, P.~J. and {Dorland}, W. and {Hammett}, G.~W.},
	year         = 2016,
	journal      = JPP,
	volume       = 82,
	pages        = 905820212,
	doi          = {10.1017/S0022377816000374}
}

@article{sch22,
	title        = {{MHD turbulence: a biased review}},
	author       = {{Schekochihin}, A.~A.},
	year         = 2022,
	journal      = JPP,
	volume       = 88,
	pages        = 155880501,
	doi          = {10.1017/S0022377822000721}
}

@article{shimomura01,
	title        = {{ITER: opportunity of burning plasma studies}},
	author       = {{Shimomura}, Y. and {Murakami}, Y. and {Polevoi}, A.~R. and {Barabaschi}, P. and {Mukhovatov}, V. and {Shimada}, M.},
	year         = 2001,
	journal      = PPCF,
	volume       = 43,
	pages        = 385,
	doi          = {10.1088/0741-3335/43/12A/329}
}

@article{sips05,
	title        = {{Advanced scenarios for ITER operation}},
	author       = {{Sips}, A.~C.~C.},
	year         = 2005,
	journal      = PPCF,
	volume       = 47,
	pages        = {A19}
}

@phdthesis{snyder99thesis,
	title        = {Gyrofluid theory and simulation of electromagnetic turbulence and transport in tokamak plasmas},
	author       = {{Snyder}, P. B.},
	year         = 1999,
	school       = {Princeton University (URL: \url{https://w3.pppl.gov/~hammett/gyrofluid/papers/1999/thesis.pdf)}},
	adsurl       = {https://ui.adsabs.harvard.edu/abs/1999PhDT.......228S}
}

@article{snyder01gf,
	title        = {{A Landau fluid model for electromagnetic plasma microturbulence}},
	author       = {{Snyder}, P.~B. and {Hammett}, G.~W.},
	year         = 2001,
	journal      = PoP,
	volume       = 8,
	pages        = 3199,
	doi          = {10.1063/1.1374238}
}

@article{snyder01,
	title        = {{Electromagnetic effects on plasma microturbulence and transport}},
	author       = {{Snyder}, P.~B. and {Hammett}, G.~W.},
	year         = 2001,
	journal      = PoP,
	volume       = 8,
	pages        = 744,
	doi          = {10.1063/1.1342029}
}

@article{tang80,
	title        = {{Kinetic-ballooning-mode theory in general geometry}},
	author       = {{Tang}, W.~M. and {Connor}, J.~W. and {Hastie}, R.~J.},
	year         = 1980,
	journal      = NF,
	volume       = 20,
	pages        = 1439,
	doi          = {10.1088/0029-5515/20/11/011}
}

@article{terry15,
	title        = {{Overview of gyrokinetic studies of finite-beta microturbulence}},
	author       = {{Terry}, P.~W. and {Carmody}, D. and {Doerk}, H. and {Guttenfelder}, W. and {Hatch}, D.~R. and {Hegna}, C.~C. and {Ishizawa}, A. and {Jenko}, F. and {Nevins}, W.~M. and {Predebon}, I. and {Pueschel}, M.~J. and {Sarff}, J.~S. and {Whelan}, G.~G.},
	year         = 2015,
	journal      = NF,
	volume       = 55,
	pages        = 104011,
	doi          = {10.1088/0029-5515/55/10/104011}
}

@article{villard13,
	title        = {{Global gyrokinetic ion temperature gradient turbulence simulations of ITER}},
	author       = {{Villard}, L. and {Angelino}, P. and {Bottino}, A. and {Brunner}, S. and {Jolliet}, S. and {McMillan}, B.~F. and {Tran}, T.~M. and {Vernay}, T.},
	year         = 2013,
	journal      = PPCF,
	volume       = 55,
	pages        = {074017},
	doi          = {10.1088/0741-3335/55/7/074017}
}

@article{villard14,
	title        = {{Turbulence and zonal flow structures in the core and L-mode pedestal of tokamak plasmas}},
	author       = {{Villard}, L. and {McMillan}, B.~F. and {Sauter}, O. and {Hariri}, F. and {Dominski}, J. and {Merlo}, G. and {Brunner}, S. and {Tran}, T.~M.},
	year         = 2014,
	journal      = {J.\ Phys.\ Conf. Ser.},
	volume       = 561,
	pages        = {012022},
	doi          = {10.1088/1742-6596/561/1/012022}
}

@article{waltz88,
	title        = {{Three-dimensional global numerical simulation of ion temperature gradient mode turbulence}},
	author       = {{Waltz}, R.~E.},
	year         = 1988,
	journal      = PoF,
	volume       = 31,
	pages        = 1962,
	doi          = {10.1063/1.866643}
}

@article{waltz10,
	title        = {{Nonlinear subcritical magnetohydrodynamic beta limit}},
	author       = {{Waltz}, R.~E.},
	year         = 2010,
	journal      = PoP,
	volume       = 17,
	pages        = {072501},
	doi          = {10.1063/1.3449075}
}

@article{wan13,
	title        = {Global gyrokinetic simulations of the H-mode tokamak edge pedestal},
	author       = {{Wan}, W and {Parker}, S. E. and {Chen}, Y. and {Groebner}, R. J. and {Yan}, Z. and {Pankin}, A. Y. and {Kruger}, S. E.},
	year         = 2013,
	journal      = PoP,
	volume       = 20,
	pages        = {055902},
	doi          = {10.1063/1.4803890}
}

@article{wan12,
	title        = {{Global Gyrokinetic Simulation of Tokamak Edge Pedestal Instabilities}},
	author       = {{Wan}, W. and {Parker}, S. E. and {Chen}, Y. and {Yan}, Z. and {Groebner}, R. J. and {Snyder}, P. B.},
	year         = 2012,
	journal      = PRL,
	volume       = 109,
	pages        = 185004,
	doi          = {10.1103/PhysRevLett.109.185004}
}

@article{wolf03,
	title        = {{Internal transport barriers in tokamak plasmas}},
	author       = {{Wolf}, R.~C.},
	year         = 2003,
	journal      = PPCF,
	volume       = 45,
	pages        = {R1},
	doi          = {10.1088/0741-3335/45/1/201}
}

@article{zocco15,
	title        = {{Kinetic microtearing modes and reconnecting modes in strongly magnetised slab plasmas}},
	author       = {{Zocco}, A. and {Loureiro}, N.~F. and {Dickinson}, D. and {Numata}, R. and {Roach}, C.~M.},
	year         = 2015,
	journal      = PPCF,
	volume       = 57,
	pages        = {065008},
	doi          = {10.1088/0741-3335/57/6/065008}
}

@article{robergclark22,
	title        = {{Coarse-grained gyrokinetics for the critical ion temperature gradient in stellarators}},
	author       = {{Roberg-Clark}, G.~T. and {Plunk}, G.~G. and {Xanthopoulos}, P.},
	year         = 2022,
	journal      = PRR,
	volume       = 4,
	pages        = {L032028},
	doi          = {10.1103/PhysRevResearch.4.L032028}
}

@article{tirkas23,
	title        = {Zonal flow excitation in electron-scale tokamak turbulence},
	author       = {{Tirkas}, S. and {Chen}, H. an {Merlo}, G. and {Jenko}, F. and {Parker}, P.},
	year         = 2023,
	journal      = NF,
	volume       = 63,
	pages        = {026015},
	doi          = {10.1088/1741-4326/acab15}
}

@article{field23,
	title        = {Comparing pedestal structure in {JET-ILW} H-mode plasmas with a model for stiff ETG turbulent heat transport},
	author       = {{Field}, A.~R. and {Chapman-Oplopoiou}, B. and {Connor}, J.~W. and {Frassinetti}, L. and {Hatch}, D.~R. and {Roach}, C.~M. and {Saarelma},~S. and {JET contributors}},
	year         = 2023,
	journal      = PTRS,
	volume       = 381,
	pages        = 20210228,
	doi          = {10.1098/rsta.2021.0228}
}

@article{catto19,
	title        = {Practical gyrokinetics},
	author       = {Catto, P. J.},
	year         = 2019,
	journal      = {{J.~Plasma Phys.}},
	volume       = 85,
	pages        = 925850301,
	doi          = {10.1017/S002237781900031X}
}

@article{smolyakov02,
	title        = {Short Wavelength Temperature Gradient Driven Modes in Tokamak Plasmas},
	author       = {Smolyakov, A. I. and Yagi, M. and Kishimoto, Y.},
	year         = 2002,
	journal      = PRL,
	volume       = 89,
	pages        = 125005,
	doi          = {10.1103/PhysRevLett.89.125005}
}

@article{beer95,
	title        = {Field‐aligned coordinates for nonlinear simulations of tokamak turbulence},
	author       = {Beer, M. A. and Cowley, S. C. and Hammett, G. W.},
	year         = 1995,
	journal      = POP,
	volume       = 2,
	pages        = 2687,
	doi          = {10.1063/1.871232}
}

@article{kobayashi12,
	title        = {The quench rule, {Dimits} shift, and eigenmode localization by small-scale zonal flows},
	author       = {Kobayashi, S  and Rogers, BN},
	year         = 2012,
	journal      = {Phys. Plasmas},
	volume       = 19,
	pages        = {012315},
	doi          = {10.1063/1.3677355}
}

@article{adam76,
	title        = {Destabilization of the trapped‐electron mode by magnetic curvature drift resonances},
	author       = {{Adam}, J.~C.  and {Tang}, W.~M. and {Rutherford}, P.~H.},
	year         = 1976,
	journal      = PoF,
	volume       = 19,
	pages        = 561,
	doi          = {10.1063/1.861489}
}

@article{chandran22,
	title        = {{A gyrokinetic dispersion relation for microtearing modes in toroidal plasmas with cold ions}},
	author       = {{Chandran}, B.~D.~G. and {Schekochihin}, A.~A.},
	year         = 2024,
	journal      = JPP,
	volume       = 90,
	number       = 905900204,
	doi          = {10.1017/S0022377824000175}
}

@article{kennedy23,
	title        = "{Electromagnetic gyrokinetic instabilities in STEP}",
	author       = {{Kennedy}, D. and {Giacomin}, M. and {Casson}, F.~J. and {Dickinson}, D. and {Hornsby}, W.~A. and {Patel}, B.~S. and {Roach}, C.~M.},
	year         = 2023,
	journal      = NF,
	volume       = {63},
	pages        = {126061},
	doi          = {10.1088/1741-4326/ad08e7}
}

@ARTICLE{kennedy24,
	title        = "{On the importance of parallel magnetic-field fluctuations for electromagnetic instabilities in STEP}",
	author       = {{Kennedy}, D. and {Roach}, C.~M. and {Giacomin}, M. and {Ivanov}, P.~G. and {Adkins}, T. and {Sheffield}, F. and {G{\"o}rler}, T. and {Bokshi}, A. and {Dickinson}, D. and {Dudding}, H.~G. and {Patel}, B.~S.},
    journal      = NF,
	year         = 2024,
	volume       = {64},
	pages        = {086049},
	doi          = {10.1088/1741-4326/ad58f3}
}

@article{giacomin24,
	title        = "{On electromagnetic turbulence and transport in STEP}",
	author       = {{Giacomin}, M. and {Kennedy}, D. and {Casson}, F. ~J. and {Ajay C.}, J. and {Dickinson}, D. and {Patel}, B.~S. and {Roach}, C. ~M.},
	year         = 2023,
	journal      = NF,
	volume       = {66},
	pages        = {055010},
	doi          = {10.1088/1361-6587/ad366f}
}

@ARTICLE{helander13,
	title        = "{Collisionless microinstabilities in stellarators. I. Analytical theory of trapped-particle modes}",
	author       = {{Helander}, P. and {Proll}, J.~H.~E. and {Plunk}, G.~G.},
	year         = 2013,
	journal      = PoP,
	volume       = {20},
	pages        = {122505},
	doi          = {10.1063/1.4846818}
}

@article{ivanov25,
	title         = "{Suppression of temperature-gradient-driven turbulence by sheared flows in fusion plasmas}",
	author        = {Ivanov, P.~G. and Adkins, T. and Kennedy, D. and Giacomin, M. and Barnes, M. and Schekochihin, A.~A.},
	journal       = JPP,
	year          = 2025,
	volume        = {91},
	pages         = {e58},
	doi           = {10.1017/S0022377825000054}
}

@article{nies26,
	title         = "{Saturation of magnetized plasma turbulence by propagating zonal flows}",
	author        = {{Nies}, R. and {Parra}, F. and {Barnes}, M. and {Mandell}, N. and {Dorland}, W.},
	journal       = PRR,
	year          = 2026,
	volume        = {8},
	pages         = {013295},
	doi           = {10.1103/yf4h-4hvw}
}

@article{nies26tsm,
	title         = "{Theory of zonal flow growth and propagation in toroidal geometry}",
	author        = {{Nies}, R. and {Parra}, F.~I.},
	journal       = PPCF,
	year          = 2026,
	volume        = {68},
	pages         = {045028},
	doi           = {10.1088/1361-6587/ae56b8}
}

@article{lin99,
	title         = "{Effects of Collisional Zonal Flow Damping on Turbulent Transport}",
	author        = {Lin, Z. and Hahm, T.~S. and Lee, W.~W. and Tang, W.~M. and Diamond, P.~H.},
	journal       = PRL,
	year          = 1999,
	volume        = {83},
	pages         = {3645},
	doi           = {10.1103/PhysRevLett.83.3645}
}

@article{plunk14itg,
	title         = "{Collisionless microinstabilities in stellarators. III. The ion-temperature-gradient mode}",
	author        = {Plunk, G.~G. and Helander, P. and Xanthopoulos, P. and Connor, J.~W.},
	journal       = PoP,
	year          = 2014,
	volume        = {21},
	pages         = {032112},
	doi           = {10.1063/1.4868412}
}

@article{ivanov25invariant,
	title         = "{The gyrokinetic field invariant and electromagnetic temperature-gradient instabilities in `good-curvature' plasmas}",
	author        = {Ivanov, P.~G. and Luhadiya, P. and Adkins, T. and Schekochihin, A.~A.},
	journal       = JPP,
	year          = 2025,
	volume        = {91},
	pages         = {E95},
	doi           = {10.1017/S0022377825000510}
}

@article{zarnstorff01,
	title         = "{Physics of the compact advanced stellarator NCSX}",
	author        = {Zarnstorff, M.~C. and Berry, L.~A. and Brooks, A. and Fredrickson, E. and Fu, G.-Y. and Hirshman, S. and Hudson, S. and Ku, L.-P. and Lazarus, E. and Mikkelsen, D. and Monticello, D. and Neilson, G.~H. and Pomphrey, N. and Reiman, A. and Spong, D. and Strickler, D. and Boozer, A. and Cooper, W.~A. and Goldston, R. and Hatcher, R. and Isaev, M. and Kessel, C. and Lewandowski, J. and Lyon, J.~F. and Merkel, P. and Mynick, H. and Nelson, B.~E. and Nuehrenberg, C. and Redi, M. and Reiersen, W. and Rutherford, P. and Sanchez, R. and Schmidt, J. and White, R.~B.},
	journal       = PPCF,
	year          = 2001,
	volume        = {43},
	pages         = {237},
	doi           = {10.1088/0741-3335/43/12A/318}
}

@article{geiger15,
	title         = "{Physics in the magnetic configuration space of W7-X}",
	author        = {Geiger, J. and Beidler, C.~D. and Feng, Y. and Maa{\ss}berg, H. and Marushchenko, N.~B. and Turkin, Y.},
	journal       = PPCF,
	year          = 2014,
	volume        = {57},
	pages         = {014004},
	doi           = {10.1088/0741-3335/57/1/014004}
}

@article{thienpondt25,
	title         = "{Influence of the density gradient on turbulent heat transport at ion-scales: an inter-machine study with the gyrokinetic code stella}",
	author        = {Thienpondt, H. and Garc{\'\i}a-Rega{\~n}a, J.~M. and Calvo, I. and Acton, G. and Barnes, M.},
	journal       = NF,
	year          = 2025,
	volume        = {65},
	pages         = {016062},
	doi           = {10.1088/1741-4326/ad9ab9}
}

@article{lion25,
	title         = "{Stellaris: A high-field quasi-isodynamic stellarator for a prototypical fusion power plant}",
	author        = {Lion, J. and Angl{\`e}s, J.-C. and Bonauer, L. and Ba{\~n}{\'o}n Navarro, A. and Cadena Ceron, S.~A. and Davies, R. and Drevlak, M. and Foppiani, N. and Geiger, J. and Goodman, A. and Guo, W. and Guiraud, E. and Hern{\'a}ndez, F. and Henneberg, S. and Herrero, R. and Hintze, C. and H{\"o}chter, H. and Jelonnek, J. and Jenko, F. and Jorge, R. and Kaiser, M. and Kubie, M. and Lascas Neto, E. and Laqua, H. and Leoni, M. and Lobsien, J.~F. and Maurin, V. and Merlo, A. and Middleton-Gear, D. and Pascu, M. and Plunk, G.~G. and Riva, N. and Savtchouk, M. and Sciortino, F. and Schilling, J. and Shimwell, J. and Di Siena, A. and Slade, R. and Stange, T. and Todd, T.~N. and Wegener, L. and Wilms, F. and Xanthopoulos, P. and Zheng, M.},
	journal       = FED,
	year          = 2025,
	volume        = {214},
	pages         = {114868},
	doi           = {10.1016/j.fusengdes.2025.114868}
}

@article{hegna25,
	title         = "{The Infinity Two fusion pilot plant baseline plasma physics design}",
	author        = {Hegna, C.~C. and Anderson, D.~T. and Andrew, E.~C. and Ayilaran, A. and Bader, A. and Bohm, T.~D. and Camacho Mata, K. and Canik, J.~M. and Carbajal, L. and Cerfon, A. and others},
	journal       = JPP,
	year          = 2025,
	volume        = {91},
	pages         = {E76},
	doi           = {10.1017/S0022377825000364}
}

@article{xanthopoulos20,
	title         = "{Turbulence Mechanisms of Enhanced Performance Stellarator Plasmas}",
	author        = {Xanthopoulos, P. and Bozhenkov, S.~A. and Beurskens, M.~N. and Smith, H.~M. and Plunk, G.~G. and Helander, P. and Beidler, C.~D. and Alcus{\'o}n, J.~A. and Alonso, A. and Dinklage, A. and Ford, O. and Fuchert, G. and Geiger, J. and Proll, J.~H.~E. and Pueschel, M.~J. and Turkin, Y. and Warmer, F. and the W7-X Team},
	journal       = PRL,
	year          = 2020,
	volume        = {125},
	pages         = {075001},
	doi           = {10.1103/PhysRevLett.125.075001}
}

@article{plunk19,
	title         = "{Stellarators Resist Turbulent Transport on the Electron Larmor Scale}",
	author        = {Plunk, G.~G. and Xanthopoulos, P. and Weir, G.~M. and Bozhenkov, S.~A. and Dinklage, A. and Fuchert, G. and Geiger, J. and Hirsch, M. and Hoefel, U. and Jakubowski, M. and Langenberg, A. and Pablant, N. and Pasch, E. and Stange, T. and Zhang, D. and the W7-X Team},
	journal       = PRL,
	year          = 2019,
	volume        = {122},
	pages         = {035002},
	doi           = {10.1103/PhysRevLett.122.035002}
}

@article{zhu20PRL,
	title         = "{Theory of the Tertiary Instability and the Dimits Shift from Reduced Drift-Wave Models}",
	author        = {Zhu, H. and Zhou, Y. and Dodin, I.~Y.},
	journal       = PRL,
	year          = 2020,
	volume        = {124},
	pages         = {055002},
	doi           = {10.1103/PhysRevLett.124.055002}
}

@article{zhu20JPP,
	title         = "{Theory of the tertiary instability and the Dimits shift within a scalar model}",
	author        = {Zhu, H. and Zhou, Y. and Dodin, I.~Y.},
	journal       = JPP,
	year          = 2020,
	volume        = {86},
	pages         = {905860405},
	doi           = {10.1017/S0022377820000823}
}

@article{hatch22,
	title         = "{Reduced models for ETG transport in the tokamak pedestal}",
	author        = {Hatch, D.~R. and Michoski, C. and Kuang, D. and Chapman-Oplopoiou, B. and Curie, M. and Halfmoon, M. and Hassan, E. and Kotschenreuther, M. and Mahajan, S.~M. and Merlo, G. and Pueschel, M.~J. and Walker, J. and Stephens, C.~D.},
	journal       = PoP,
	year          = 2022,
	volume        = {29},
	pages         = {062501},
	doi           = {10.1063/5.0087403}
}

@article{turica25,
	title         = "{Reconstructions of electron-temperature profiles from EUROfusion Pedestal Database using turbulence models and machine learning}",
	author        = {Turica, L.-P. and Field, A.~R. and Frassinetti, L. and Schekochihin, A.~A. and JET Contributors and the EUROfusion Tokamak Exploitation Team},
	journal       = {arXiv e-prints},
	year          = 2025,
	volume        = {arXiv:2504.17486}
}

@article{cheng81,
	title         = "{Ballooning-mode theory of trapped-electron instabilities in tokamaks}",
	author        = {Cheng, C.~Z. and Chen, L.},
	journal       = NF,
	year          = 1981,
	volume        = {21},
	pages         = {403},
	doi           = {10.1088/0029-5515/21/3/009}
}

@article{adam73,
	title         = "{Localized drift dissipative modes in Tokamaks}",
	author        = {Adam, J.~C. and Laval, G. and Pellat, R.},
	journal       = NF,
	year          = 1973,
	volume        = {13},
	pages         = {47},
	doi           = {10.1088/0029-5515/13/1/006}
}

@article{rodriguez24,
	title         = "{The maximum-J property in quasi-isodynamic stellarators}",
	author        = {Rodr{\'\i}guez, E. and Helander, P. and Goodman, A.~G.},
	journal       = JPP,
	year          = 2024,
	volume        = {90},
	pages         = {905900212},
	doi           = {10.1017/S0022377824000345}
}

@article{proll12,
	title         = "{Resilience of Quasi-Isodynamic Stellarators against Trapped-Particle Instabilities}",
	author        = {Proll, J.~H.~E. and Helander, P. and Connor, J.~W. and Plunk, G.~G.},
	journal       = PRL,
	year          = 2012,
	volume        = {108},
	pages         = {245002},
	doi           = {10.1103/PhysRevLett.108.245002}
}

@article{patel25,
	title         = "{The impact of $\vE\times \vB$ shear on microtearing-based transport in spherical tokamaks}",
	author        = {Patel, B. and Hardman, M.~R. and Kennedy, D. and Giacomin, M. and Dickinson, D. and Roach, C.~M.},
	journal       = NF,
	year          = 2025,
	volume        = {65},
	pages         = {026063},
	doi           = {10.1088/1741-4326/ada627}
}

@article{rodriguez25,
	title         = "{The kinetic ion-temperature-gradient-driven instability and its localisation}",
	author        = {Rodr{\'\i}guez, E. and Zocco, A.},
	journal       = JPP,
	year          = 2025,
	volume        = {91},
	pages         = {E21},
	doi           = {10.1017/S0022377824001120}
}

@article{rogers05,
	title        = {{Noncurvature-driven modes in a transport barrier}},
	author       = {{Rogers}, B.~N. and {Dorland}, W. and {Kotschenreuther}, M.},
	year         = 2005,
	journal      = PoP,
	volume       = 12,
	pages        = 062511,
	doi          = {TO BE ADDED!!!!}
}

@phdthesis{acton25thesis,
	title        = {Modelling and optimising turbulence in 3D magnetic geometries for enhanced microstability},
	author       = {{Acton}, G. O.},
	year         = 2025,
	school       = {University of Oxford (URL: \url{https://ora.ox.ac.uk/objects/uuid:78daaa1f-6954-4e00-b801-0882691e5997})}
}

@article{martin18,
	title         = "{The parallel boundary condition for turbulence simulations in low magnetic shear devices}",
	author        = {Martin, M.~F. and Landreman, M. and Xanthopoulos, P. and Mandell, N.~R. and Dorland, W.},
	journal       = PPCF,
	year          = 2018,
	volume        = {60},
	pages         = {095008},
	doi           = {10.1088/1361-6587/aad38a}
}

@article{sanchez20,
	title         = "{Nonlinear gyrokinetic PIC simulations in stellarators with the code EUTERPE}",
	author        = {S{\'a}nchez, E. and Mishchenko, A. and Garc{\'\i}a-Rega{\~n}a, J.~M. and Kleiber, R. and Bottino, A. and Villard, L. and the W7-X Team},
	journal       = JPP,
	year          = 2020,
	volume        = {86},
	pages         = {855860501},
	doi           = {10.1017/S0022377820000926}
}

@article{alonso17,
	title         = "{Observation of Oscillatory Radial Electric Field Relaxation in a Helical Plasma}",
	author        = {Alonso, J.~A. and S{\'a}nchez, E. and Calvo, I. and Velasco, J.~L. and McCarthy, K.~J. and Chmyga, A. and Eliseev, L.~G. and Estrada, T. and Kleiber, R. and Krupnik, L.~I. and Melnikov, A.~V. and Monreal, P. and Parra, F.~I. and Perfilov, S. and Zhezhera, A.~I. and the TJ-II Team},
	journal       = PRL,
	year          = 2017,
	volume        = {118},
	pages         = {185002},
	doi           = {10.1103/PhysRevLett.118.185002}
}

@unpublished{gupta25,
	title        = {Scaling laws for the cutoff wavenumber of the short-wavelength, ion-temperature-gradient mode in a Z-pinch},
	author       = {Gupta, O. and Barnes, M. and Parra, F.~I. and Podavini, L. and Zocco, A. and Adkins, T. and Ivanov, P.~G.},
	note         = {in preparation for J. Plasma Phys.},
	year         = 2025
}

@article{hassam90,
	title         = "{Theory of ion temperature gradient instabilities: Thresholds and transport}",
	author        = {Hassam, A.~B. and Antonsen, T.~M., Jr. and Drake, J.~F. and Guzdar, P.~N.},
	journal       = PFB,
	year          = 1990,
	volume        = {2},
	pages         = {1822},
	doi           = {10.1063/1.859454}
}

@article{guo93,
	title         = "{The linear threshold of the ion-temperature-gradient-driven mode}",
	author        = {Guo, S.~C. and Romanelli, F.},
	journal       = PFB,
	year          = 1993,
	volume        = {5},
	pages         = {520},
	doi           = {10.1063/1.860537}
}

@article{kim02,
	title         = "{Dynamics of zonal flow saturation in strong collisionless drift wave turbulence}",
	author        = {Kim, E. and Diamond, P.~H.},
	journal       = PoP,
	year          = 2002,
	volume        = {9},
	pages         = {4530},
	doi           = {10.1063/1.1514641}
}

@article{stonge17,
	title         = "{On non-local energy transfer via zonal flow in the Dimits shift}",
	author        = {St-Onge, D.~A.},
	journal       = JPP,
	year          = 2017,
	volume        = {83},
	pages         = {905830504},
	doi           = {10.1017/S0022377817000708}
}

@article{staebler07,
	title         = "{A theory-based transport model with comprehensive physics}",
	author        = {Staebler, G.~M. and Kinsey, J.~E. and Waltz, R.~E.},
	journal       = PoP,
	year          = 2007,
	volume        = {14},
	pages         = {055909},
	doi           = {10.1063/1.2436852}
}

@article{staebler10,
	title         = "{Electron collisions in the trapped gyro-Landau fluid transport model}",
	author        = {Staebler, G.~M. and Kinsey, J.~E.},
	journal       = PoP,
	year          = 2010,
	volume        = {17},
	pages         = {122309},
	doi           = {10.1063/1.3505308}
}

@article{garbet04,
	title         = "{Physics of transport in tokamaks}",
	author        = {Garbet, X. and Mantica, P. and Angioni, C. and Asp, E. and Baranov, Y. and Bourdelle, C. and Budny, R. and Crisanti, F. and Cordey, G. and Garzotti, L. and Kirneva, N. and Hogeweij, D. and Hoang, T. and Imbeaux, F. and Joffrin, E. and Litaudon, X. and Manini, A. and McDonald, D.~C. and Nordman, H. and Parail, V. and Peeters, A. and Ryter, F. and Sozzi, C. and Valovic, M. and Tala, T. and Thyagaraja, A. and Voitsekhovitch, I. and Weiland, J. and Weisen, H. and Zabolotsky, A. and the JET EFDA Contributors},
	journal       = PPCF,
	year          = 2004,
	volume        = {46},
	pages         = {B557},
	doi           = {10.1088/0741-3335/46/12B/045}
}

@article{mantica09,
	title         = "{Experimental Study of the Ion Critical-Gradient Length and Stiffness Level and the Impact of Rotation in the JET Tokamak}",
	author        = {Mantica, P. and Strintzi, D. and Tala, T. and Giroud, C. and Johnson, T. and Leggate, H. and Lerche, E. and Loarer, T. and Peeters, A.~G. and Salmi, A. and Sharapov, S. and Van Eester, D. and de Vries, P.~C. and Zabeo, L. and Zastrow, K.-D.},
	journal       = PRL,
	year          = 2009,
	volume        = {102},
	pages         = {175002},
	doi           = {10.1103/PhysRevLett.102.175002}
}

@article{bourdelle07,
	title         = "{A new gyrokinetic quasilinear transport model applied to particle transport in tokamak plasmas}",
	author        = {Bourdelle, C. and Garbet, X. and Imbeaux, F. and Casati, A. and Dubuit, N. and Guirlet, R. and Parisot, T.},
	journal       = PoP,
	year          = 2007,
	volume        = {14},
	pages         = {112501},
	doi           = {10.1063/1.2800869}
}

@article{bourdelle16,
	title         = "{Core turbulent transport in tokamak plasmas: bridging theory and experiment with {QuaLiKiz}}",
	author        = {Bourdelle, C. and Citrin, J. and Baiocchi, B. and Casati, A. and Cottier, P. and Garbet, X. and Imbeaux, F. and the JET Contributors},
	journal       = PPCF,
	year          = 2016,
	volume        = {58},
	pages         = {014036},
	doi           = {10.1088/0741-3335/58/1/014036}
}

@article{goodman24,
	title         = "{Quasi-isodynamic stellarators with low turbulence as fusion reactor candidates}",
	author        = {Goodman, A.~G. and Xanthopoulos, P. and Plunk, G.~G. and Smith, H{\aa}kan and N{\"u}hrenberg, C. and Beidler, C.~D. and Henneberg, S.~A. and Roberg-Clark, G. and Drevlak, M. and Helander, P.},
	journal       = PRXE,
	year          = 2024,
	volume        = {3},
	pages         = {023010},
	doi           = {10.1103/PRXEnergy.3.023010}
}

@article{kim24,
	title         = "{Optimization of nonlinear turbulence in stellarators}",
	author        = {Kim, P. and Buller, S. and Conlin, R. and Dorland, W. and Dudt, D.~W. and Gaur, R. and Jorge, R. and Kolemen, E. and Landreman, M. and Mandell, N.~R. and Panici, D.},
	journal       = JPP,
	year          = 2024,
	volume        = {90},
	pages         = {905900210},
	doi           = {10.1017/S0022377824000369}
}

@article{landreman25,
	title         = "{How does ion temperature gradient turbulence depend on magnetic geometry? Insights from data and machine learning}",
	author        = {Landreman, M. and Choi, J.~Y. and Alves, C. and Balaprakash, P. and Churchill, M. and Conlin, R. and Roberg-Clark, G.},
	journal       = JPP,
	year          = 2025,
	volume        = {91},
	pages         = {E120},
	doi           = {10.1017/S0022377825100536}
}

@article{weir21,
	title         = "{Heat pulse propagation and anomalous electron heat transport measurements on the optimized stellarator W7-X}",
	author        = {Weir, G.~M. and Xanthopoulos, P. and Hirsch, M. and H{\"o}fel, U. and Stange, T. and Pablant, N. and Grulke, O. and {\"A}k{\"a}slompolo, S. and Alcus{\'o}n, J. and Bozhenkov, S. and Beurskens, M. and Dinklage, A. and Fuchert, G. and Geiger, J. and Landreman, M. and Langenberg, A. and Lazerson, S. and Marushchenko, N. and Pasch, E. and Schilling, J. and Scott, E.~R. and Turkin, Y. and Klinger, T. and the W7-X Team},
	journal       = NF,
	year          = 2021,
	volume        = {61},
	pages         = {056001},
	doi           = {10.1088/1741-4326/abea55}
}

@article{wilms24,
	title         = "{Global gyrokinetic analysis of {Wendelstein~7-X} discharge: unveiling the importance of trapped-electron-mode and electron-temperature-gradient turbulence}",
	author        = {Wilms, F. and Ba{\~n}{\'o}n Navarro, A. and Windisch, T. and Bozhenkov, S. and Warmer, F. and Fuchert, G. and Ford, O. and Zhang, D. and Stange, T. and Jenko, F. and the W7-X Team},
	journal       = NF,
	year          = 2024,
	volume        = {64},
	pages         = {096040},
	doi           = {10.1088/1741-4326/ad6675}
}

@article{zhang92,
	title         = "{Edge turbulence scaling with shear flow}",
	author        = {{Zhang}, Y.~Z. and {Mahajan}, S.~M.},
	journal       = PFB,
	year          = 1992,
	volume        = {4},
	pages         = {1385},
	doi           = {10.1063/1.860095}
}

@article{hatch18,
	title         = "{Flow shear suppression of pedestal ion temperature gradient turbulence---A first principles theoretical framework}",
	author        = {{Hatch}, D.~R. and {Hazeltine}, R.~D. and {Kotschenreuther}, M.~K. and {Mahajan}, S.~M.},
	journal       = PPCF,
	year          = 2018,
	volume        = {60},
	pages         = {084003},
	doi           = {10.1088/1361-6587/aac7a7}
}

@article{casson09,
	title         = "{Anomalous parallel momentum transport due to $E \times B$ flow shear in a tokamak plasma}",
	author        = {{Casson}, F.~J. and {Peeters}, A.~G. and {Camenen}, Y. and {Hornsby}, W.~A. and {Snodin}, A.~P. and {Strintzi}, D. and {Szepesi}, G.},
	journal       = PoP,
	year          = 2009,
	volume        = {16},
	pages         = {092303},
	doi           = {10.1063/1.3227650}
}

@article{kinsey05,
	title         = "{Nonlinear gyrokinetic turbulence simulations of $E \times B$ shear quenching of transport}",
	author        = {{Kinsey}, J.~E. and {Waltz}, R.~E. and {Candy}, J.},
	journal       = PoP,
	year          = 2005,
	volume        = {12},
	pages         = {062302},
	doi           = {10.1063/1.1920327}
}

@article{waltz07,
	title         = "{Gyrokinetic theory and simulation of angular momentum transport}",
	author        = {{Waltz}, R.~E. and {Staebler}, G.~M. and {Candy}, J. and {Hinton}, F.~L.},
	journal       = PoP,
	year          = 2007,
	volume        = {14},
	pages         = {122507},
	doi           = {10.1063/1.2824376}
}

@article{mcmillan19,
	title         = "{Simulating background shear flow in local gyrokinetic simulations}",
	author        = {{McMillan}, B.~F. and {Ball}, J. and {Brunner}, S.},
	journal       = PPCF,
	year          = 2019,
	volume        = {61},
	pages         = {055006},
	doi           = {10.1088/1361-6587/ab06a4}
}

@article{sugama05,
	title         = "{Dynamics of zonal flows in helical systems}",
	author        = {{Sugama}, H. and {Watanabe}, T.-H.},
	journal       = PRL,
	year          = 2005,
	volume        = {94},
	pages         = {115001},
	doi           = {10.1103/PhysRevLett.94.115001}
}

@article{sugama06,
	title         = "{Collisionless damping of zonal flows in helical systems}",
	author        = {{Sugama}, H. and {Watanabe}, T.-H.},
	journal       = PoP,
	year          = 2006,
	volume        = {13},
	pages         = {012501},
	doi           = {10.1063/1.2149311}
}

@article{mishchenko08,
	title         = "{Collisionless dynamics of zonal flows in stellarator geometry}",
	author        = {{Mishchenko}, A. and {Helander}, P. and {K{\"o}nies}, A.},
	journal       = PoP,
	year          = 2008,
	volume        = {15},
	pages         = {072309},
	doi           = {10.1063/1.2963085}
}

@article{watanabe08,
	title         = "{Reduction of turbulent transport with zonal flows enhanced in helical systems}",
	author        = {{Watanabe}, T.-H. and {Sugama}, H. and {Ferrando-Margalet}, S.},
	journal       = PRL,
	year          = 2008,
	volume        = {100},
	pages         = {195002},
	doi           = {10.1103/PhysRevLett.100.195002}
}

@article{monreal16,
	title         = "{Residual zonal flows in tokamaks and stellarators at arbitrary wavelengths}",
	author        = {{Monreal}, P. and {Calvo}, I. and {S{\'a}nchez}, E. and {Parra}, F.~I. and {Bustos}, A. and {K{\"o}nies}, A. and {Kleiber}, R. and {G{\"o}rler}, T.},
	journal       = PPCF,
	year          = 2016,
	volume        = {58},
	pages         = {045018},
	doi           = {10.1088/0741-3335/58/4/045018}
}

@article{nicolau21,
	title         = "{Global gyrokinetic simulation with kinetic electron for collisionless damping of zonal flow in stellarators}",
	author        = {{Nicolau}, J.~H. and {Choi}, G. and {Fu}, J. and {Liu}, P. and {Wei}, X. and {Lin}, Z.},
	journal       = NF,
	year          = 2021,
	volume        = {61},
	pages         = {126041},
	doi           = {10.1088/1741-4326/ac31da}
}

@article{singh22,
	title         = "{Global gyrokinetic simulations of electrostatic microturbulent transport using kinetic electrons in {LHD} stellarator}",
	author        = {{Singh}, T. and {Nicolau}, J.~H. and {Lin}, Z. and {Sharma}, S. and {Sen}, A. and {Kuley}, A.},
	journal       = NF,
	year          = 2022,
	volume        = {62},
	pages         = {126006},
	doi           = {10.1088/1741-4326/ac906d}
}

@article{xanthopoulos11,
	title         = "{Zonal flow dynamics and control of turbulent transport in stellarators}",
	author        = {{Xanthopoulos}, P. and {Mishchenko}, A. and {Helander}, P. and {Sugama}, H. and {Watanabe}, T.-H.},
	journal       = PRL,
	year          = 2011,
	volume        = {107},
	pages         = {245002},
	doi           = {10.1103/PhysRevLett.107.245002}
}

@electronic{adkins26data,
	title         = "{Data for ``Asymptotic scaling theory of electrostatic turbulent transport in magnetised fusion plasmas''}",
	author        = {{Adkins}, T. and {Abel}, I.~G. and {Barnes}, M. and {Buller}, S. and {Dorland}, W. and {Ivanov}, P.~G. and {Meyrand}, R. and {Parra}, F.~I. and {Nies}, R. and {Schekochihin}, A.~A. and {Squire}, J.},
	year          = 2026,
	publisher     = {Princeton Plasma Physics Laboratory, Princeton University},
	doi           = {10.34770/j3ee-2g49}
}

@article{mackenbach25,
	title         = "{On the curvature-driven ion-temperature-gradient instability and its available energy}",
	author        = {{Mackenbach}, R. and {Helander}, P. and {Landreman}, M. and {Brunner}, S. and {Proll}, J.},
	journal       = JPP,
	year          = 2025,
	volume        = {91},
	pages         = {E144},
	doi           = {10.1017/S0022377825100846}
}

@article{turnbull99,
	title         = "{Improved magnetohydrodynamic stability through optimization of higher order moments in cross-section shape of tokamaks}",
	author        = {{Turnbull}, A.~D. and {Lin-Liu}, Y.~R. and {Miller}, R.~L. and {Taylor}, T.~S. and {Todd}, T.~N.},
	journal       = PoP,
	year          = 1999,
	volume        = {6},
	pages         = {1113},
	doi           = {10.1063/1.873380}
}
